\documentclass[apj,twocolumn]{openjournal}
\usepackage{amsmath}
\usepackage{booktabs}
\usepackage{multirow}
\usepackage{color}
\usepackage{soul}

\usepackage[dvipsnames]{xcolor} 

\usepackage[breaklinks,colorlinks,citecolor=blue,urlcolor=blue]{hyperref}

\newcommand{\red}[1]{\textcolor{red}{#1}}
\newcommand{\green}[1]{\textcolor{green}{#1}}
\newcommand{\blue}[1]{\textcolor{blue}{#1}}

\defcitealias{Shen2018}{S18}

\renewcommand{\arraystretch}{1.2}  
\usepackage{listings}
\usepackage{color}
\definecolor{dkgreen}{rgb}{0,0.6,0}
\definecolor{gray}{rgb}{0.5,0.5,0.5}
\definecolor{mauve}{rgb}{0.58,0,0.82}
\definecolor{golden}{rgb}{0.86,0.65,0.01}
\usepackage{orcidlink}

\begin{document}


\title{The fastest stars in the Galaxy} 

\author{\vspace{-1.2cm}Kareem El-Badry\,\orcidlink{0000-0002-6871-1752}$^{1,2,3}$}
\author{Ken J. Shen\,\orcidlink{0000-0002-9632-6106}$^{4}$}
\author{Vedant Chandra\,\orcidlink{0000-0002-0572-8012}$^{1}$}
\author{Evan B. Bauer\,\orcidlink{0000-0002-4791-6724}$^{1}$}
\author{Jim Fuller\,\orcidlink{0000-0002-4544-0750}$^3$}
\author{Jay Strader\,\orcidlink{0000-0002-1468-9668}$^{5}$}
\author{Laura Chomiuk\,\orcidlink{0000-0002-8400-3705}$^{5}$}
\author{Rohan P. Naidu\,\orcidlink{0000-0003-3997-5705}$^{6,\dagger}$}
\author{Ilaria Caiazzo\,\orcidlink{0000-0002-4770-5388}$^{3}$}
\author{Antonio C. Rodriguez\,\orcidlink{0000-0003-4189-9668}$^{3}$}
\author{Pranav Nagarajan\,\orcidlink{0000-0002-1386-0603}$^{3}$}
\author{Natsuko Yamaguchi\,\orcidlink{0000-0001-6970-1014}$^{3}$}
\author{Zachary P. Vanderbosch\,\orcidlink{0000-0002-0853-3464}$^{3}$}
\author{Benjamin R. Roulston\,\orcidlink{0000-0002-9453-7735}$^{3}$}
\author{Jan van Roestel\,\orcidlink{0000-0002-2626-2872}$^{7}$}
\author{Boris G\"ansicke\,\orcidlink{0000-0002-2761-3005}$^{8}$}
\author{Jiwon Jesse Han\,\orcidlink{0000-0002-6800-5778}$^{1}$}
\author{Kevin B. Burdge\,\orcidlink{0000-0002-7226-836X}$^{6}$}
\author{Alexei V. Filippenko\,\orcidlink{0000-0003-3460-0103}$^{9}$}
\author{Thomas G. Brink\,\orcidlink{0000-0001-5955-2502}$^{9}$}
\author{WeiKang Zheng\,\orcidlink{0000-0002-2636-6508}$^{9}$}

\affiliation{$^1$Center for Astrophysics $|$ Harvard \& Smithsonian, 60 Garden Street, Cambridge, MA 02138, USA}
\affiliation{$^2$Max-Planck Institute for Astronomy, K\"onigstuhl 17, D-69117 Heidelberg, Germany}
\affiliation{$^3$Department of Astronomy, California Institute of Technology, 1200 E. California Blvd., Pasadena, CA 91125, USA}
\affiliation{$^4$Department of Astronomy, and Theoretical Astrophysics Center, University of California, Berkeley, CA 94720-3411, USA}
\affiliation{$^5$Center for Data Intensive and Time Domain Astronomy, Department of Physics and Astronomy, Michigan State University, East Lansing, MI 48824, USA}
\affiliation{$^6$MIT Kavli Institute for Astrophysics and Space Research, 77 Massachusetts Ave., Cambridge, MA 02139, USA}
\affiliation{$^7$Department of Physics, University of Warwick, Gibbet Hill, Coventry CV4 7AL, UK}
\affiliation{$^8$Anton Pannekoek Institute for Astronomy, University of Amsterdam, 1090 GE Amsterdam, The Netherlands}
\affiliation{$^9$Department of Astronomy, University of California, Berkeley, CA 94720-3411, USA}

\thanks{$\dagger$NASA Hubble Fellow}
\email{Corresponding author: kelbadry@caltech.edu}




\begin{abstract}
We report a spectroscopic search for hypervelocity white dwarfs (WDs) that are runaways from Type Ia supernovae (SNe Ia) and related thermonuclear explosions. Candidates are selected from {\it Gaia} data with high tangential velocities and blue colors. We find six new runaways, including four stars with radial velocities (RVs) $>1000\,\rm km\,s^{-1}$ and total space velocities $\gtrsim 1300\,\rm km\,s^{-1}$. These are most likely the surviving donors from double-degenerate binaries in which the other WD exploded. The other two objects have lower minimum velocities, $\gtrsim 600\,\rm km\,s^{-1}$, and may have formed through a different mechanism, such as pure deflagration of a WD in a Type Iax supernova. The four fastest stars are hotter and smaller than the previously known ``D$^6$ stars,'' with effective temperatures ranging from $\sim$20,000 to $\sim$130,000\,$\rm K$ and radii of $\sim 0.02$--0.10\,$R_{\odot}$. Three of these have carbon-dominated atmospheres, and one has a helium-dominated atmosphere. Two stars have RVs of $-1694$ and $-2285\,\rm km\,s^{-1}$ -- the fastest systemic stellar RVs ever measured. Their inferred birth velocities, $\sim 2200$--2500\,$\rm km\,s^{-1}$, imply that both WDs in the progenitor binary had masses $>1.0\,M_{\odot}$. The high observed velocities suggest that a dominant fraction of the  observed hypervelocity WD population comes from double-degenerate binaries whose total mass significantly exceeds the Chandrasekhar limit. However, the two nearest and faintest D$^6$ stars have the lowest velocities and masses, suggesting that observational selection effects favor rarer, higher-mass stars. A significant population of fainter low-mass runaways may still await discovery. We infer a birth rate of $\rm D^6$ stars that is consistent with the SN Ia rate. The birth rate is poorly constrained, however, because the luminosities and lifetimes of $\rm D^6$ stars are uncertain.

\end{abstract}

\maketitle

\section{Introduction}
\label{sec:intro}

Despite several decades of investigation, the dominant progenitor channel of Type Ia supernovae (SNe~Ia) remains uncertain \citep[e.g.,][]{Maoz2014, Livio2018}. A promising class of models involves the gravitational wave-driven inspiral of a double white dwarf (WD) binary, culminating in the detonation of carbon in one component's core. Such a detonation can plausibly be achieved in a wide range of double WD binaries if it is preceded by detonation of helium near the surface of the more massive WD (the ``accretor''), which produces a converging shock that detonates carbon in the core \citep{Livne1990, Fink2010, Shen2014}. Such a ``double detonation'' (first helium, then carbon)  could occur either after the build-up of a helium shell on the accreting WD through stable mass transfer, or dynamically during the coalescence of a WD binary in which mass transfer becomes unstable \citep{Guillochon2010, Dan2011}. An attractive feature of the double-detonation scenario is that it does not necessarily require the total mass of the WD binary to approach the Chandrasekhar limit, but can potentially lead to a SN~Ia as long as the accreting WD has a CO core with mass $\gtrsim 0.8\,M_{\odot}$ \citep[e.g.,][]{Sim2010, Shen2018b}. This is attractive because super-Chandrasekhar-mass WD binaries are expected to be rare, with a predicted coalescence rate an order of magnitude lower than that of all WD+WD binaries \citep[e.g.,][]{Nelemans2001, Yungelson2017}. 

The fate of the lower-mass WD  (the ``donor'') after the accretor's detonation is uncertain. It is possible that it will be destroyed, either in a second double detonation triggered by the first, or, if the accretor only detonates late in the merger, by tides. However, most models predict that the lower-mass WD will survive in some or all binary configurations \citep{Pakmor2013, Papish2015, Tanikawa2019, Pakmor2022, Burmester2023}.
In this case, that WD will flee the scene with a velocity similar to its pre-explosion orbital velocity. This is predicted to be quite large -- in the range of 1000--2000\,$\rm km\,s^{-1}$ for $\sim 1\,M_{\odot}$ accretors and 0.2--0.8\,$M_{\odot}$ donors. These velocities significantly exceed the Milky Way's escape velocity, and so the runaway WDs will be launched into intergalactic space, traveling at a rate of 1--2 kpc\,Myr$^{-1}$. These runaway WDs are smoking guns of double-degenerate detonations. Constraints on their birth rate hold promise to determine the fraction of SNe~Ia that come from a double-degenerate channel, and measurement of their surface abundances can potentially constrain the yields of SNe~Ia. 

\citet{Shen2018b} termed the scenario in which unstable mass transfer in a WD+WD binary leads to a helium-shell detonation followed by carbon detonation in the more massive WD the ``D$^6$ scenario'' (dynamically driven, double-degenerate, double-detonation). Using astrometry from {\it Gaia} DR2, \citet[][hereafter S18]{Shen2018} selected candidate $\rm D^6$ stars with large apparent tangential velocities and well-constrained parallaxes. Their spectroscopic follow-up observations revealed three sources with unusual and very similar spectra, which they named D6-1, D6-2, and D6-3. All three objects have atmospheres devoid of hydrogen and spectra dominated by metal lines. They fall in a tight clump in the color-magnitude diagram between the main sequence and the WD cooling track, corresponding to temperatures of order 7000\,K and radii of $\sim 0.2\,R_\odot$. These objects are much larger and puffier than normal WDs, possibly because they were inflated by tidal heating during the merger and/or energy injected from the explosion of their companions. In the $\rm D^6$ scenario, the runaway WDs are expected to be free of hydrogen (and possibly also helium) because their outer layers were stripped and transferred to the companion prior to its detonation \citep[e.g.,][]{Shen2013}.
All three objects have inferred tangential velocities above $1000\,\rm km\,s^{-1}$. One of them has a radial velocity (RV) of $\sim 1200\,\rm km\,s^{-1}$; puzzlingly, the other two have RVs near 0. However, the unusual and very similar spectra of the three objects, as well as the high quality of their astrometric solutions, strongly suggests that they are all genuine hypervelocity WDs. 

The three $\rm D^6$ stars discovered by \citetalias{Shen2018} are all relatively bright and nearby, with $G$-band apparent magnitudes of 17.0, 17.4, and 18.2, distances of 0.8--2.5\,kpc, and well-constrained parallaxes. 
Given this and the fact that the objects were discovered immediately after {\it Gaia } DR2, it seemed likely that a larger population of $\rm D^6$ stars would soon be discovered. However, despite deeper searches in the intervening 5\,yr \citep[e.g.,][]{Raddi2019, Igoshev2023} and improved astrometry from {\it Gaia} DR3, no additional $\rm D^6$ stars have been discovered. A few other high-velocity WDs with unusual spectra {\it have} been discovered \citep{Raddi2019}, but these have inferred birth velocities of $\sim 600\,\rm km\,s^{-1}$ -- slower than expected in the $\rm D^6$ scenario -- and spectra that differ from those of the three D$^6$ stars. A proposed explanation for these objects is that they are remnants of deflagrations of near-Chandrasekhar-mass WDs, perhaps from single-degenerate binaries \citep{Foley2013, Vennes2017, Raddi2018, Raddi2019}.
In this scenario it is the accretor, not the donor, that is detected as a high-velocity star.

Here we present results from a new search for $\rm D^6$ stars. Since previous work has investigated most of the candidates with well-constrained parallaxes and distances, we expand our search to include objects with large proper motions and parallaxes consistent with zero. We rely on spectroscopic analysis and RVs to distinguish true $\rm D^6$ stars from lower-velocity interlopers. The remainder of this paper is organized as follows. Section~\ref{sec:sample} describes our strategy for selecting candidate hypervelocity WDs from {\it Gaia} astrometry and summarizes our follow-up spectroscopy. In Section~\ref{sec:spec_analysis}, we discuss our spectroscopic analysis of the most interesting objects. Section~\ref{sec:seds} is concerned with estimating radii from broadband SEDs, while Section~\ref{sec:kinematic_modeling} focuses on modeling the stars' trajectories through the Galaxy. In Section~\ref{sec:census}, we compare the newly discovered objects to other known hypervelocity WDs, and Section~\ref{sec:d6_masses} estimates their masses from their inferred birth velocities. We discuss the implications of our results in Section~\ref{sec:disc} and conclude in Section~\ref{sec:conclusions}.

\section{Search for hypervelocity WDs with {\it Gaia} }
\label{sec:sample}

\subsection{Candidate sample}
\label{sec:candidates}
We wish to select stars with large tangential velocities as inferred by their proper motions and parallaxes. We use astrometry from {\it Gaia} DR3 \citep{GaiaCollaboration2021, GaiaCollaboration2022}. Given a measured proper motion, $\mu$, and parallax, $\varpi$, the implied tangential velocity is 
\begin{equation}
    \label{vperp}
    v_{\perp}=4.74\,{\rm km\,s^{-1}}\times\left(\frac{\mu}{{\rm mas}\,{\rm yr}^{-1}}\right)\left(\frac{\varpi}{{\rm mas}}\right)^{-1}.
\end{equation}
For random trajectories, we expect the tangential velocity to be of comparable magnitude to the three-dimensional (3D) velocity.\footnote{In particular, for random trajectories the ratio $x=v_{\perp}/v_{\rm 3D}$ should be distributed as $p(x) = x/\sqrt{1-x^2}$, with a mean of $\pi/4\approx 0.79$ and a median of $\sqrt{3}/2=0.87$. The ratio $|{\rm RV}|/v_{\rm 3D}$ should be uniformly distributed between 0 and 1 \citep[e.g.,][]{Nottale2018}. We do not expect trajectories to be completely random since all stars are launched from the Galaxy and the Galaxy is rotating, but these ratios still provide a useful heuristic for interpreting observed radial and tangential velocities.} With precise measurements of $v_\perp$, most hypervelocity WDs could thus be selected with a  simple cut of  $v_\perp > 800\,\rm km\,s^{-1}$: fast enough to exclude normal stars bound to the Milky Way and slow enough to include typical stars produced by the $\rm D^6$ scenario unless their trajectories happen to be aligned with our line of sight.  

Astrometric errors can cause contamination of high-velocity candidate samples with foreground lower-velocity false-positives. 
For the objects of interest here, the proper motions are always well-constrained, so parallax uncertainties are the limiting factor in measuring tangential velocities. The simplest way to limit contamination is to only consider sources with well-constrained parallaxes. This was the strategy employed by \citetalias{Shen2018}, who required $\varpi/\sigma_{\varpi} >3$. The three $\rm D^6$ stars they identified have $\varpi/\sigma_{\varpi} = 18.2$, 7.6, and 4.3 in {\it Gaia} DR3, and thus even the $3\sigma$ lower limits on their tangential velocities exceed $800\,\rm km\,s^{-1}$.

We expect additional $\rm D^6$ stars to exist at larger distance than the objects discovered by \citetalias{Shen2018}. These objects will likely both be fainter and have smaller parallaxes than the known $\rm D^6$ stars, such that their tangential velocities have large uncertainties. However, they can still be distinguished from contaminants on the basis of their hydrogen-free spectra and high RVs. 

Candidates for such objects are selected as follows. We require that the best-fit tangential velocity, $v_\perp$, exceeds $600\,\rm km\,s^{-1}$, and that a proxy for its $1\sigma$ lower limit, $v_{\perp,{\rm lower}}=4.74\mu/\left(\varpi+\sigma_{\varpi}\right)$, exceeds $400\,\rm km\,s^{-1}$.\footnote{We neglect the uncertainty in $\mu$ in calculating $v_{\perp,\rm \,lower}$ here, since its contribution to the uncertainty in $v_{\perp}$ is always small compared to the uncertainty in $\varpi$. $v_{\perp,\rm lower}$ is not, strictly speaking, the $1\sigma$ lower limit on $v_{\perp}$, because the inversion of parallax is a nonlinear transformation \citep[e.g.][]{Bailer-Jones2015, Igoshev2016}. In the limit of low-significance parallax measurements, the inferred distance and tangential velocity depend unavoidably on the adopted distance prior. Our kinematic modeling of the spectroscopically confirmed candidates (Section~\ref{sec:kinematic_modeling}) uses a Bayesian analysis, including exploration of the sensitivity of our constraint to the adopted distance prior. For simplicity, we use $v_{\perp,\rm lower}$ to select initial candidates.} We require an apparent magnitude $G< 20$ in the interest of efficient follow-up spectroscopy. Most sources with well-constrained parallaxes and apparently large proper motions have already been investigated in other work \citep[e.g.,][]{Raddi2019, Igoshev2023}; they are primarily normal halo stars with underestimated parallaxes. We therefore explicitly target sources with $\varpi/\sigma_{\varpi} < 5$, which are primarily WDs.  To minimize contamination from sources with spurious astrometry, we require \texttt{ruwe} $<1.4$; this filters out sources with poor astrometric goodness-of-fit compared to typical {\it Gaia} sources with similar magnitude and color. We also adopt a proper-motion lower limit of $\mu > 50\,\rm mas\,yr^{-1}$, which corresponds to $v_\perp > 710\,\rm km\,s^{-1}$ at a distance of 3\,kpc, or $v_\perp > 1185\,\rm km\,s^{-1}$ at a distance of 5\,kpc. The proper motion cut dramatically reduces contamination from distant sources with parallaxes near zero.

\subsubsection{Color cut}
In addition to the astrometric cuts described above, we also use a color cut, $G_{\rm BP}-G_{\rm RP} < 0.5$\,mag. This is motivated by several considerations. First, the three $\rm D^6$ stars discovered by \citetalias{Shen2018} are sufficiently blue to pass this cut. Evolutionary models for temporarily inflated WDs suggest that after a thermonuclear transient, these objects will first expand and cool, and then contract and heat up \citep{Zhang2019, Bauer2019}. This suggests that older analogs of the three known $\rm D^6$ stars should be bluer than they are. We cut on measured color without any extinction correction, because the poorly constrained distances to stars in our sample make extinction corrections nontrivial.

Most importantly, blue stars are relatively rare: among the 1058 million stars in the {\it Gaia} archive with $G<20$\,mag, fewer than 6 million have   $G_{\rm BP}-G_{\rm RP} < 0.5$\,mag. This means that focusing on blue stars dramatically reduces the number of contaminants. The price to pay for this increased efficiency is that the search is not sensitive to red sources, including intrinsically blue stars with significant foreground extinction, stars that are red because they are cool ($T_{\rm eff} \lesssim 6000\,\rm K$), and stars enshrouded in dust. 

\subsubsection{ADQL query}
\label{sec:adql}
The selection described above is implemented in the following ADQL query:

\begin{lstlisting}
select * from gaiadr3.gaia_source
where phot_g_mean_mag < 20
and ruwe < 1.4
and pm > 50
and (4.74*pm/(parallax + parallax_error) > 400 or (parallax + parallax_error) < 0)
and (4.74*pm/parallax > 600 or parallax < 0)
and bp_rp < 0.5
and parallax_over_error < 5
\end{lstlisting}
This query returns 25 sources, which are listed in Table~\ref{tab:query}. Two were rejected because they have a bright neighbor and likely have spurious colors and parallaxes \citep{El-Badry2021, Rybizki2022}. We consider the remaining 23 stars the highest-priority candidates, and we have obtained spectroscopy of 22 of them in order to have high completeness within the parameter space delineated by the cuts described here. 

We also observed an additional 21 candidates with somewhat lower proper motions and/or more significant parallaxes. These candidates are summarized in Table~\ref{tab:query2}. All of our high-confidence, hydrogen-free candidates come from the query above. We expect the false-positive rate to be higher among the sample with lower proper motions (even at fixed $v_\perp$), because there is a larger pool of contaminants with underestimated parallaxes at large distances, and these sources necessarily have lower proper motions.

\begin{table*}
\begin{tabular}{llllllllll}
{\it Gaia} DR3 Source ID & Name & $G$  & $\varpi$ & $\mu$ & $4.74\mu/\varpi$ & $\frac{4.74\mu}{\varpi+\sigma_{\varpi}}$ & RV & verdict & instrument \\
 &  & [mag]  & [mas] & [$\rm mas\,yr^{-1}$] & [$\rm km\,s^{-1}$] & [$\rm km\,s^{-1}$] & [$\rm km\,s^{-1}$] &  &  \\
\hline
6156470924553703552 & J1235-3752 & 19.05 & $-0.10 \pm 0.24$ & 95.2 & -4685 & 3183 & $-1694\pm 10$ & \green{hot $\rm D^6$ star } & MagE \\
2156908318076164224 &            & 18.25 & $0.42 \pm 0.10$ & 212.0 & 2374 & 1922 & $-20\pm 80$ & D6-3  & \citetalias{Shen2018} \\
3335306915849417984 & J0546+0836 & 19.06 & $0.07 \pm 0.31$ & 76.1 & 5289 & 942  & $1200\pm 20$  & \green{hot $\rm D^6$ star} & LRIS/ESI \\
5250394728194220800 & J0927-6335 & 19.37 & $0.13 \pm 0.21$ & 54.9 & 2062 & 764  & $-2285\pm 20$ & \green{hot $\rm D^6$ star} & MagE \\
6164642052589392512 & J1332-3541 & 19.42 & $0.66 \pm 0.54$ & 155.5 & 1112 & 613 & $1090\pm 50$ & \green{hot $\rm D^6$ star} & MagE/LRIS \\
3542263595793124480 &            & 19.18 & $0.67 \pm 0.46$ & 144.3 & 1016 & 604  &   & \red{bright neighbor} &   \\
3804182280735442560 & J1109+0001 & 19.09 & $0.40 \pm 0.31$ & 87.6 & 1051 & 592  & $100\pm 10$ & \blue{LP 40-365 star} & LRIS \\
5703888058542880896 &            & 19.60 & $1.36 \pm 0.32$ & 207.9 & 723 & 586   & $280 \pm 50$  & \red{DA WD} & LRIS \\
5517276097516408576 &            & 19.33 & $0.10 \pm 0.35$ & 55.3 & 2735 & 582   &  & \red{bright neighbor} &   \\
4771417432717575680 &            & 19.93 & $0.13 \pm 0.32$ & 54.2 & 1998 & 571  & $100\pm 50$ & \red{DA WD} & SOAR \\
4546525523392712064 &            & 19.39 & $0.52 \pm 0.32$ & 93.6 & 860 & 534    & $-180\pm 50$ & \red{DA WD} & DBSP \\
6853349473073333632 &            & 19.56 & $0.50 \pm 0.46$ & 107.7 & 1027 & 532 & $120\pm 50$ & \red{DA WD} & SOAR/GMOS  \\
2393804867149529856 &            & 19.78 & $0.13 \pm 0.41$ & 58.8 & 2167 & 513  & $-10\pm 50$ &  \red{DA WD} & SOAR \\
3507697866498687232 & J1311-1846 & 18.26 & $0.60 \pm 0.19$ & 83.1 & 653 & 496    & $55\pm 10$ & \blue{LP 40-365 star} & DBSP/LRIS \\
5998866829060560768 &            & 19.49 & $0.73 \pm 0.44$ & 115.6 & 754 & 468   & $-65\pm 10$ & \red{MS star} & MagE \\
5183592902806605824 &            & 19.81 & $0.73 \pm 0.58$ & 128.2 & 836 & 463   &  &   &   \\
4129413800145771776 &            & 19.30 & $0.65 \pm 0.35$ & 96.2 & 697 & 452    & $-290\pm 80$ & \red{DA WD} & LRIS \\
1512757030058082304 &            & 19.94 & $0.35 \pm 0.33$ & 64.3 & 879 & 450   & $-70\pm 50$ & \red{DA WD} & LRIS \\
1271663056690275712 &            & 19.20 & $0.46 \pm 0.22$ & 65.0 & 669 & 450   & $-120\pm 30$ & \red{DA WD} & SDSS \\
6536783647184624640 &            & 19.94 & $0.07 \pm 0.53$ & 55.9 & 3844 & 444  & $-20\pm 50$  &   \red{DA WD} & SOAR/GMOS  \\
4096984529352659584 &            & 19.60 & $0.40 \pm 0.49$ & 83.0 & 983 & 440   & $60 \pm 30$ & \red{MS star} & LRIS \\
6592388973858801920 &            & 18.45 & $0.39 \pm 0.19$ & 52.3 & 638 & 425   & $160\pm 50$  & \red{DA WD} & SOAR  \\
6688913592127235584 &            & 19.19 & $0.41 \pm 0.29$ & 59.8 & 686 & 404   &  $200 \pm 50 $ & \red{DA WD} & SOAR/GMOS \\
4096331041464263296 &            & 19.99 & $-1.42 \pm 0.82$ & 58.6 & -195 & -463 & $-210\pm 30$ & \red{MS star} & LRIS \\
6703717691563155968 &            & 19.24 & $-0.53 \pm 0.53$ & 52.7 & -471 & -79269 & $-10\pm 50$ & \red{DA WD} & SOAR \\
\end{tabular}
\caption{All {\it Gaia} sources returned by the ADQL query from Section~\ref{sec:adql}, together with results from our follow-up spectroscopy. Sources are sorted by $4.74\mu/\left(\varpi+\sigma_{\varpi}\right)$, the $1\sigma$ lower limit on their tangential velocity.  ``D$^6$ stars'' are suspected runaway donors from WD+WD binaries with velocities $\gtrsim 1000\,\rm km\,s^{-1}$. ``LP 40-365 stars'' are suspected to be partially burned runaway accretors; they have lower velocities and different abundance patterns (see Section~\ref{sec:lp40}). DA~WDs and main-sequence (MS) stars are false positives. }
\label{tab:query}
\end{table*}
\subsection{Follow-up spectroscopy}
We obtained spectra of 39 runaway WD candidates, prioritizing the brightest targets with the highest inferred tangential velocities. Tables~\ref{tab:query} and~\ref{tab:query2} summarize our observations, most of which employed low-resolution spectrographs to check for unusual spectra and/or high RVs. Details about the observing setup and data reduction for each instrument are provided in Appendix~\ref{sec:appendix}.

Several objects turned out to have unusual spectra and/or high RVs; these are discussed in detail individually below. Not surprisingly, the false-positive rate is lowest among the stars with the highest tangential velocities; indeed, 100\% of the 5 sources with the highest $v_{\perp, \rm \, lower}$ are $\rm D^6$ stars!

The most common false positives were normal DA WDs and sdO/B stars. A few main-sequence stars are also among the contaminants, mostly with low metallicity. A significant fraction of the false positives do have moderately large RVs, up to $\sim 450\,\rm km\,s^{-1}$. We suspect that these are mostly stars on halo-like orbits (which have high space velocities compared to most stars in the Solar neighborhood) with underestimated parallaxes, such that their true tangential velocities are smaller than $v_\perp$. These do not necessarily have ``wrong'' parallaxes, but may simply represent the noise tail of the parallax distribution. 

A few of our rejected candidates have been considered as runaway WD candidates in other work on the basis of their {\it Gaia} astrometry. For example, \citet{Igoshev2023} argued that the source {\it Gaia} DR3 5703888058542880896 -- one of our rejected candidates -- has a velocity above 700\,$\rm km\,s^{-1}$ and is likely unbound from the Galaxy. They also classified the sources 6368583523760274176, 6640949596389193856, and 3537042874067950336 as candidates for being unbound. Since we find these sources to have fairly normal spectra that in all cases display hydrogen lines, we consider it more likely that they are the high-velocity (or high-noise) tail of the normal halo WD and sdO/B star population. It is also possible that some high-velocity WDs that are not hydrogen-free were accelerated by other mechanisms besides thermonuclear events, such as dynamical few-body interactions.

\subsection{Completeness of our search}
\label{sec:completeness}
For population modeling, it is important that our search be described by a selection function that can be modeled. The {\it Gaia} proper motion and parallax uncertainties are primarily functions of apparent magnitude and position in the sky \citep[e.g.,][]{Lindegren2021b, Castro-Ginard2023}. It is thus straightforward to ask, given the intrinsic properties of a hypothetical star, whether it would have been detected by our search. The basic parameters that must be satisfied are as follows.

\begin{enumerate}
    \item $G < 20$\,mag,
    \item $G_{\rm BP}-G_{\rm RP} < 0.5$\,mag,
    \item $\mu > 50\,\rm mas\,yr^{-1}$,
    \item $v_{\perp} > 600\,\rm km\,s^{-1}$,
    \item $v_{\perp, \rm lower} \rm > 400\,\rm km\,s^{-1}$.
\end{enumerate}

We can consider a $\rm D^6$ star with absolute magnitude $M_{G,0} = 6$\,mag, similar to the objects discovered by \citetalias{Shen2018}. Such a source will have $G < 20$\,mag to a distance of 6.4\,kpc. If its tangential velocity is 1000 (1500)\,$\rm km\,s^{-1}$, it will satisfy $\mu > 50\,\rm mas\,yr^{-1}$ to a distance of 4.2 (6.3)\,kpc. Given the {\it Gaia} DR3 sky-averaged parallax uncertainties as a function of apparent magnitude, the cut of $v_{\perp, \rm lower} > 400\,\rm km\,s^{-1}$ translates to a distance limit of 4.7 (6.1)\,kpc for a source with tangential velocity 1000 (1500)\,$\rm km\,s^{-1}$. 

Given these considerations, our typical search volume for unreddened $\rm D^6$ stars with $M_{G,0} = 6$\,mag is $\sim 5\, \rm kpc$. For stars with $M_{G,0} = 8\,(10)$\,mag, it is $\sim 2.5$ ($\sim 1.0$)\,kpc. Our search sensitivity is much lower in the Galactic plane, where extinction will both redden sources to $G_{\rm BP}-G_{\rm RP} > 0.5$\,mag and make them fainter than $G < 20$\,mag. This can be forward-modeled using a dust map; here we simply note that we expect the sensitivity to fall precipitously in most locations with Galactic latitude $|b| < 10^\circ$.

\section{Spectroscopic Analysis}
\label{sec:spec_analysis}
\subsection{Spectral models}
Our confirmed hypervelocity WDs have spectra that are both heterogeneous and unusual. This made the process of identifying spectral features and measuring RVs nontrivial. Following \citetalias{Shen2018}, we first tried cross-correlating them with all the spectra in the empirical MILES library \citep{Sanchez-Blazquez2006}. This yielded no plausible matches for any of our objects of interest, because the library contains no objects that even approximately resemble their spectra. 

We therefore calculated synthetic spectra with a variety of atmospheric parameters and abundance patterns and compared them to the observed spectra.
We used ATLAS 12 \citep{Kurucz1970SAOSR, kurucz_model_1979, Kurucz1992} to compute the atmosphere structure and SYNTHE \citep{Kurucz1993sssp} for the radiative-transfer calculations, self-consistently re-computing the atmosphere structure for each set of abundances. These codes model one-dimensional (1D) plane-parallel atmospheres and assume local thermodynamic equilibrium (LTE). We use the linelist maintained by R. Kurucz\footnote{http://kurucz.harvard.edu/linelists.html} and assumed a microturbulent velocity of $2\,\rm km\,s^{-1}$. Spectra were generated at resolution $R=300,000$ and applied instrumental broadening to match the observed data.  We did not attempt to model rotational broadening because the resolution and signal-to-noise ratio (SNR) of our spectra is generally inadequate to constrain it. 

We experimented with a broad but nonexhaustive variety of bulk atmosphere compositions. We expect the approximations made by the Kurucz codes to be suboptimal for the objects in our sample, which have nonstandard compositions and temperatures and surface gravities outside the regime where the codes are well-tested.  Given the complexities involved in the spectral modeling, we do not attempt a quantitative abundance analysis here. Our goals are to (a) measure RVs of the newly discovered objects, and (b) determine the elements responsible for the most obvious features in the observed spectra. This approach allows us to estimate the effective temperatures and bulk atmospheric compositions of the stars, while deferring a more detailed abundance analysis to future work.

In the D$^6$ scenario, the hydrogen and helium layers on the donor's surface will most likely be transferred to the companion before it explodes. The default expectation for these stars' surface compositions is thus an atmosphere consisting mostly of carbon and oxygen, likely with no hydrogen or helium, and possibly contaminated with ejecta from the exploded companion. These considerations motivated us to calculate a grid of atmospheres and model spectra that are $\sim 70$\% carbon and $\sim 30$\% oxygen by mass, consistent with the predictions of \citet{Zhang2019}. After identifying Mg and Si lines in the observed spectrum of one object, we also added 0.2\% by mass of these elements to approximately match the observed line strengths. 

If the donor in the D$^6$ scenario has a helium core -- or if the accretor explodes before all of the donor's surface helium is transferred to the companion -- one might expect the runaway WD to have a helium-dominated atmosphere. This motivated us to calculate another set of spectra with pure helium atmospheres. We additionally calculated models for carbon/oxygen atmospheres with a small amount of helium added in order to estimate upper limits on the helium abundances of the observed objects. 

We cross-correlated each object in our sample with all the synthetic spectra over an RV grid spanning $\pm 5000\,\rm km\,s^{-1}$. We adopted the effective temperature of the model that most closely matches the data in our subsequent analysis. 

\subsection{Discussion of individual objects}
\label{sec:individual}

\subsubsection{J1235-3752:}
\begin{figure*}
    \centering
    \includegraphics[width=\textwidth]{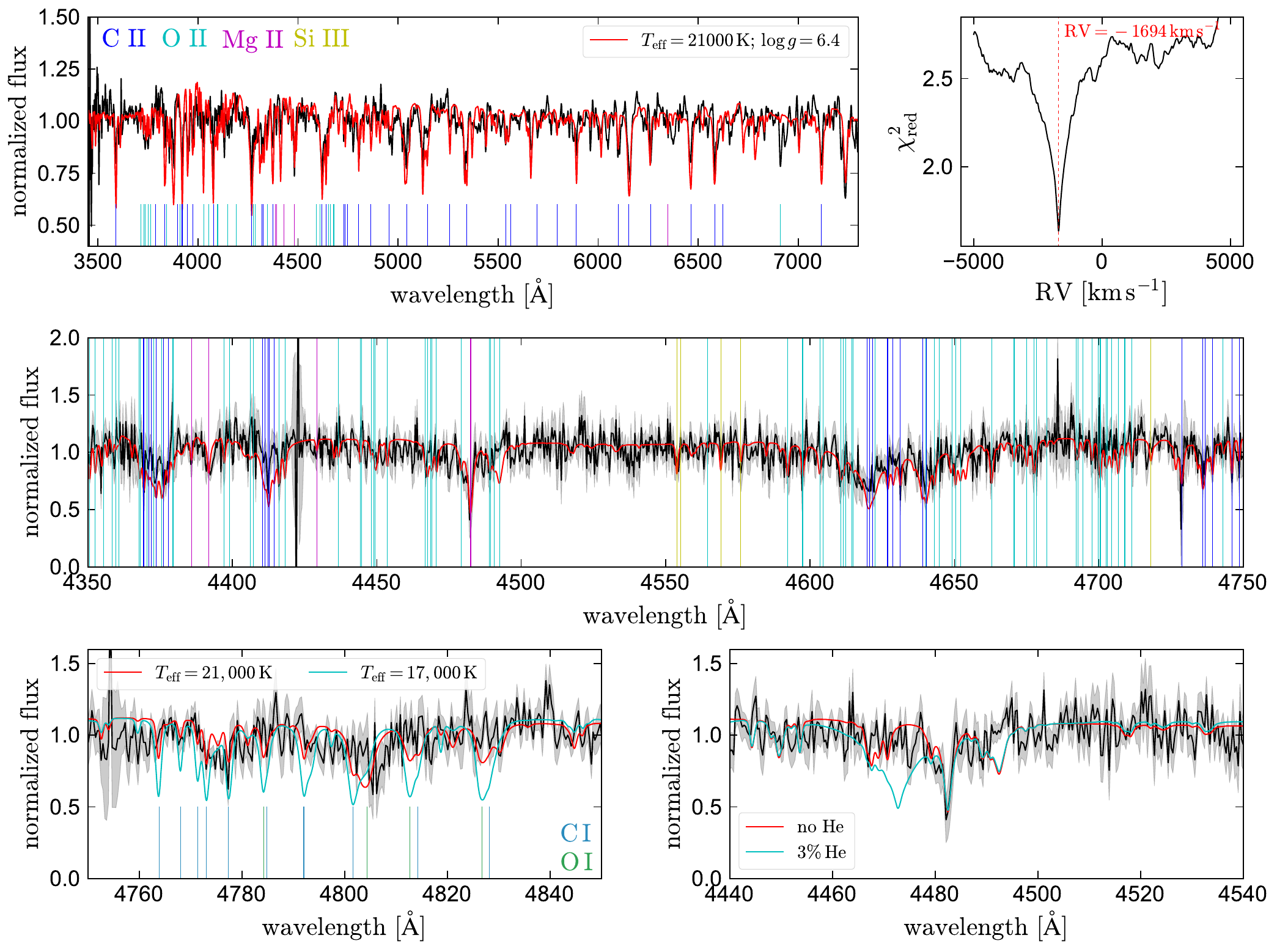}
    \caption{MagE spectrum of J1235-3752, shifted to the rest frame. Upper-left panel shows the full spectrum, smoothed to a resolution $R=1000$ (black line). Lower panels show the spectrum before smoothing, with gray shading showing 1$\sigma$ uncertainties. The red line shows a model spectrum with $T_{\rm eff}=21,000\,\rm K$ and $\log g = 6.4$, calculated for an atmosphere that by mass is $72\%$ C, $28\%$ O, and $0.2\%$ each of Mg and Si. The strongest lines in the observed spectrum are due to these elements; in the middle panel, colored vertical lines indicate which species is responsible for each line in the model spectrum. Cross-correlation with the model spectrum yields an RV of $-1694\,\rm km\,s^{-1}$ (upper right). A significantly lower $T_{\rm eff}$ is ruled out by the weak observed C~I and O~I lines (lower left). The lack of He~I lines rules out a significant helium contribution to the atmosphere (lower right). }
    \label{fig:spec6156}
\end{figure*}

The MagE spectrum of J1235-3752 is shown in Figure~\ref{fig:spec6156}. The strongest lines are due to C~II, O~II, and Mg~II. Both Si~II and Si~III lines are detected, suggesting an effective temperature in the range of 15--25\,\rm kK. Cross-correlation yields a well-measured RV of $-1694\pm 10\,\rm km\,s^{-1}$. We do not attempt to correct for gravitational redshift, here or elsewhere in the paper. The expected gravitational redshifts for objects in our confirmed sample are of order $10\,\rm km\,s^{-1}$, and corrections would make the measured RVs more negative. 

Comparison of the observed normalized spectrum with a grid of Kurucz models yielded a best-fit effective temperature near 21,000\,K. The surface gravity is not well constrained spectroscopically; we adopt $\log g = 6.4$ (cgs units) as motivated by an assumed mass of $\sim 1\,M_{\odot}$ (Section~\ref{sec:d6_masses}) and constraints on the source's radius from the spectral energy distribution (SED; Section~\ref{sec:seds}). The fact that the object shows no He~I lines rules out a surface helium mass fraction above $\sim 1\%$. The model spectrum shown in Figure~\ref{fig:spec6156} contains {\it only} C, O, Mg, and Si. Not all of the metal lines in the observed spectrum are well-accounted for in the best-fit spectral model; other elements are almost certainly present and detectable, but we do not attempt to model them given the relatively low SNR of the data. In the $\rm D^6$ scenario, we expect a significant quantity of iron-peak elements to be deposited on the donor's surface, but at the relevant temperatures, these elements produce only weak lines in the optical unless they dominate the atmosphere.  

The zeropoint-corrected {\it Gaia} parallax, $\varpi = -0.10 \pm 0.24$, implies  a 1$\sigma$ lower limit on distance of $d > 7.4\,\rm kpc$, and a 2$\sigma$ lower limit of $d > 2.68$\,kpc. These limits are respectively calculated as $1/\left(\varpi+\sigma_{\varpi}\right)$ and $1/\left(\varpi+2\sigma_{\varpi}\right)$, and do not include any distance prior. The proper motion is well-constrained by {\it Gaia} at $\mu=95.17 \pm 0.19\,\rm mas\,yr^{-1}$.  Our kinematic modeling with a $3000\,\rm km\,s^{-1}$ upper limit on the birth velocity (Section~\ref{sec:kinematic_modeling}) implies a maximum distance of $\sim 5.5$\,kpc. The total extinction to infinity from the \citet[][hereafter SFD]{Schlegel1998} dust map is $E(B-V)\approx 0.08$\,mag, and the \citet{Lallement2022} 3D dust map suggests that almost all of this extinction is within 1\,kpc of the Sun. This leads to a reasonably well-constrained $M_{G,0} \approx 5.8\pm 0.5$\,mag and a radius of $R\approx 0.1\,R_{\odot}$, corresponding to a luminosity of order $1\,L_{\odot}$.

Samples from the object's trajectory (as modeled in Section~\ref{sec:kinematic_modeling}) are shown in Figure~\ref{fig:trajectory6156}. It is currently above the disk and flying away from it; the best-fit flight time back to the disk midplane is $z/v_z = 1.87_{-0.25}^{+0.86}$\,Myr.

\subsubsection{J0927-6335:}
\begin{figure*}
    \centering
    \includegraphics[width=\textwidth]{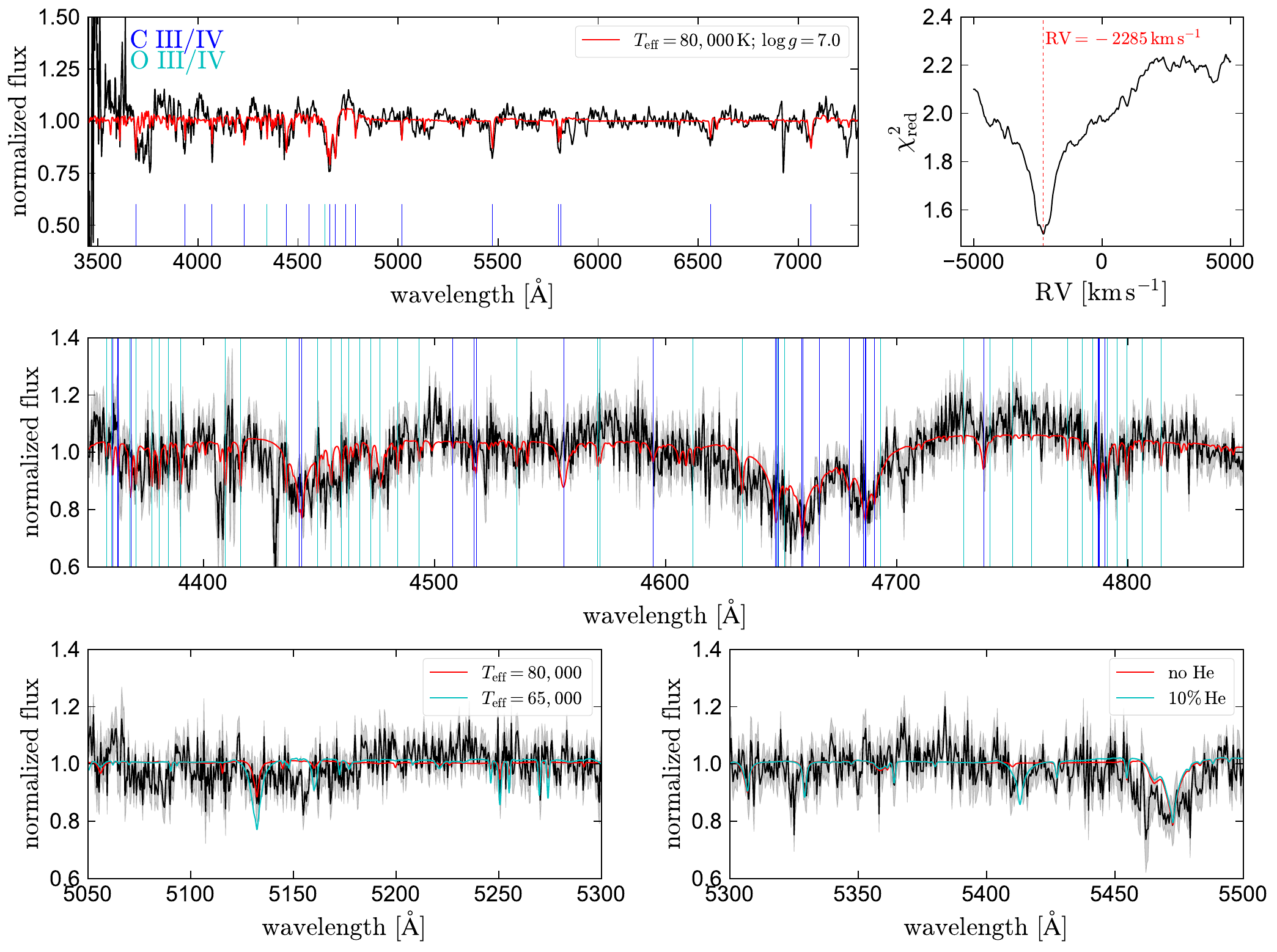}
    \caption{MagE spectrum of J0927-6335, shifted to the rest frame. Upper-left panel shows the full spectrum, smoothed to a resolution $R=1000$. Lower panels show the spectrum and uncertainties before smoothing. The red line shows a model spectrum with $T_{\rm eff}=80000\,\rm K$ and $\log g = 7.0$, calculated for an atmosphere that by mass is $72\%$ C, $28\%$ O, and $0.2\%$ each of Mg and Si (same abundances as Figure~\ref{fig:spec6156}). The inferred RV is $-2285\,\rm km\,s^{-1}$ (upper right). The same spectrum (without smoothing) is compared to model spectra in the bottom three panels. Most of the strongest lines are due to C~III, C~IV, O~III, and O~IV, but there are numerous other lines yet to be identified. The weak C~III lines in the observed spectrum rule out a temperature below $\sim 70,000\,\rm K$ (lower left); the lack of He~II lines rules out a helium mass fraction above $\sim 10\%$ (lower right). }
    \label{fig:spec5250}
\end{figure*}

The MagE spectrum of J0927-6335 is shown in Figure~\ref{fig:spec5250}; it is dominated by C~IV and O~IV lines.  The presence of these higher-ionization states and the weakness of C~III lines allows us to set a lower limit of $T_{\rm eff} \gtrsim 70\,000\,\rm K$.  The best-fit Kurucz model has $T_{\rm eff} = 80,000$\,K, but the agreement between data and model remains comparably good up to temperatures of $\sim 95,000$\,K. Higher temperatures are ruled out by the lack of O~V lines, particularly  O~V $\lambda$4931.65. 

The model spectrum shown in Figure~\ref{fig:spec5250} assumes an abundance pattern identical to the one shown in Figure~\ref{fig:spec6156} for J1235-3752. The reasonably good agreement with the observed spectra suggests that both objects indeed have atmospheres dominated by carbon and oxygen. We considered atmospheres dominated by He, Ne, and Fe and can rule these out since the predicted spectra are in much worse agreement with the data for any $T_{\rm eff}$. Most other elements besides C and O are not predicted to have any strong lines in the optical at these temperatures, so an ultraviolet (UV) spectrum is required to measure abundances of trace elements. Helium is not detected, and from the lack of a detected Pickering series, we can place a limit of $\lesssim 10\%$ on the surface helium mass fraction. 

Cross-correlation of the observed and model spectra leads to a reasonably unambiguous RV measurement of $-2285\pm 20\,\rm km\,s^{-1}$. The cross-correlation function is less sharply peaked than in J1235-3752, mainly because most of the observed lines are broad and relatively weak, and a single complex of C~IV lines at 4600--4700\,\AA\ dominates the total signal. We confirmed that we obtain consistent RVs when we analyze different portions of the spectrum independently. To our knowledge, this is the most negative bulk velocity ever measured for an astronomical object, exceeding even the RVs measured for the S stars orbiting the Galactic Center black hole \citep{GRAVITYCollaboration2018, Do2019}. If we exclude objects at cosmological distances, stars orbiting Sag A*, and the other objects presented in this work, J0927-6335 has the fastest RV measured for a star by more than a factor of 2 \citep[e.g.,][]{Brown2018, Koposov2020}.

The zeropoint-corrected {\it Gaia} parallax, $\varpi = 0.14 \pm 0.21$, implies a 1$\sigma$ distance lower limit of $d > 2.86 \,\rm kpc$, and a 2$\sigma$ lower limit of $d > 1.77$\,kpc. The proper motion is well-constrained by {\it Gaia}, with $\mu= 54.91 \pm 0.25\,\rm mas\,yr^{-1}$. The \citet{Lallement2022} 3D dust map predicts an integrated extinction $E(B-V)\approx 0.13$\,mag along this line of sight to a distance of 2.5\,kpc, while the SFD dust map predicts $E(B-V)\approx 0.2$\,mag to infinity. Since the source is relatively close to the Galactic plane ($b=-9^\circ$), there may indeed be additional foreground extinction beyond the reach of the \citet{Lallement2022} map. We adopt $E(B-V)=0.15$\,mag. 

Our kinematic fitting (Section~\ref{sec:kinematic_modeling}) yields a best-fit distance of $\sim 4.5\,\rm kpc$, which would correspond to an absolute magnitude $M_{G,0}\approx 5.7$. Distances beyond 8\,kpc are ruled out because they would imply $v_{\rm ejection} > 3000\,\rm km\,s^{-1}$. Samples from the posterior are shown in Figure~\ref{fig:trajectory5250}. This object is unique among those in our sample because it is moving {\it toward} the Galactic disk. With a Galactic latitude of $b=-9.1^\circ$, distances of 3--6\,kpc correspond to perpendicular distances from the disk midplane of 0.47--0.95\,kpc. The midplane distance at the time of explosion must have been at least this large. This suggests that the object was born from a kinematically hot stellar population -- most likely, the thick disk.

\subsubsection{J0546+0836:}
\begin{figure*}
    \centering
    \includegraphics[width=\textwidth]{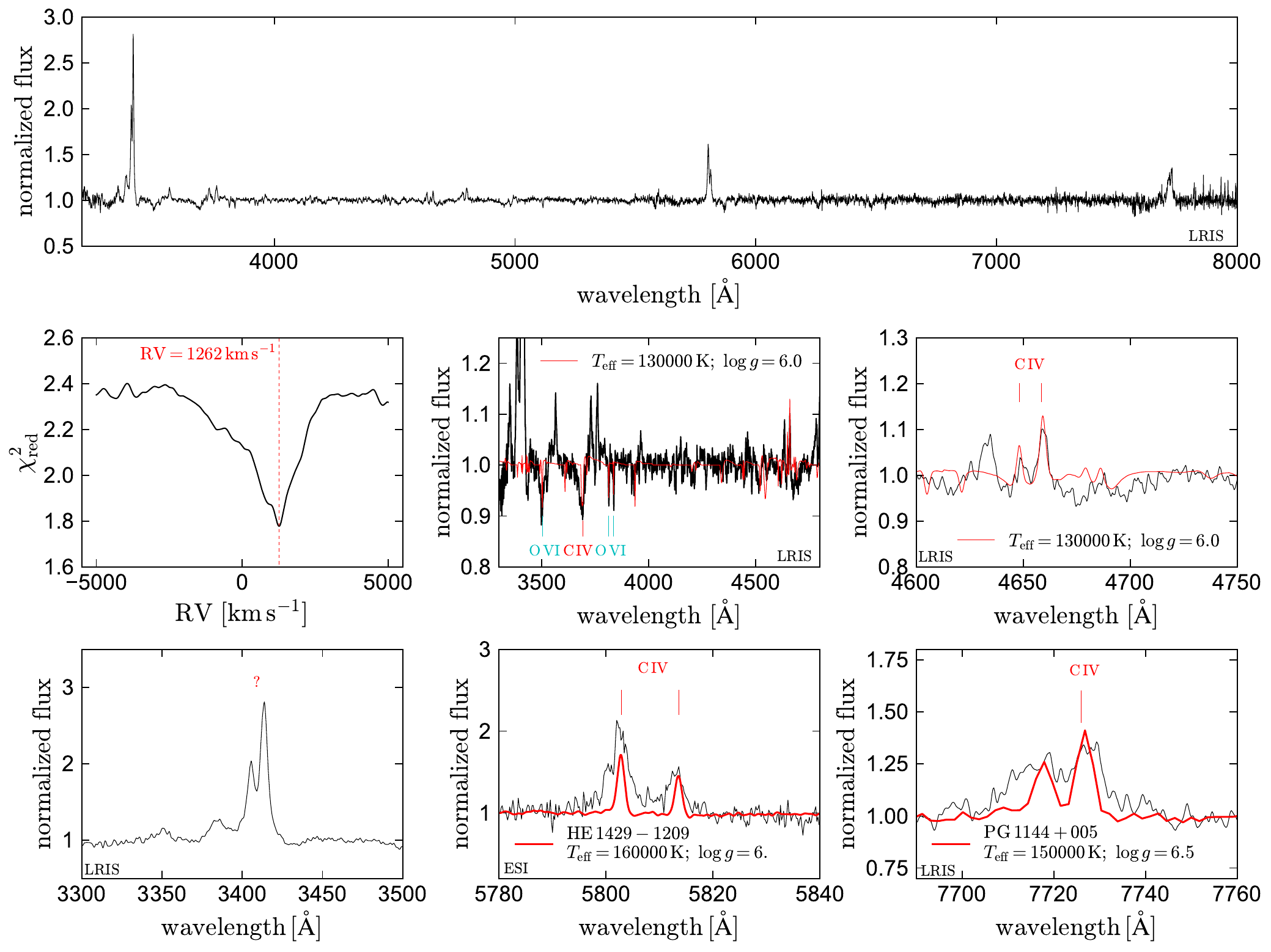}
    \caption{Rest-frame Keck/LRIS and Keck/ESI spectra of J0546+0836. Top panel shows the full LRIS spectrum, which is dominated by a few strong emission lines and also exhibits many weak absorption lines.  Middle-left panel shows cross-correlation of the spectrum with a model for a PG~1159 star with $T_{\rm eff}=130,000\,\rm K$ and $\log g = 6$, which suggests an RV of $\sim +1200\,\rm km\,s^{-1}$. That model (middle center) lacks many of the features of the observed spectrum, but it matches several lines of C~IV and O~VI in both absorption and emission (e.g., middle right). We compare the doublets at $\sim 5800$ and $\sim 7720$\,\AA\ to observed spectra of two PG~1159 stars at the same resolution in the bottom-middle and right panels. The emission lines in J0546+0836 are stronger and broader than those found in typical PG~1159 stars.  Many of the emission lines, including the strongest doublet near  3400\,\AA\ (bottom left), are unidentified.}
    \label{fig:spec3335}
\end{figure*}

Spectra of J0546+0836 obtained with Keck/LRIS and Keck/ESI are shown in Figure~\ref{fig:spec3335}.  Unlike any other objects in our sample, this object's spectrum is dominated by emission lines, though some absorption lines are also visible. The absorption lines are weak, with a maximum depth of $\sim 0.1$ relative to the continuum. 

Cross-correlation with our grids of Kurucz LTE model spectra yielded no plausible matches. We noticed, however, that the spectrum is visually reminiscent of PG~1159 stars, which are very hot \mbox{(pre-)WDs} with little hydrogen and large amounts of carbon in their photospheres \citep[e.g.,][]{Werner1991}. In particular, we identified three doublets in emission at $\sim 4660$\,\AA,  $\sim 5810$\,\AA, and $\sim 7720$\,\AA, whose wavelengths match three C~IV doublets commonly found in PG~1159 stars if the stellar RV is $\sim 1200\,\rm km\,s^{-1}$. This prompted us to compare the spectrum to empirical and synthetic spectra of PG~1159 stars, from which we also identified several C~IV and O~IV absorption lines. 

The emission lines in PG~1159 stars are a result of non-LTE (NLTE) effects. Since the Kurucz codes assume LTE, we compared the spectrum to a grid of NLTE model spectra calculated with TMAP \citep{Werner1999, Werner2003, Rauch2003} and available through TheoSSA \citep{Rauch2018}. We used the grid of models with surface mass fractions of helium (33\%), carbon (50\%), oxygen (15\%), and nitrogen (2\%). The closest match within this grid was the model with $T_{\rm eff} = 130,000\,\rm K$ and $\log g = 6.0$, which we show in Figure~\ref{fig:spec3335}. Cross-correlation with the LRIS spectrum yields an RV of $\rm 1262\, km\,s^{-1}$. Since the LRIS wavelength solution is not very stable, we instead measured the RV from the C~IV $\lambda\lambda$5801.33, 5811.98 emission-line doublet in the ESI spectrum. This yielded an RV of $\rm 1200 \pm 20\,\rm km\,s^{-1}$, which we adopt in our subsequent analysis. 

Although the closest-matching model spectrum has a surface gravity of $\log g = 6.0$, we suspect a higher surface gravity of  $\log g \approx 7$ on the basis of the SED-inferred radius $R \approx 0.05\,R_{\odot}$ (Section~\ref{sec:seds}). None of the model spectra with $T_{\rm eff} \lesssim 100,000\,\rm K$ have the strong C~IV emission lines found in the observed spectrum, so we adopt 100,000\,K as a rough lower limit on the star's effective temperature.

The model spectrum shown in Figure~\ref{fig:spec3335} assumes a 33\% helium mass fraction, but we find that the helium lines in the model spectra are absent in the data. We conclude that J0546+0836 is most likely helium free. Because spectral models still fail to reproduce several lines in normal PG~1159-star spectra, we also compare portions of the spectra to empirical spectra of other PG~1159 stars. This yielded a few additional line identifications. For example, the blue component of the doublet at 7717\,\AA\ is not present in the model spectra and has long evaded identification \citep[e.g.,][]{Werner1991, Werner2014}, but it is clearly present in the spectra of many PG~1159 stars. 

There are also plenty of lines in the spectrum of J0546+0836 that are not found in most PG~1159-star spectra and remain to be identified. Indeed, the most conspicuous feature in the LRIS spectrum is a strong emission doublet, with rest wavelength near 3410\,\AA\ and equivalent width $\sim 20$\,\AA; this feature is not present in the model spectrum at all, and we were unable to identify a comparable feature in the spectra of known PG~1159 or [WC] stars. The same is true for several other emission lines; a more detailed analysis of this spectrum is  warranted. 

``Normal'' PG~1159 stars are thought to be post-asymptotic-giant-branch (post-AGB) stars that experienced a late helium flash; such objects are frequently found at the centers of planetary nebulae \citep[e.g.][]{Herwig1999}. Given its high space velocity, we think that J0546+0836 is a thermonuclear runaway, and is not directly related to normal PG~1159 stars in an evolutionary sense. However, it is very hot and has an atmosphere dominated by carbon -- hence, the resemblance to PG~1159 stars. 

The ESI spectrum of J0546+0836 has a resolution $R\approx 8000$ -- high enough that the emission lines are reasonably well resolved. A cutout of the spectrum is shown in the lower-middle panel of Figure~\ref{fig:spec3335}, where we compare the observed C~IV $\lambda\lambda$5801, 5812 doublet to that of the PG~1159 star HE~1429-1209 observed with UVES \citep{Werner2004} and degraded to $R = 8000$. The doublet in J0546+0836 is both stronger (with an equivalent width of 7\,\AA, compared to $\sim 3$\,\AA\ for HE~1429-1209) and broader. If the observed broadening is due to rotation, this would imply $v\sin\,i \approx 180\,\rm km\,s^{-1}$, which implies a rotation period $P_{{\rm rot}} \approx 20\,{\rm min}\times\left(\frac{R}{0.05\,R_{\odot}}\right)\sin\,i$. This is plausibly within the expected range for a synchronously rotating $\sim 0.01\,R_{\odot}$ WD that expanded following the detonation of its companion at an orbital period of 2--3\,min. It is also a significantly shorter rotation period than expected for a normal young WD \citep{Hermes2017}. That being said, we cannot exclude the possibility that processes besides rotation dominate the emission-line broadening, and deeper high-resolution spectra are needed to confirm whether the absorption lines imply a similar $v\sin\,i$.

The \citet{Green2019} 3D dust map predicts an extinction $E(B-V) = 0.19\pm 0.03$\,mag at all distances beyond 2.5\,kpc, while the \citet{Lallement2022} map predicts $E(B-V) = 0.29$\,mag at 2.5\,kpc and the SFD map predicts $E(B-V) = 0.27$\,mag at infinity. We adopt $E(B-V)=0.25$\,mag. 

The {\it Gaia} parallax, $\varpi = 0.06 \pm 0.31$, implies a 1$\sigma$ distance lower limit of $d > 2.64$\,kpc, and a 2$\sigma$ lower limit of $d > 1.44$\,kpc. The proper motion is well-constrained by {\it Gaia}, with $\mu= 76.07 \pm 0.25\,\rm mas\,yr^{-1}$. Sample trajectories of J0546+0836 are shown in Figure~\ref{fig:trajectory3335}. The source is located toward the Galactic anticenter and is moving in a counterrotating trajectory, meaning that it was likely slowed by Galactic rotation. It is below the disk midplane and is moving away from it; the midplane crossing would have occurred $0.61^{+0.05}_{-0.08}$\,Myr ago. 

\subsubsection{J1332-3541:}
\begin{figure*}
    \centering
    \includegraphics[width=\textwidth]{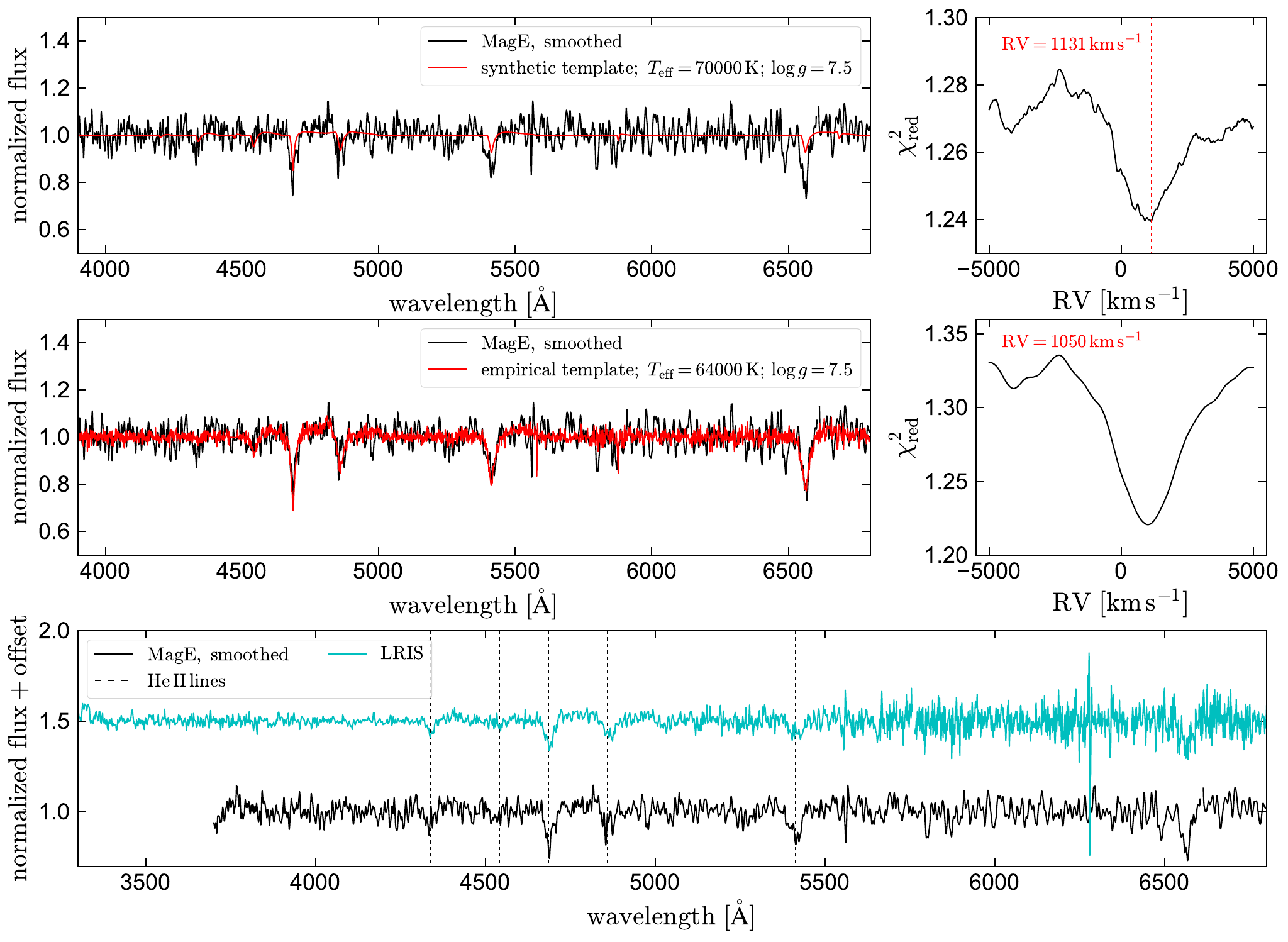}
    \caption{Rest-frame MagE and LRIS spectra of J1332-3541. Top panel compares the smoothed MagE spectrum  ($R\approx 1000$) to a model spectrum of a pure He atmosphere with $T_{\rm eff}\approx 70,000\,\rm K$ and $\log g = 7.5$. The strongest lines in the model match the observed spectrum at an RV of $\sim 1100\,\rm km\,s^{-1}$, but the observed lines are significantly stronger than those in the model spectrum. In the middle panels, we compare the spectrum to an empirical template from SDSS of a DO~WD. Like J1332-3541, this object has deeper He~II lines than predicted by models. Using this spectrum as an empirical spectrum, we infer an RV difference of $\sim +1050\,\rm km\,s^{-1}$. Bottom panel compares the MagE spectrum to an LRIS spectrum (also with $R\approx 1000$), which has higher SNR at blue wavelengths. All of the features that appear in both spectra are due to He~II. }
    \label{fig:spec6164}
\end{figure*}

The MagE and LRIS spectra of J1332-3541 are shown in Figure~\ref{fig:spec6164}. Unlike the other objects in our sample, the spectrum is dominated by He~II lines; i.e., it is not obviously different from that of a normal DO~WD. We nevertheless consider it a strong candidate for being a thermonuclear runaway owing to its high RV.

We compared the spectrum to a grid of pure He models calculated with TMAP and retrieved from TheoSSA. We use these, rather than spectra generated with the Kurucz codes, because they account for NLTE effects and were generated with a code designed for hot WDs. The lack of a strong He~I $\lambda$5876 line sets a lower limit on the effective temperature of $T_{\rm eff}\gtrsim 70,000\,\rm K$. The observed He~II lines are stronger than predicted in any of the model spectra (upper-left panel of Figure~\ref{fig:spec6164}). This ``He~II line problem'' is in fact rather common in DO~WDs, with about half of all hydrogen-deficient WDs at $T_{\rm eff}>60,000\,\rm K$ exhibiting it \citep{Werner1995, Dreizler1995, Werner2014}. The reason for the phenomenon is not understood; one possibility is that the excess absorption originates in a wind-fed magnetosphere \citep{Reindl2019}. 

To assess whether the excess absorption found in J1332-3541 is within the normal range for DO~WDs, we compared its spectrum to the SDSS spectra of several DO~WDs known to exhibit the He~II line problem. One example is shown in the middle-left panel of Figure~\ref{fig:spec6164}, where we compare the object to the SDSS spectrum of Ton~519 (SDSS J102907.31+254008.3), a DO~WD with the He~II line problem \citep{Kepler2019}. For this star \citet{Bedard2020} estimated $T_{\rm eff} = 64,000\pm 1300\,\rm K$ and $\log g=7.47\pm 0.14$. However, we caution that their spectral models also could not fit the strong He~II lines, so the uncertainties are likely underestimated. The spectrum matches that of J1332-3541 fairly well. The most significant difference between the two objects is that He~II $\lambda$4541 is stronger than He~II $\lambda$4339 in Ton~519, while the opposite is true in J1332-3541. There is no strong evidence of any lines in the J1332-3541 spectrum that are not present in Ton~519. The star has the smallest radius and likely the highest surface gravity in our sample (Section~\ref{sec:seds}), so metals deposited on its surface may already have diffused out of the atmosphere. UV spectroscopy is required to asses the atmospheric composition more robustly.

The RV difference between Ton~519 and J1332-3541 is $\sim 1050\,\rm km\,s^{-1}$; this includes an unknown component that represents the RV of Ton 519 and the difference in the stars' gravitational redshifts. Fitting a Gaussian profile to the He~II $\lambda$4686  line, we find an RV of $+1090\pm 50\,\rm km\,s^{-1}$, consistent with the best-fit value inferred from cross-correlation with the synthetic template.

The zeropoint-corrected {\it Gaia} parallax, $\varpi = 0.65 \pm 0.54$, implies a distance of $\sim 1.5$\,kpc, with a 1$\sigma$ lower limit of distance $d > 0.84\,\rm kpc$, and a 2$\sigma$ lower limit of $d > 0.58$\,kpc. J1332-3541 is thus likely the nearest of our newly discovered objects. The proper motion is well-constrained by {\it Gaia} at $\mu= 155.54 \pm 0.44\,\rm mas\,yr^{-1}$. The \citet{Lallement2022} dust map predicts a foreground extinction of $E(B-V)=0.05$\,mag at a distance of 1\,kpc, and the SFD map predicts $E(B-V)=0.05$\,mag at infinity, so we adopt $E(B-V)=0.05$\,mag.

We classify J1332-3541 as a $\rm D^6$ star on the basis of its high velocity. The fact that its atmosphere is dominated by He rather than C could indicate that it formed from a He-core WD. 
Alternatively, J1332-3541 could be a CO~WD whose He shell was not fully stripped before its companion exploded. 

\begin{figure*}
    \centering
    \includegraphics[width=\textwidth]{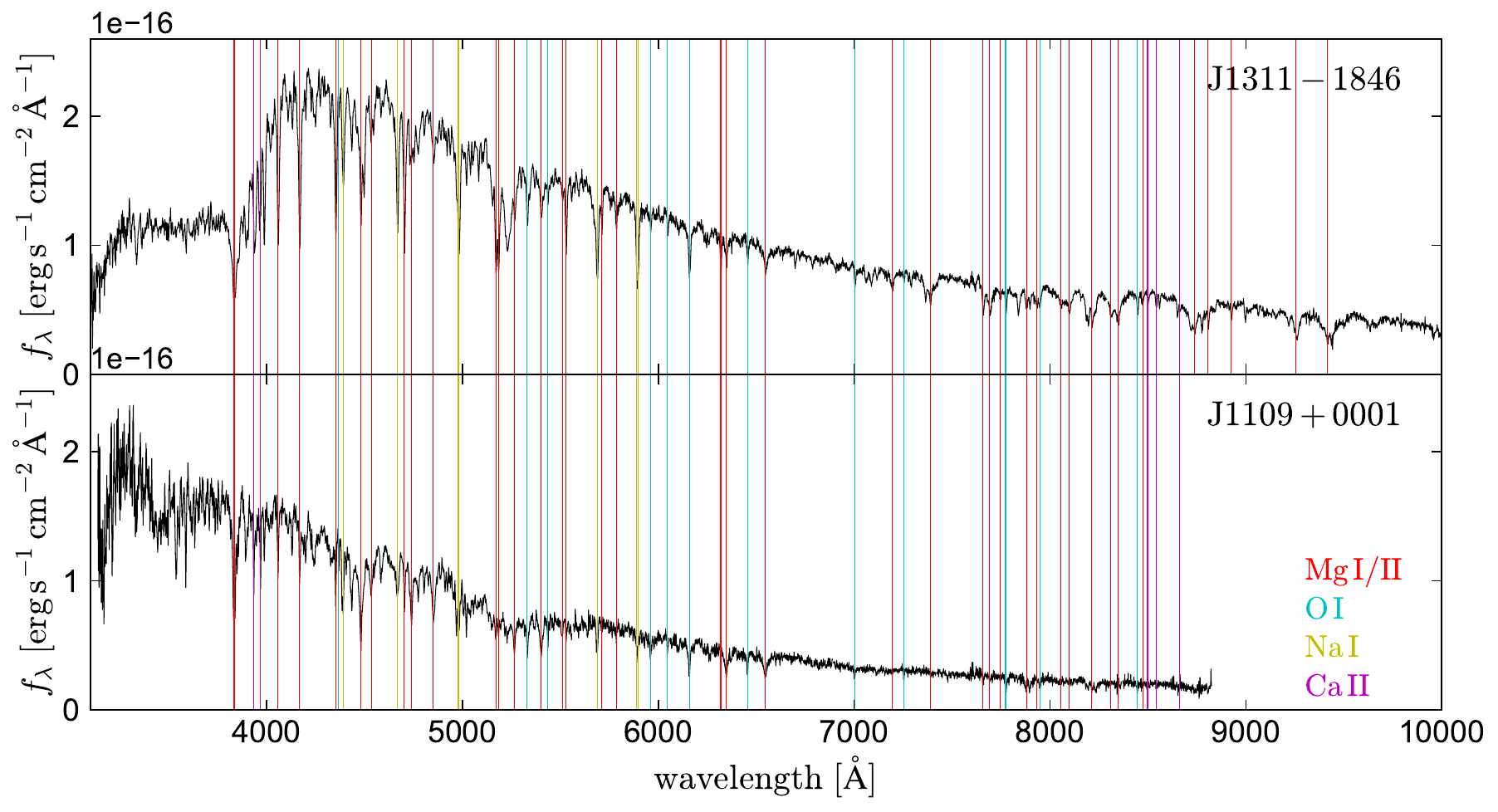}
    \caption{LRIS spectra of the two newly discovered LP~40-365 stars, shifted to the rest frame. Strong lines of Mg, O, Na, and Ca are labeled. Many other metal lines are also detected, but they are mostly blended at this resolution. No H or He lines are detected. While these objects have some features in common with $\rm D^6$ stars, their somewhat lower velocities and different abundance patterns suggest formation through a different mechanism, likely involving underluminous thermonuclear supernovae \citep{Raddi2019}. } 
    \label{fig:speclp40}
\end{figure*}

\subsubsection{Two LP~40-365 stars:  J1311-1846 and J1109+0001}
\label{sec:lp40}

Two objects we observed, J1311-1846 and J1109+0001, have relatively slow RVs of $-55$ and +100\,$\rm km\,s^{-1}$. Their spectra, however, are clearly different from those of WDs, sdO/B stars, or main-sequence stars (Figure~\ref{fig:speclp40}).  The strongest lines in both objects are due to Mg, O, Na, and Ca. Many other metal lines are likely present, but most are blended at $R\approx 1000$. Mg~II lines are stronger than Mg~I in J1109+0001; the opposite is true in J1311-1846, suggesting J1109+0001 is the hotter of the two stars. Their dereddened colors, $G_{\rm BP}-G_{\rm RP}=0.23$\,mag and $G_{\rm BP}-G_{\rm RP}=-0.09$\,mag, suggest temperatures of order 7000\,K and 10,000\,K. H and He are not detected in either star. The nondetection of He~I $\lambda$5876 in J1109+0001 rules out a surface helium mass fraction of $\gtrsim 10\%$. The lower temperature of J1311-1846 prevents a spectroscopic limit on its He abundance. 

Modeling their 3D kinematics (Section~\ref{sec:kinematic_modeling}), we find 1$\sigma$ lower limits on their birth  velocities of $610\,\rm km\,s^{-1}$ and $661\,\rm km\,s^{-1}$. These lower limits are slower than those we infer for the four objects discussed above. 

On the basis of their velocities, spectra, and positions in the color-magnitude diagram (CMD), we classify these objects as LP~40-365 stars. 
The LP 40-365 stars are a small group of high-velocity stars (with only four members until now) proposed to be the partially burned accretors left behind by underluminous thermonuclear supernovae with nondegenerate donors \citep{Vennes2017, Raddi2019}. In this case, their high velocities (typically $\sim 600\,\rm km\,s^{-1}$; see Section~\ref{sec:kinematic_modeling}) may be the result of pre-explosion orbital velocities and/or kicks due to asymmetric explosions.  The spectrum of J1109+0001 is very similar to that of the star J1825-3757, while the spectrum of J1311-1846 is very similar to the prototype LP~40-365, as well as to J1603-6613 and J0905+2510 \citep[see][]{Raddi2019}.

\citet{Raddi2019} have proposed that LP~40-365 stars have Ne-dominated atmospheres. This hypothesis is difficult for us to test because Ne lines are predicted to be weak even if it is the dominant component of the atmosphere. In any case, the similar spectra, CMD positions, and velocities of J1311-1846 and J1109+0001 to other LP~40-365 stars suggest that they belong to the group. It is important to note that the boundary between D$^6$ stars and LP~40-365 stars is observationally somewhat murky. Classification to date has focused primarily on optical spectra, which suggest very similar abundance patterns among LP~40-365 stars. No abundance analysis has been published for D$^6$ stars, but their spectra are more similar to one another than to the LP~40-365 stars. The abundance patterns of LP~40-365 stars, particularly the detection of Mn, have been interpreted as indicative of Chandrasekhar-mass explosions. \citet{Raddi2018, Raddi2019} proposed that these objects are the partially burned runaway {\it accretors} from single-degenerate binaries, while the D$^6$ stars are the runaway donors from double-degenerate binaries. While this interpretation is still somewhat uncertain, it is supported by the clearest difference between the two groups: LP~40-365 stars have significantly lower velocities. 

J1109+0001 was previously observed by the 2dF QSO redshift survey \citep{Croom2004}, in which it was classified as a DB~WD. It was also observed as a color-selected AM~CVn candidate by \citet{Carter2013}, who did not offer a classification. The misclassification as a DB~WD is most likely a result of confusion of He~I $\lambda$4471, which is strong in DB~WDs, with the strong Mg~II $\lambda$4481 line found in this object.

\section{Spectral energy distributions}
\label{sec:seds}
To constrain the radii of the objects in our sample, we constructed their broadband SEDs by combining photometry from several surveys, including Pan-STARRS1 \citep{Chambers2016}, SkyMapper \citep{Keller2007}, {\it GALEX} \citep{Martin2005}, and DECaPS2 \citep{Saydjari2023}.

\begin{table*}
\begin{tabular}{lllllllll}
Name & $\alpha$  &  $\delta$  &  Classification & Elements detected & $T_{\rm eff}$ &  $R$ & $E(B-V)$ & $\log\left[g/\left({\rm cm\,s^{-2}}\right)\right]$ \\
  & J2016.0 &  J2016.0 &   &  & $[\rm K]$ &  $[R_{\odot}]$ & [mag] & assuming $M=1\,M_{\odot}$ \\

\hline
J1235-3752 & 12:35:31.8 & -37:52:35.53 & $\rm D^6$ star & C, O, Mg, Si & 21,000 & $0.104_{-0.030}^{+0.026}$ & 0.08 & $6.41_{-0.19}^{+0.30}$ \\ 
J0927-6335 & 09:27:42.68 & -63:35:40.33 &  $\rm D^6$ star & C, O & 80,000 & $0.052_{-0.020}^{+0.025}$ & 0.15 & $7.01_{-0.34}^{+0.41}$ \\ 
J0546+0836 & 05:46:17.6 & 08:36:6.65 &  $\rm D^6$ star & C, O & 130,000 & $0.051_{-0.021}^{+0.029}$ & 0.25 & $7.02_{-0.39}^{+0.47}$ \\
J1332-3541 & 13:32:57.98 & -35:41:10.68 &  $\rm D^6$ star  & He & 70,000 & $0.017_{-0.007}^{+0.013}$ & 0.05 & $7.98_{-0.48}^{+0.47}$ \\
J1311-1846 & 13:11:57.18 & -18:46:11.78 & LP 40-365 star & Mg, O, Na, Ca & 7000 & $0.234_{-0.064}^{+0.137}$ & 0.09 & $5.70_{-0.40}^{+0.28}$ \\
J1109+0001 & 11:09:7.95 & 00:01:34.31 & LP 40-365 star & Mg, O, Na, Ca & 10,000 & $0.133_{-0.057}^{+0.106}$ & 0.04 & $6.19_{-0.51}^{+0.48}$ \\
\end{tabular}
\caption{Constraints from our analysis of the spectra and broadband SEDs. Uncertainties are $1\sigma$ (middle 68\%).}
\label{tab:sed}
\end{table*}

\begin{figure*}
    \centering
    \includegraphics[width=\textwidth]{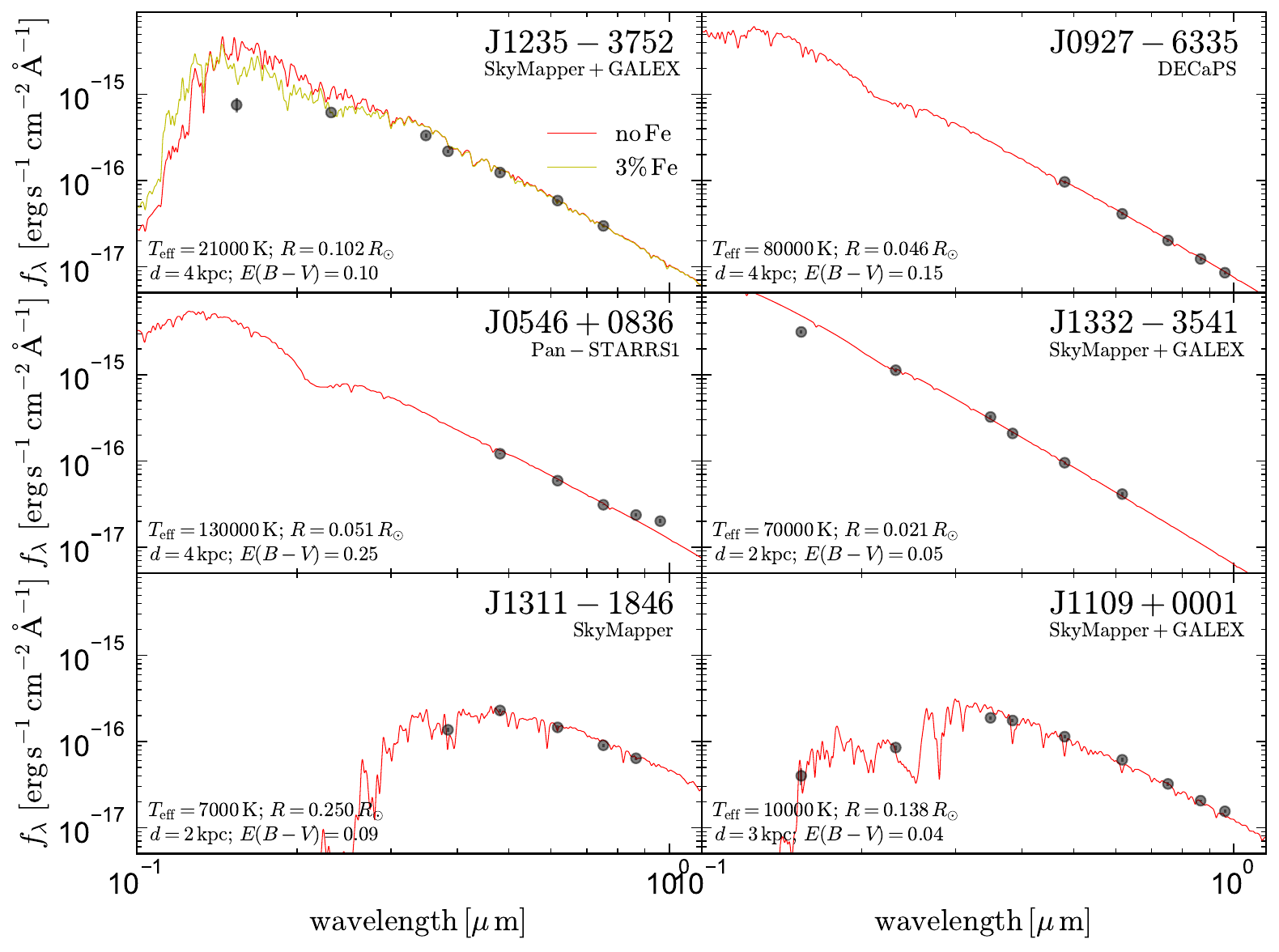}
    \caption{Spectral energy distributions of candidate $\rm D^6$ stars (top-two rows) and LP~40-365 stars (bottom row). Black points show observed broadband flux densities, while lines show models. The temperatures and compositions of the models are set to the values estimated from spectra (Figures~\ref{fig:spec6156}--\ref{fig:speclp40} and summarized in Table~\ref{tab:sed}), not fits to the SEDs. Radii are set to match the observed flux in the optical, for the distance and extinction listed in each panel. The thus-inferred radii (accounting for distance uncertainties; see Section~\ref{sec:kinematic_modeling}) are listed in Table~\ref{tab:sed}. For J1235-3752, the fiducial model overpredicts the UV flux. This may be a result of line blanketing in the UV by elements not included in the model.}
    \label{fig:seds}
\end{figure*}

Figure~\ref{fig:seds} compares the  SEDs of the six objects discussed in Section~\ref{sec:individual} to models. The model temperatures and compositions are set to the values we inferred from analysis of the spectra; they are not fit to the observed SEDs. Extinction is set to the values taken from the 3D dust maps. For the LP 40-365 stars, we assume abundances similar to those inferred by \citet{Raddi2019} and take the extinction from the \citet{Green2019} map. In each panel, we assume a ``round number'' distance consistent with the results of our kinematic modeling (Section~\ref{sec:kinematic_modeling}) and then scale the assumed radius of the SED model to match the observed flux in the optical. To obtain constraints on the stars' radii that account for distance uncertainty, we take Monte Carlo draws from the distance posteriors inferred in Section~\ref{sec:kinematic_modeling} and re-estimate the radius for each draw. The resulting constraints on the stars' radii are reported in Table~\ref{tab:sed}. We also list the predicted surface gravities corresponding to these radii, assuming a mass of $1\,M_{\odot}$. This mass estimate is motivated by the velocities of the $\rm D^6$ stars (Section~\ref{sec:d6_masses}); the masses of the LP~40-365 stars are more uncertain. 

These estimates neglect uncertainties in the source's temperatures, which are difficult to reliably quantify. However, we expect that distance uncertainties are the dominant source of uncertainty in radii. For all the $\rm D^6$ stars, the optical photometry is in the Rayleigh–Jeans tail of the SED, where $f_{\lambda}\propto T_{{\rm eff}}R^{2}d^{-2}\lambda^{-4}$. This means that given  a measured flux density $f_\lambda$ at wavelength $\lambda$, the inferred radius scales as $R\propto d/\sqrt{T_{{\rm eff}}}$. The fractional radius uncertainty thus scales with distance as $\frac{\sigma_{R}}{R}\propto\frac{\sigma_{d}}{d}$ and with $T_{\rm eff}$ as $\frac{\sigma_{R}}{R}\propto\frac{1}{2}\frac{\sigma_{T_{{\rm eff}}}}{T_{{\rm eff}}}$. We find typical fractional distance errors of $>30$\% for the $\rm D^6$ stars (Table~\ref{tab:all_systems}), so uncertainties in $T_{\rm eff}$ would become a dominant source of uncertainty in $R$ only if the objects had fractional temperature uncertainties at the 50\% level.

\begin{table*}
\begin{tabular}{llllllllll}
Name & {\it Gaia} DR3 Source ID &  $G$  & $G_{\rm BP}-G_{\rm RP}$ & $\rm RV$  & 
$d$ & $v_{\rm tot}$ & $v_{\rm ejection}$ & $z$ & $t_{\rm disk} = z/v_z$ \\
  &   &  [mag] &  [mag] & $\rm [km\,s^{-1}]$  & 
$[\rm kpc]$ & $[\rm\,km\,s^{-1}]$ & $[\rm\,km\,s^{-1}]$ & $[\rm kpc]$ & $[\rm Myr]$ \\
\hline
\hline 
\multicolumn{10}{l}{\underline{Suspected $\rm D^6$ stars}}   \\ 
D6-1 & 5805243926609660032 & 17.4 & 0.48 & $1200 \pm 40$ & $1.91^{+0.29}_{-0.22}$ & $2045^{+251}_{-187}$ & $2254^{+248}_{-185}$ & $-0.57^{+0.07}_{-0.09}$ & $0.63^{+0.04}_{-0.04}$ \\ 
D6-2 & 1798008584396457088 & 17.0 & 0.41 & $80 \pm 10$ & $0.84^{+0.05}_{-0.04}$ & $1151^{+59}_{-51}$ & $1051^{+62}_{-54}$ & $-0.27^{+0.01}_{-0.02}$ & $-0.72^{+0.01}_{-0.01}$ \\ 
D6-3 & 2156908318076164224 & 18.2 & 0.43 & $-20 \pm 80$ & $2.26^{+0.34}_{-0.35}$ & $2248^{+340}_{-353}$ & $2393^{+377}_{-391}$ & $0.93^{+0.14}_{-0.14}$ & $2.24^{+0.20}_{-0.17}$ \\ 
{\bf J1235} & 6156470924553703552 & 19.0 & -0.28 & $-1694 \pm 10$ & $4.07^{+1.01}_{-1.19}$ & $2670^{+339}_{-350}$ & $2471^{+351}_{-345}$ & $1.73^{+0.42}_{-0.50}$ & $1.87^{+0.86}_{-0.25}$ \\ 
{\bf J0927} & 5250394728194220800 & 19.4 & -0.32 & $-2285 \pm 20$ & $4.53^{+2.17}_{-1.71}$ & $2753^{+271}_{-148}$ & $2519^{+271}_{-147}$ & $-0.70^{+0.27}_{-0.35}$ & $-1.32^{+0.42}_{-0.42}$ \\ 
{\bf J0546} & 3335306915849417984 & 19.1 & -0.25 & $1200 \pm 20$ & $4.02^{+2.25}_{-1.67}$ & $1699^{+670}_{-390}$ & $1864^{+682}_{-416}$ & $-0.69^{+0.29}_{-0.40}$ & $0.61^{+0.05}_{-0.08}$ \\ 
{\bf J1332} & 6164642052589392512 & 19.4 & -0.55 & $1090 \pm 50$ & $1.63^{+1.19}_{-0.68}$ & $1464^{+740}_{-331}$ & $1619^{+707}_{-320}$ & $0.74^{+0.53}_{-0.30}$ & $0.64^{+0.13}_{-0.14}$ \\ 
\multicolumn{10}{l}{\underline{Suspected LP 40-365 stars}}   \\ 
LP 40-365 & 1711956376295435520 & 15.6 & 0.23 & $498 \pm 5$ & $0.61^{+0.01}_{-0.01}$ & $837^{+5}_{-5}$ & $607^{+5}_{-5}$ & $0.43^{+0.01}_{-0.01}$ & $4.87^{+0.39}_{-0.33}$ \\ 
J1603 & 5822236741381879040 & 17.8 & 0.16 & $-480 \pm 5$ & $2.07^{+0.49}_{-0.32}$ & $833^{+62}_{-38}$ & $606^{+73}_{-42}$ & $-0.35^{+0.06}_{-0.09}$ & $1.58^{+0.12}_{-0.10}$ \\ 
J0905 & 688380457508503040 & 19.6 & 0.24 & $300 \pm 50$ & $4.49^{+7.59}_{-2.76}$ & $519^{+1122}_{-271}$ & $737^{+1123}_{-341}$ & $2.90^{+4.86}_{-1.77}$ & $-13.58^{+32.35}_{-15.41}$ \\ 
J1825 & 6727110900983876096 & 13.3 & -0.02 & $-47 \pm 5$ & $0.95^{+0.03}_{-0.03}$ & $429^{+18}_{-17}$ & $662^{+18}_{-17}$ & $-0.17^{+0.01}_{-0.01}$ & $1.60^{+0.02}_{-0.02}$ \\ 
{\bf J1311} & 3507697866498687232 & 18.3 & 0.35 & $55 \pm 10$ & $1.88^{+1.09}_{-0.51}$ & $952^{+430}_{-201}$ & $827^{+461}_{-217}$ & $1.32^{+0.76}_{-0.35}$ & $3.88^{+0.20}_{-0.20}$ \\ 
{\bf J1109} & 3804182280735442560 & 19.1 & -0.09 & $100 \pm 10$ & $2.90^{+2.31}_{-1.24}$ & $1378^{+957}_{-510}$ & $1171^{+993}_{-510}$ & $2.35^{+1.86}_{-0.99}$ & $3.92^{+0.26}_{-0.38}$ \\ 
\multicolumn{10}{l}{\underline{Suspected runaway helium star donors}}   \\ 
US 708 & 815106177700219392 & 18.9 & -0.44 & $917 \pm 7$ & $8.38^{+1.00}_{-1.01}$ & $994^{+10}_{-10}$ & $833^{+17}_{-16}$ & $6.17^{+0.74}_{-0.74}$ & $11.39^{+1.82}_{-1.70}$ \\ 
\multicolumn{10}{l}{\underline{Other/unknown}}   \\ 
J1240 & 1682129610835350400 & 18.4 & -0.29 & $-177 \pm 10$ & $0.42^{+0.02}_{-0.02}$ & $239^{+20}_{-18}$ & $446^{+21}_{-18}$ & $0.35^{+0.02}_{-0.02}$ & $-$ \\ 
J1637 & 1327920737357113088 & 20.3 & -0.13 & $300 \pm 50$ & $3.28^{+3.18}_{-1.74}$ & $1202^{+962}_{-463}$ & $1030^{+1205}_{-467}$ & $2.20^{+2.12}_{-1.16}$ & $4.26^{+1.03}_{-1.33}$ \\ 
\end{tabular}
\caption{Known high-velocity objects suspected to be runaways from thermonuclear supernovae. Uncertainties are $1\sigma$ (middle 68\%). Objects in bold are new discoveries. We also list previously known $\rm D^6$ stars (\citetalias{Shen2018}) and LP~40-365 stars \citep{Raddi2019}, as well as the one known high-velocity helium star US~708 \citep{Geier2015}, the oxygen-atmosphere WD~J1240 \citep{Gansicke2020}, and the object J1637 \citep{Raddi2019}, which may be a $\rm D^6$ star or in the same class as J1240. }
\label{tab:all_systems}
\end{table*}

Only J1235-3752, J1332-3541, and J1109+0001 are in the {\it GALEX} footprint; the other sources do not have published UV photometry. The overall shapes of the model SEDs are in reasonable agreement with the data for all sources except J1235-3752, for which the observed far-UV flux is a factor of 5 lower than predicted by the fiducial SED model scaled to match the flux in the optical. The simplest explanation would be an over-estimated effective temperature. However, matching the observed SED shape with the same composition would require a much lower temperature of $T_{\rm eff} \approx 13,000$\,K, and we are unable to satisfactorily reproduce the optical spectrum (Figure~\ref{fig:spec6156}) 
 with such a low $T_{\rm eff}$ for any plausible $\log g$. In particular, the lack of C~I and O~I lines, and the relative strength of Si~II and Si~III lines, strongly suggest $T_{\rm eff}\gtrsim 20,000\,\rm K$.

 An alternative explanation for the fainter-than-predicted UV flux in J1235-3752 is that there is additional opacity in the UV due to elements not included in the spectral model. Recall that the model only includes C, O, Mg, and Si, as these are the elements responsible for the strongest lines in the optical. However, there are many lines in the UV due to elements that have no strong lines in the optical. To illustrate this, we show in Figure~\ref{fig:seds} a model spectrum with the same $T_{\rm eff}$ as the fiducial model but a 3\% Fe mass fraction. This model is almost identical to the fiducial model in the optical but has a significantly shallower slope in the UV. Obtaining UV spectra of objects in our sample would allow us to measure abundances of many other elements, testing this hypothesis. 

\section{Kinematic modeling}
\label{sec:kinematic_modeling}
We infer the 3D trajectory of each star based on its measured parallax, proper motion, and RV. For each target, we draw samples from the posterior of $\left(M_{G,0},\mu_{\alpha}^{*},\mu_{\delta},{\rm RV}\right)$, where $M_{G,0}$ is the absolute magnitude in the {\it Gaia} bandpass, $\mu_{\alpha}^*$ and $\mu_\delta$ are the proper motions in the RA and Dec directions, and RV is the radial velocity. From the absolute magnitude, we predict the parallax in milliarcseconds (mas) as $\varpi=10^{\left(M_{G,0}-G+10+A_{G}\right)/5}$, where $G$ is the apparent magnitude and $A_G=2.67E(B-V)$. The likelihood function then compares the vector of predicted observables, $\boldsymbol  \theta_{{\rm pred}} = \left(\varpi, \mu_{\alpha}^{*},\mu_{\delta},{\rm RV}\right)$, to the corresponding vector of observed quantities, $\boldsymbol  \theta_{{\rm obs}}$:
\begin{equation}
    \label{eq:lnL_ast}
    \ln L=-\frac{1}{2}\left(\boldsymbol  \theta_{{\rm pred}}-\boldsymbol \theta_{{\rm obs}}\right)^{\intercal}\boldsymbol\Sigma_{{\rm obs}}^{-1}\left(\boldsymbol \theta_{{\rm pred}}-\boldsymbol \theta_{{\rm obs}}\right).
\end{equation}
Here $\boldsymbol\Sigma_{{\rm obs}}$ is the observational covariance matrix, which we construct from the correlation coefficients reported in {\it Gaia} DR3 between $\varpi$, $\mu_{\alpha}^*$, and $\mu_{\delta}$, with no covariance between $\rm RV$ and the astrometric parameters. We correct the parallax for the position- and color-dependent zeropoint inferred by \citet{Lindegren2021}. These corrections have only minor effects on our results because they are small compared to the parallax uncertainties. 


We transform the posterior samples to calculate other quantities of interest, including the flight time back to the disk, $t_{\rm disk} = z/v_z$, and the location where each trajectory intersects the midplane. Given the high velocities and short flight times of the objects in our sample, we neglect the gravitational potential of the Milky Way in these calculations; i.e., we assume the stars fly in straight lines with constant velocity. If $t_{\rm disk}$ is positive (so the disk crossing is in the past), we calculate the instantaneous velocity vector of a circular orbit in the Galactic disk at the location where the trajectory intersects it. We call this quantity $\vec{v}_{\rm rot,birth}$. For simplicity, we model the Galactic rotation curve as flat, with a circular velocity of $240\,\rm km\,s^{-1}$. Finally, we calculate the star's birth ``ejection'' velocity, 
\begin{equation}
    v_{\rm ejection} = \left|\vec{v}-\vec{v}_{\rm rot,birth}\right|.
\end{equation}
This represents the change in the star's velocity at the time of the explosion; i.e., it is an attempt to subtract off the pre-explosion velocity due to Galactic rotation. If $t_{\rm disk}$ is negative, meaning that the disk-crossing will happen in the future, we take $\vec{v}_{{\rm rot,birth}}$ to be the velocity of a circular, corotating orbit at the source's current position. 

We adopt a prior that the total ejection velocity cannot exceed $3000\,\rm km\,s^{-1}$, 
\begin{equation}
    v_{\rm ejection}<3000\,{\rm km\,s^{-1}}.
\end{equation}
An upper limit of 3000\,$\rm km\,s^{-1}$ on the ejection velocity is fairly conservative. As we discuss in Section~\ref{sec:d6_masses}, the expected $v_{\rm ejection}$ depends on the masses of both stars, with higher-mass donors and accretors yielding larger $v_{\rm ejection}$. The 3000\,$\rm km\,s^{-1}$ limit corresponds roughly to  two $1.3\,M_{\odot}$ WDs, with weak dependence on their temperature and core composition. Given that (a) only WDs with CO cores are expected to give rise to a $\rm D^6$ scenario, and (b) forming a $1.3\,M_{\odot}$ CO~WD would require accretion of at least $\sim 0.25\,M_{\odot}$ from a companion -- which would itself need to become a near-Chandrasekhar-limit WD -- it is difficult to imagine any scenario in which orbital velocity of a close binary produces an ejection faster than 3000\,$\rm km\,s^{-1}$.

\begin{figure*}[!ht]
    \centering
    \includegraphics[width=\textwidth]{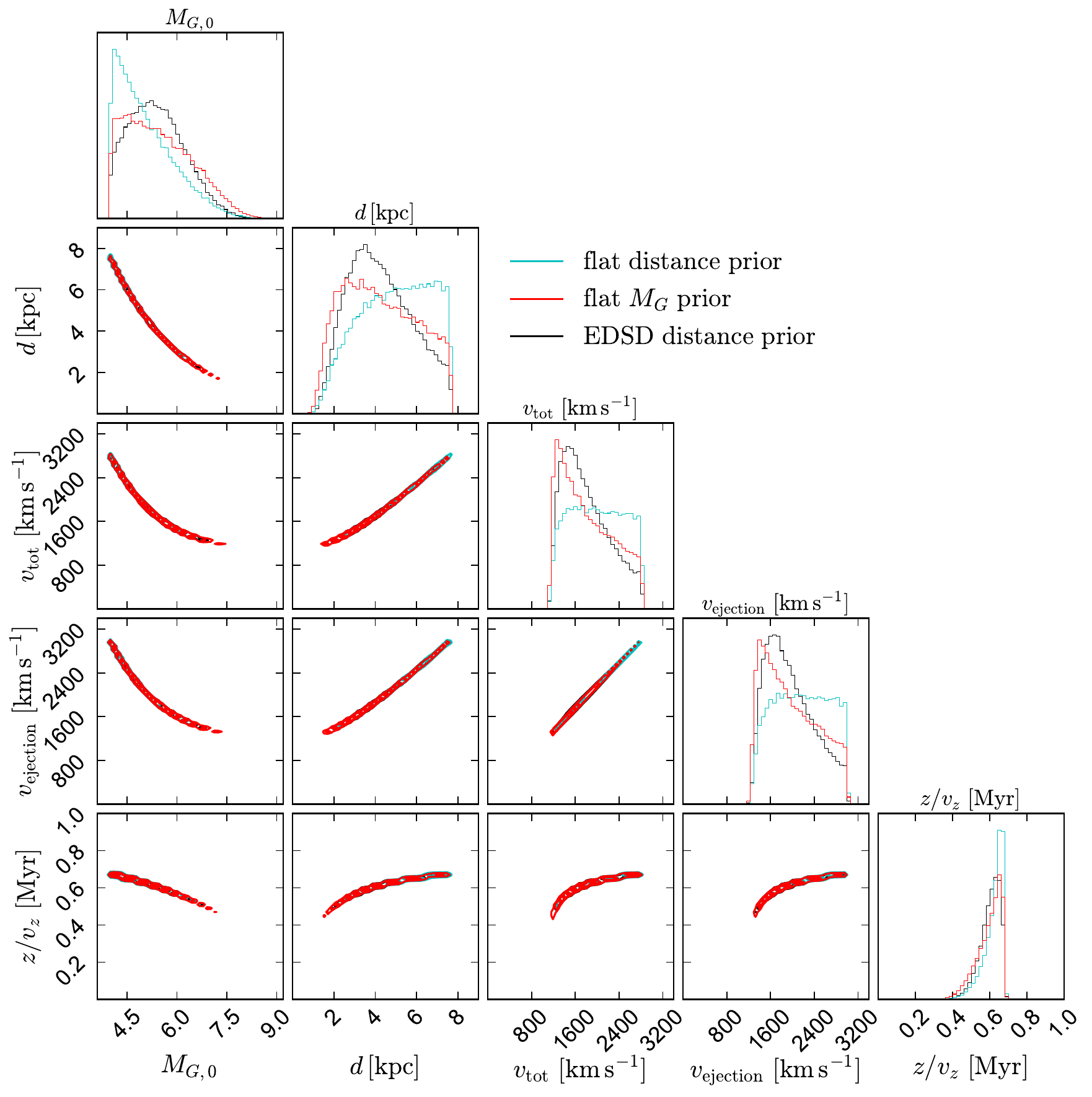}
    \caption{Sensitivity of our results to the adopted distance prior. $v_{\rm tot}$ is the total 3D velocity in a Galactocentric frame. $v_{\rm ejection}$ is the inferred velocity with which the star was launched, assuming it had a circular orbit in the Galactic disk before the supernova.  $z/v_z$ is the flight time back to the disk midplane assuming a straight-line trajectory. For a single object, J0546+0836, we repeat the kinematic modeling with a flat prior in distance (cyan), a flat prior in absolute magnitude (red), and an exponentially decreasing space density distance prior (black). The priors result in different distance posteriors, which translate to different posteriors on the absolute magnitude, total velocity, and ejection velocity. The {\it minimum} total and ejection velocities are insensitive to the prior, because these are set mainly by the RV. The maximum velocities and distance are in all cases set by the prior that $v_{\rm ejection} < 3000\,\rm km\,s^{-1}$. }
    \label{fig:distance_prior}
\end{figure*}

We assume flat priors on $M_{G,0}$, RV, and the astrometric parameters. For each call to the likelihood function, we calculate the predicted phase space vector and corresponding $\vec{v}_{\rm rot,birth}$ and $v_{\rm ejection}$ using standard coordinate transformations as implemented in \texttt{astropy} \citep{AstropyCollaboration2022}. We sample from the posterior using \texttt{emcee} \citep{Foreman-Mackey2013}, drawing 1000 samples each with 64 walkers after a burn-in period of 1000 steps.

\begin{figure*}[!ht]
    \centering
    \includegraphics[width=\textwidth]{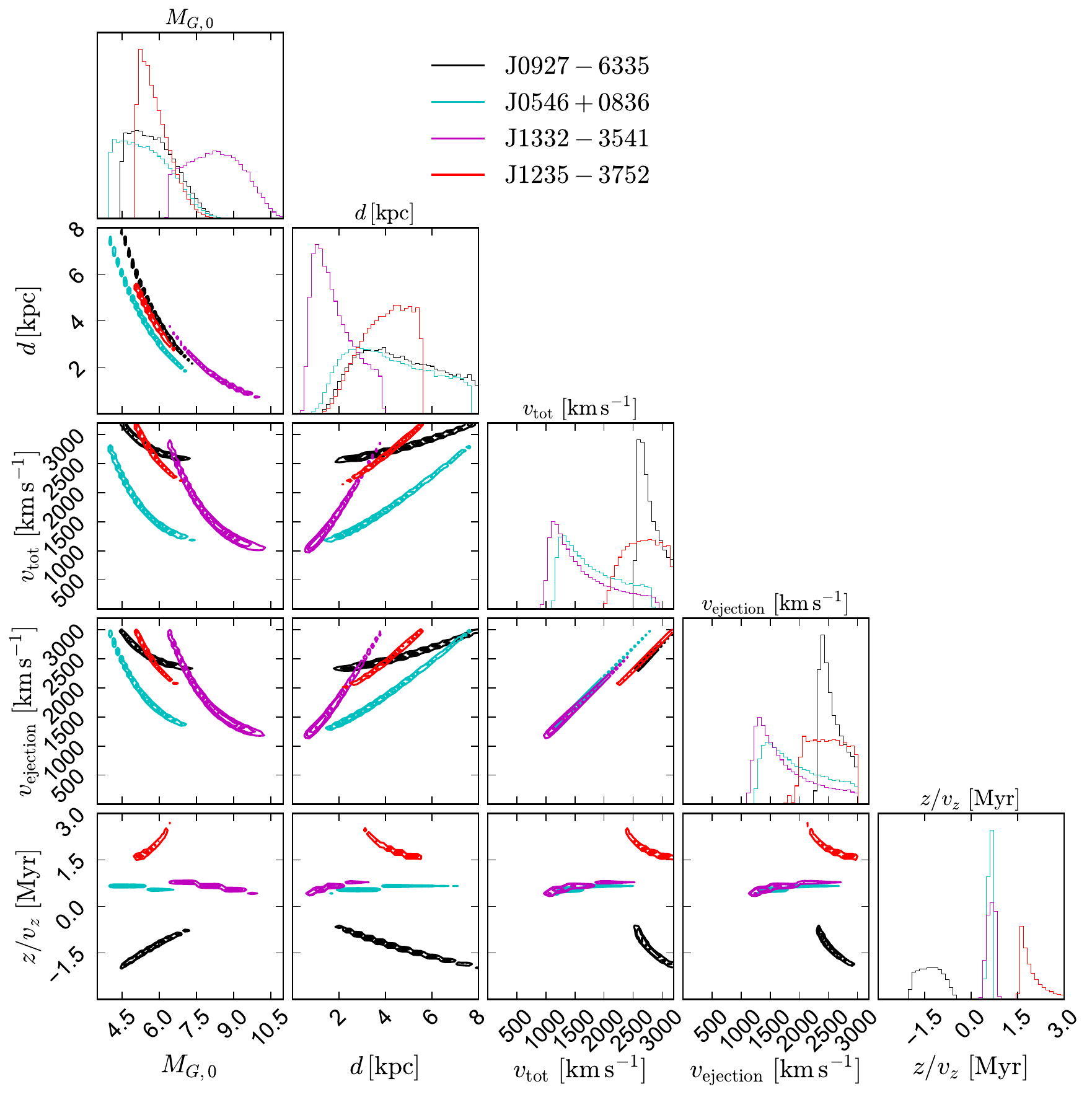}
    \caption{Constraints on the distance and kinematics of the four candidate $\rm D^6$ stars (see Section~\ref{sec:kinematic_modeling}). In all cases, the dominant source of uncertainty is the unknown distance.  For J0927, this quantity is negative, because the star is moving toward the disk (Figure~\ref{fig:trajectory5250}). The minimum ejection velocity is $> 1000 \, \rm km\,s^{-1}$ for all objects, and $> 2000 \rm \,km\,s^{-1}$ for J0927 and J1235, the two objects with the fastest RVs.   }
    \label{fig:corner}
\end{figure*}

\subsection{Distance prior}
\label{sec:distance}

For all of the objects in our sample, the RVs and proper motions are reasonably well-constrained, but the parallax is not. This means that the adopted distance prior -- be it explicit or implicit -- is important. We experimented with three different distance priors as follows.

\begin{enumerate}
    \item A flat prior on $M_{G,0}$, which is equivalent to a distance prior $p(d)\propto 1/d$.
    \item A flat prior on distance. 
    \item An exponentially decreasing space density prior, $p\left(d\right)\propto d^{2}e^{-d/L}$ \citep[e.g.,][]{Bailer-Jones2015}. We use $L=1.35$\,kpc, a value which \citet{Astraatmadja2016} found to reasonably approximate the distance distribution of detected sources in a mock {\it Gaia} catalog. This results in a prior that peaks at $d=2.7$ kpc and falls off exponentially at large distances.
\end{enumerate}

In Figure~\ref{fig:distance_prior}, we compare the results of our kinematic modeling for one object, J0546+0836, with these three priors. The maximum and minimum distance and velocity are similar for all three priors, as these are set respectively by the requirement that $v_{\rm ejection} < 3000\,\rm km\,s^{-1}$ and by the measured parallax and RV. The shape of the distance and velocity posteriors do depend non-negligibly on the prior; this is unavoidable when the fractional parallax error is large. 

\begin{figure*}
    \centering
    \includegraphics[width=\textwidth]{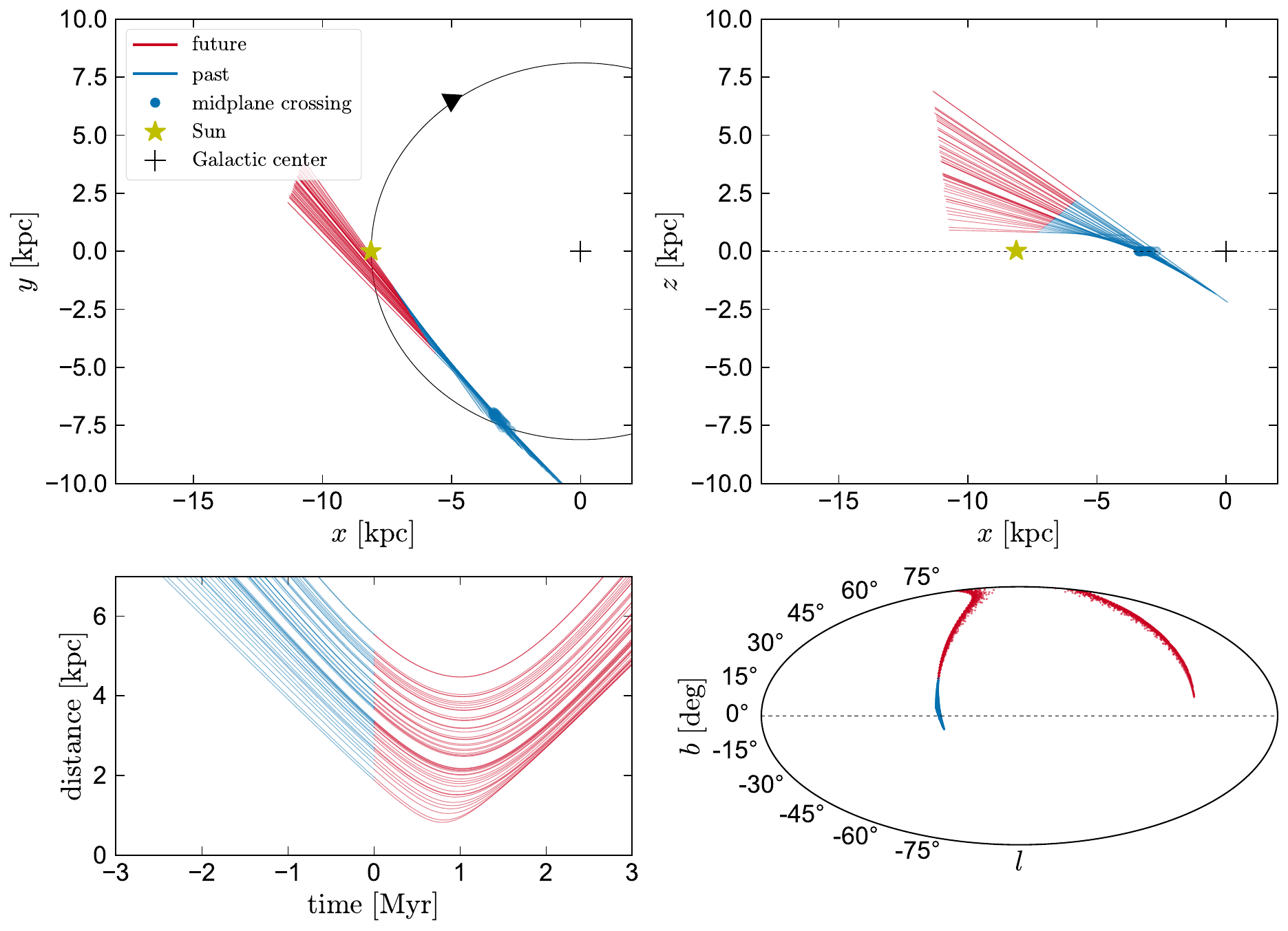}
    \caption{Constraints on the trajectory of J1235-3752. Past and future motion for $\pm 3$\,Myr are shown in blue and red; the source's current position is along the locus where blue and red lines meet. Points along each sampled trajectory in the top two panels show the predicted crossing of the disk midplane. This object is currently $\sim 1.5$--2\,kpc above the disk and its trajectory intersects the midplane $\sim 2$\,Myr ago. Its trajectory at the time was aligned with Galactic rotation, so the object most likely received a $\sim 200\,\rm km\,s^{-1}$ boost to its total velocity from Galactic rotation. We assume straight-line trajectories here and elsewhere in the paper; the apparent curvature in the upper-right panel reflects the fact that trajectories with closer distances imply lower total velocities.}
    \label{fig:trajectory6156}
\end{figure*}

No prior is truly uninformative. While a flat distance prior may seem neutral, Figure~\ref{fig:distance_prior} shows that it implied an absolute-magnitude prior that rises steeply toward bright $M_{G,0}$. A flat prior in absolute magnitude, on the other hand, implies a distance prior $p(d)\propto 1/d$. The exponentially decreasing space density prior implies a steeper falloff at large distances, and typical distance of $2L= 2.7$\,kpc. 

We do expect the distance distribution of detected objects to eventually fall off at large distances because more-distant objects are fainter. However, {\it where} it falls off depends significantly on the intrinsic absolute magnitude distribution, which is quite uncertain. This means that the distances and total velocities of objects in our sample will necessarily be uncertain until the parallaxes are measured more precisely. On the other hand, the {\it minimum} distance and velocity are fairly robust. 

We adopt the flat prior on $M_{G,0}$ as fiducial. We report constraints resulting from the other two priors in Appendix~\ref{sec:appendix_priors}. 


\subsection{Results}
 Figure~\ref{fig:corner} shows constraints on the posteriors of some parameters of interest for the four new $\rm D^6$ star candidates. All parameters are strongly correlated with one another, reflecting the fact that most of the uncertainty ultimately stems from the uncertain distances. 

\begin{figure*}
    \centering
    \includegraphics[width=\textwidth]{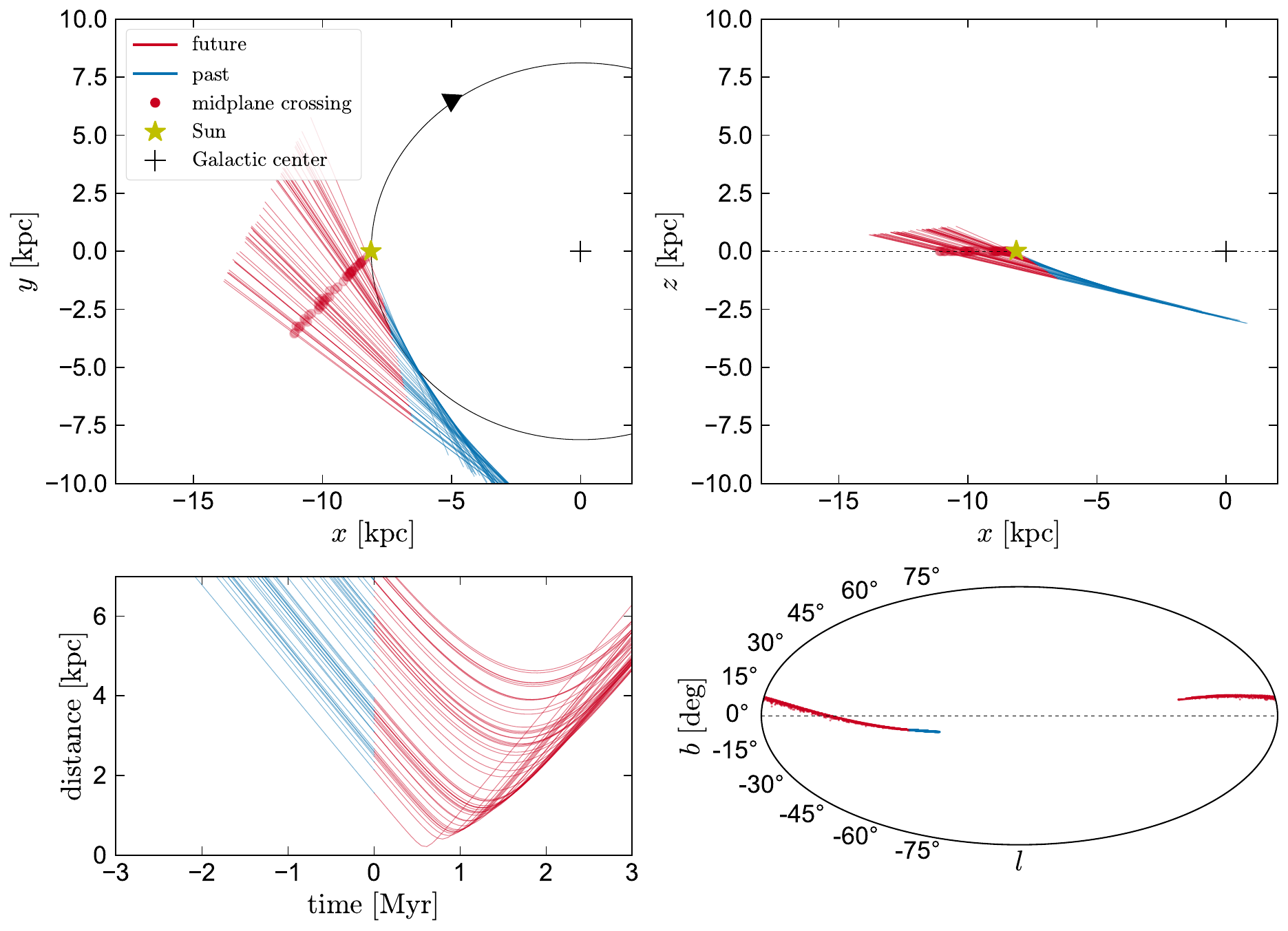}
    \caption{Similar to Figure~\ref{fig:trajectory6156}, but for J0927-6335. This object is unambiguously below the disk today, but it is traveling upward {\it toward} the disk. The star's current distance from the midplane is $z\approx -0.7\pm 0.3$\,kpc, implying that it was born at least that far from the disk. This suggest the star was born from a kinematically hot population. Like J1235-3752, it received a $\sim 200\,\rm km\,s^{-1}$ boost from Galactic rotation. Its past trajectory in the sky (blue points in bottom right panel) is quite well-constrained. }
    \label{fig:trajectory5250}
\end{figure*}

\begin{figure*}
    \centering
    \includegraphics[width=\textwidth]{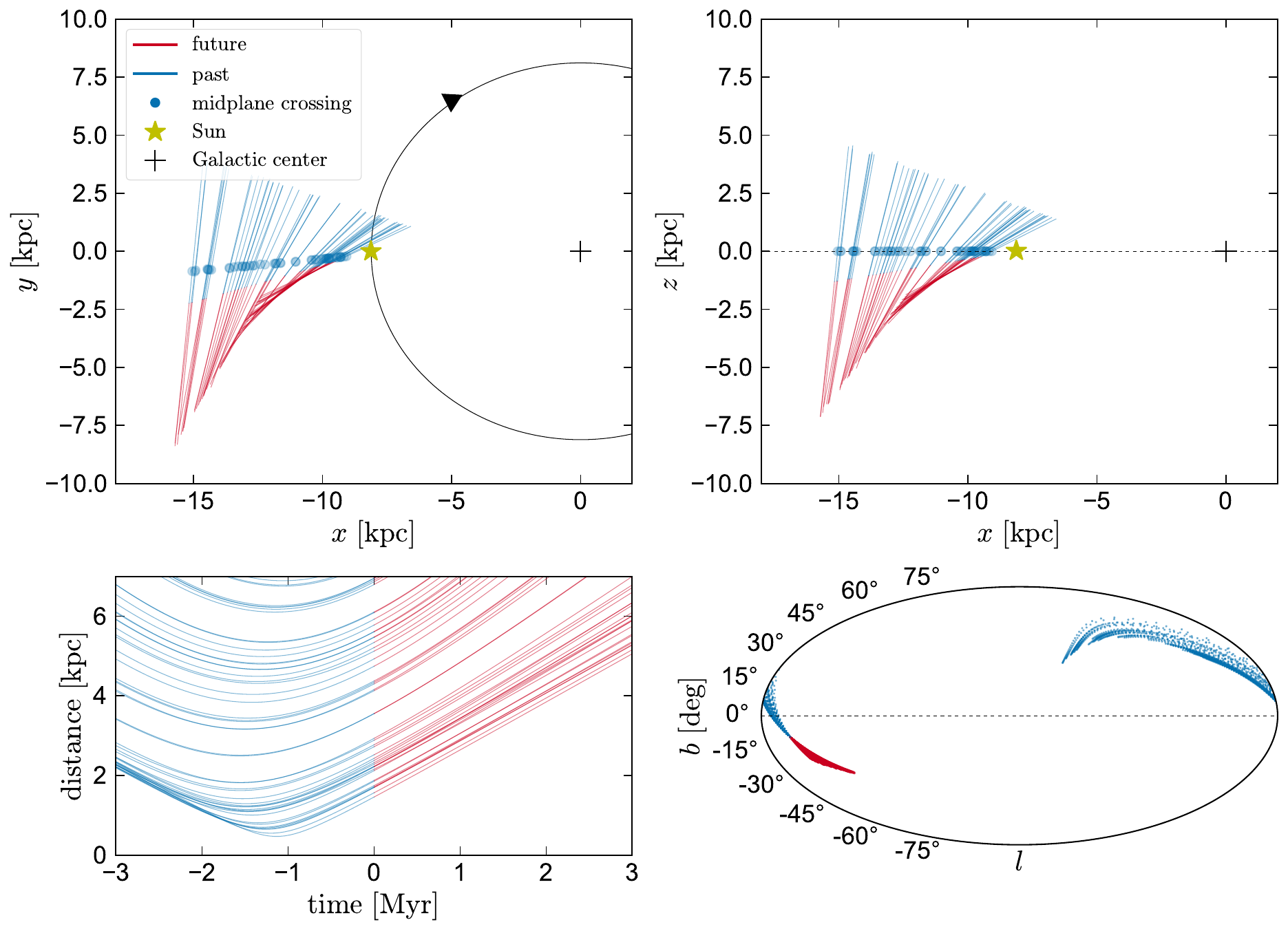}
    \caption{Similar to Figure~\ref{fig:trajectory6156}, but for J0546+0836. This object is toward the Galactic anticenter and is moving away from the disk; its trajectory intersects the midplane 0.6\,Myr ago. It is moving in the opposite direction of the local Galactic rotation where its trajectory intersects the disk and was thus slowed $\sim 200\,\rm km\,s^{-1}$ by Galactic rotation. }
    \label{fig:trajectory3335}
\end{figure*}

\begin{figure*}
    \centering
    \includegraphics[width=\textwidth]{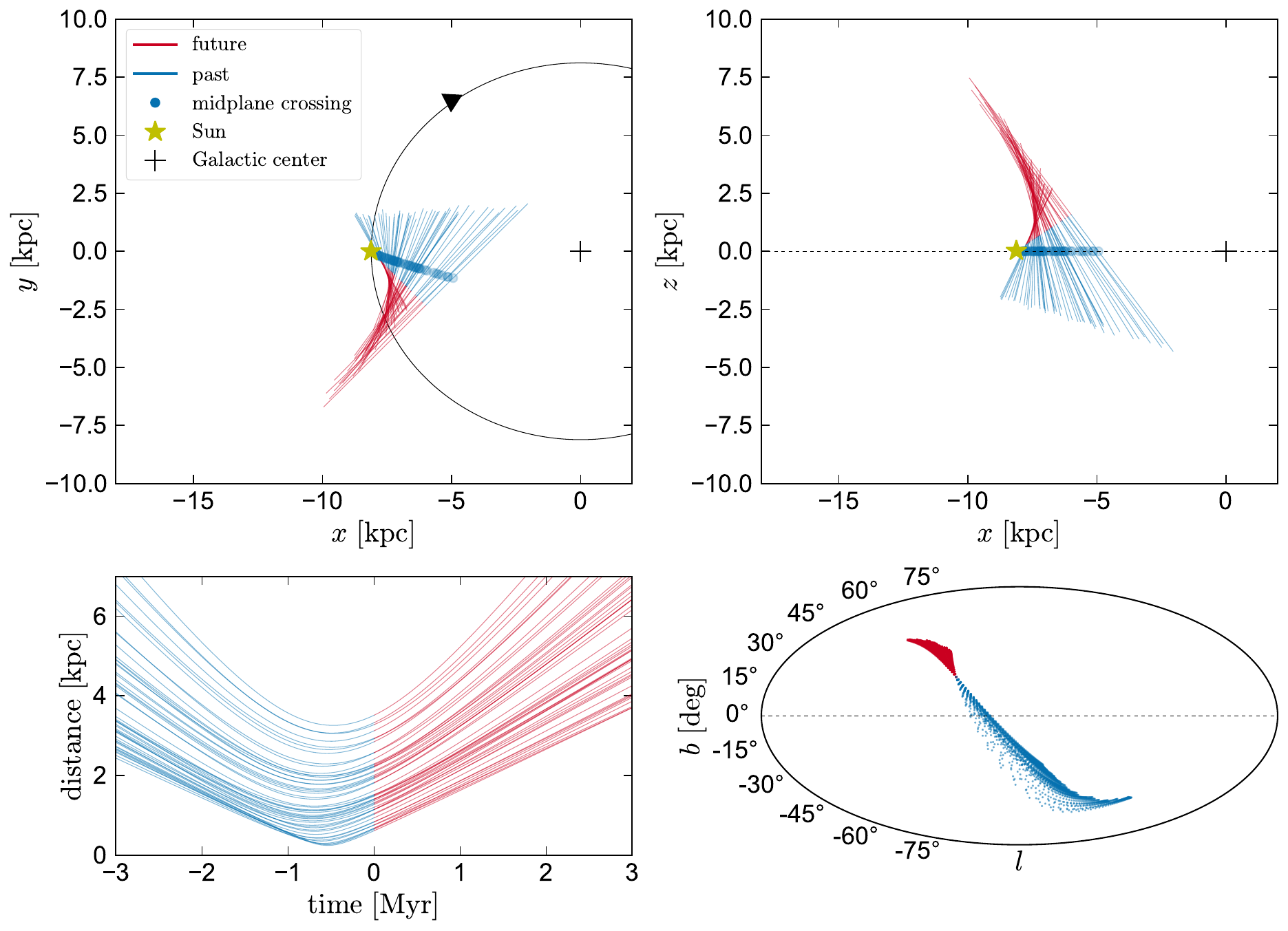}
    \caption{Similar to Figure~\ref{fig:trajectory6156}, but for J1332-3541, the only new D6 star with a helium-dominated atmosphere. This object is $\sim 0.7$\,kpc above the disk, with a flight time back to the midplane of $ 0.64_{-0.14}^{+0.13}$\,Myr. Its flight was slowed $\sim 100\,\rm km\,s^{-1}$ by Galactic rotation.  }
    \label{fig:trajectory6164}
\end{figure*}

Figures~\ref{fig:trajectory6156}--\ref{fig:trajectory6164} illustrate possible trajectories for each object generated from these posteriors. Each line shows a single sample from the posterior, and blue and red lines extrapolate the trajectory $\pm 3$\,Myr into the past and future. The star's current position is thus somewhere along the locus where red and blue lines meet. Circles along each trajectory show the point at which it intersects the midplane (i.e., $z=0$).  It is clear from these trajectories that J1235-3752 and J0927-6335 were boosted by Galactic rotation, while J0546+0836 and J1332-3541 were slowed by it. Similarly, the trajectories demonstrate that J1235-3752, J0546+0836, and J1332-3541 are moving away from the disk, as expected for an object recently launched from near the midplane, while J0927-6335 is moving toward it. As expected, none of these objects have trajectories plausibly passing through the Galactic Center.

Searching backward along these trajectories, it may be possible to identify a surviving supernova remnant. Because their distances are uncertain, the past trajectories of each star except J0927-6335 cover a significant fraction of the sky, and all three of the objects whose past trajectories cross the Galactic plane plausibly encounter multiple remnants. Establishing a high-confidence association would require a careful analysis of chance-alignment probabilities, which we defer to future work.

\begin{figure*}
    \centering
    \includegraphics[width=\textwidth]{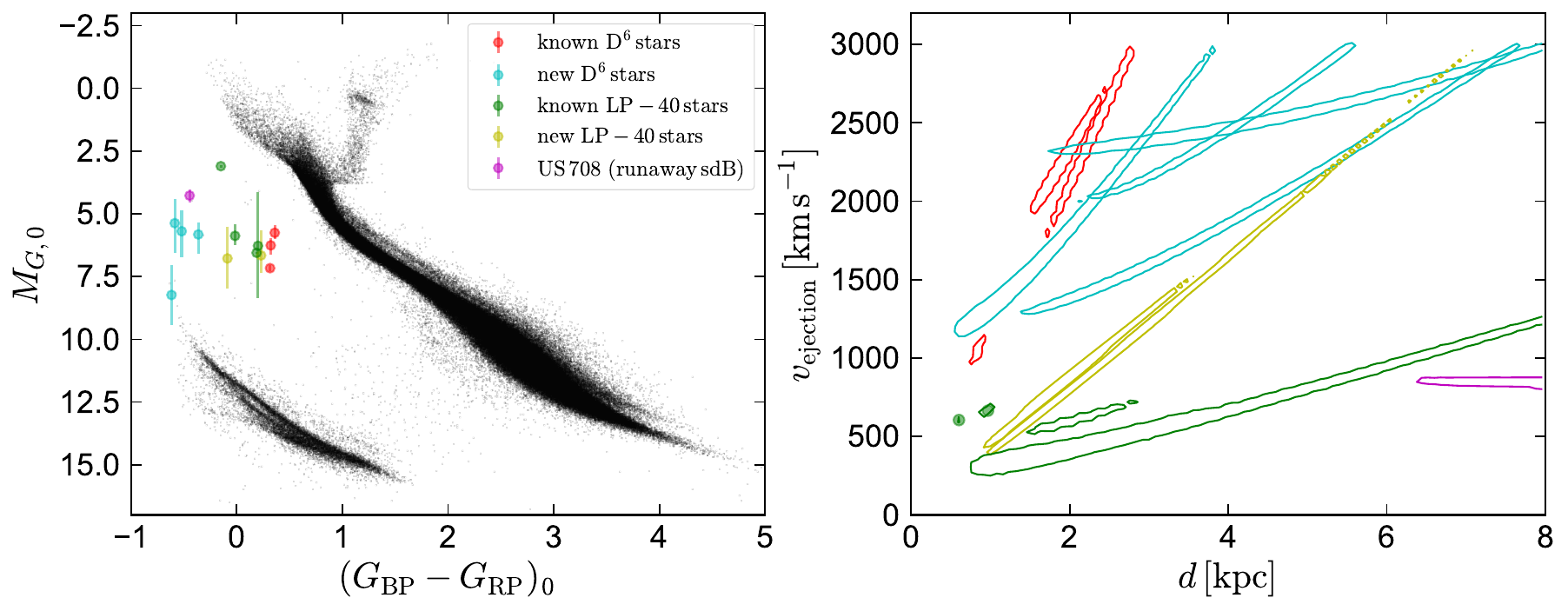}
    \caption{Comparison of our newly discovered D$^6$ stars (cyan) and LP~40-365 stars (yellow) to previously known D$^6$ stars (red), LP~40-365 stars (green), and the runaway sdB star US 708 (magenta). While the newly discovered LP~40-365 stars inhabit a region of the CMD similar to previously known LP~40-365 stars, the new D$^6$ stars are all significantly bluer than the previously known D$^6$ stars, implying that they are hotter and more compact. They are also smaller than core helium burning sdB stars like US 708. Right panel shows our constraints on the distance and ejection velocity of each object. The D$^6$ stars all have minimum ejection velocities above $1000\,\rm km\,s^{-1}$, while the LP~40-365 stars are all consistent with $v_{\rm ejection}\approx 600\,\rm km\,s^{-1}$. US 708, the only runaway sdB star, has an inferred $v_{\rm ejection}\approx 850\,\rm km\,s^{-1}$, between the two populations.} 
    \label{fig:cmd}
\end{figure*}

\section{Census of the hypervelocity WD population}
\label{sec:census}


To understand how the newly discovered objects fit in with the rest of the suspected thermonuclear supernova runaway population, we compile a list of candidates from the literature in Table~\ref{tab:all_systems}. To our knowledge, it contains all objects that have been credibly proposed to be runaways from thermonuclear events. We use the kinematic modeling described in Section~\ref{sec:kinematic_modeling} for all objects. For US~708, we adopt a distance prior of $d\,[\rm kpc]\sim \mathcal{N}(8.5, 1.0)$ following \citet{Geier2015}. 

The two largest classes discovered so far are the $\rm D^6$ and LP~40-365 stars. We plot our constraints on the CMD positions, distances, and birth kick velocities of these objects in Figure~\ref{fig:cmd}. Although most of our newly discovered objects have parallaxes consistent with 0 and large uncertainties, their absolute magnitudes are reasonably well-constrained because their parallaxes rule out close distances, while large distances would imply ejection velocities above $3000\,\rm km\,s^{-1}$. 

Almost all of the objects shown in Figure~\ref{fig:cmd} have similar absolute magnitudes in the optical, $M_{G,0}\approx 5$--7. The only exceptions are the LP~40-365 star J1825-3757, which with $M_{G,0}\approx 3$\,mag is larger and brighter in the optical than any of the other objects, and J1332-3541, which is smaller and somewhat fainter. We note that the bolometric luminosities of these objects span a wider range than their optical magnitudes, ranging from $\sim 0.1\,L_{\odot}$ for the coolest $\rm D^6$ star, to $\gtrsim 10^2\,L_{\odot}$ for the hottest. 

Curiously, the three original $\rm D^6$ stars discovered by \citetalias{Shen2018} and the new hot  $\rm D^6$ stars presented here fall in two distinct clumps in the CMD, on either side of the LP~40-365 stars. If the two sets of objects fall at different places along an evolutionary track, this would seem to imply that $\rm D^6$ stars first spend a significant period of time in the cool/bloated clump and then rapidly contract and heat up at near constant (optical, not bolometric) luminosity. While it is tempting to draw an evolutionary arrow from cool $\rm D^6$ stars to LP~40-365 stars to hot $\rm D^6$ stars, the different inferred ejection velocities of the two populations (right panel) makes this unlikely. 

The hot D$^6$ stars have overlapping temperatures and optical colors with core helium burning sdB stars; they fall below them in the CMD due to the D$^6$ stars' smaller radii. Because their high velocities require the  D$^6$ stars to have been $\gtrsim 10^4$ times denser than typical sdB stars at the time of their ejection, it is unlikely that they were ejected while still burning helium \citep[e.g.][]{Neunteufel2022}. A lack of photospheric hydrogen and helium also makes these objects spectroscopically distinguishable from sdB stars.

The right panel of Figure~\ref{fig:cmd} shows joint constraints on the ejection velocities and distances of the $\rm D^6$ and LP~40-365 stars. For the four previously known LP~40-365 stars, we find ejection velocities consistent with $\rm 600\,km\,s^{-1}$, which would allow a fairly clean separation between $\rm D^6$ and LP~40-365 stars on the basis of their velocities. The lower limits on $v_{\rm ejection}$ inferred for the new LP~40-365 stars continue this trend, though the uncertainties are significant. 

D6-2, the nearest object classified as a $\rm D^6$ star, has an inferred $v_{\rm ejection}$ that is slower than  any of the other objects with that classification, and consistent with the values we infer for the two new LP~40-365 stars. No detailed spectroscopic analysis has been published for that star. However, we compared its optical spectrum from \citet{Chandra2022} to that of J1311-1846, which has a similar effective temperature, and we find significant differences: J1311-1846 has much stronger Mg~I and Na lines, while D6-2 has stronger C and Si lines. These differences support classifying D6-2 as a $\rm D^6$ star.

Among the objects classified as $\rm D^6$ stars, the three cool, bloated objects discovered by \citetalias{Shen2018} are almost certainly closer to Earth on average than the hotter objects discovered in this work.  The simplest interpretation is that the hotter objects are rarer, at least in a magnitude-limited sample. However, J1332 has a distance posterior that overlaps significantly with D6-1 and D6-3; it is most likely the second-closest of the $\rm D^6$ stars.

Table~\ref{tab:all_systems} also lists $t_{\rm disk}= z/v_z$ for each object, which represents a flight time to the disk midplane assuming a straight-line trajectory\footnote{We do not list a flight time for J1240, because this object most likely has had multiple passages through the disk since it was launched.}. Negative values of $t_{\rm disk}$ imply that the object is moving toward the disk midplane and have little physical interpretability; positive values can be viewed as a rough estimate of time since explosion. $t_{\rm disk}$ is often reasonably well-constrained even when $z$ and $v_z$ are independently poorly constrained, because the uncertain distance factors cancel out. 

There is no obvious systematic difference between these ``kinematic ages'' of the hot and cool $\rm D^6$ stars. The LP~40-365 stars on average have somewhat older kinematic ages than the $\rm D^6$ stars. This probably simply reflects the fact that the $\rm D^6$ stars are faster, with a typical velocity of 2\,kpc\,Myr$^{-1}$, and will exit the volume within which we can detect them within 2--3\,Myr. 

Besides the $\rm D^6$ and LP~40-365 stars, we also list in Table~\ref{tab:all_systems} a few additional high-velocity stars with unusual spectra that are likely surviving remnants of thermonuclear supernova. One of them, US~708, is a core helium burning star with a $\sim 12$\,Myr flight time to the disk. This object is very likely the runaway donor from a WD + helium star binary in which the WD exploded \citep[e.g.,][]{Geier2015, Neunteufel2022}. Despite ongoing searches, no other systems in this class have been discovered \citep{Heber2023}. Our follow-up observations included several likely core helium burning stars with high apparent tangential velocities, but none turned out have high RVs (Appendix~\ref{sec:appendix}), suggesting that the sample is dominated by objects with underestimated parallaxes. Sources like US~708 are more luminous in the optical and probably longer lived than the hot $\rm D^6$ stars we have identified. Although they are slower than $\rm D^6$ stars, our search would have been sensitive to them within $\sim 4$--5\,kpc if their typical velocities are $\sim 1000\,\rm km\,s^{-1}$. The fact that only one such system has been discovered (and none within our search volume) thus suggests that they are somewhat rarer than $\rm D^6$ stars. 

There are also two objects whose origin is unclear. One of them, SDSS J1240 \citep{Gansicke2020}, is a WD relatively far down the cooling track ($M_G = 10.3$\,mag) with an oxygen-dominated atmosphere. \citet{Gansicke2020} propose that this object is the remnant of a partially burned lower-mass ($\sim 0.8\,M_{\odot}$) CO~WD that underwent a thermonuclear event. This is the closest object in the sample by a significant margin ($d\approx 0.43$\,kpc), but it is one of the few that is bound to the Milky Way and is likely significantly older than the $\rm D^6$ stars, so it is nontrivial to infer the relative birth rates of such objects and $\rm D^6$ stars. We note that it is unlikely we will find a $\rm D^6$ star so far down the cooling track -- at least, if it was born in the Milky Way -- because $\rm D^6$ stars will have escaped the Galaxy by the time they fully cool and contract. 

Finally, we comment on the object J1637 identified by \citet{Raddi2019}. They classified it as a possible LP~40-365 star, or as an analog to SDSS J1240. We note that its spectrum is quite different from that of J1109+0001 -- the LP~40-365 star we found with the most similar color -- and from that of SDSS J1240. It appears to have more Si and less Mg than either of these objects, perhaps suggesting it is a $\rm D^6$ star. Its distance is poorly constrained, so its total velocity is uncertain. 


\section{Masses of the $\rm D^6$ progenitors}
\label{sec:d6_masses}
\begin{figure*}
    \centering
    \includegraphics[width=\textwidth]{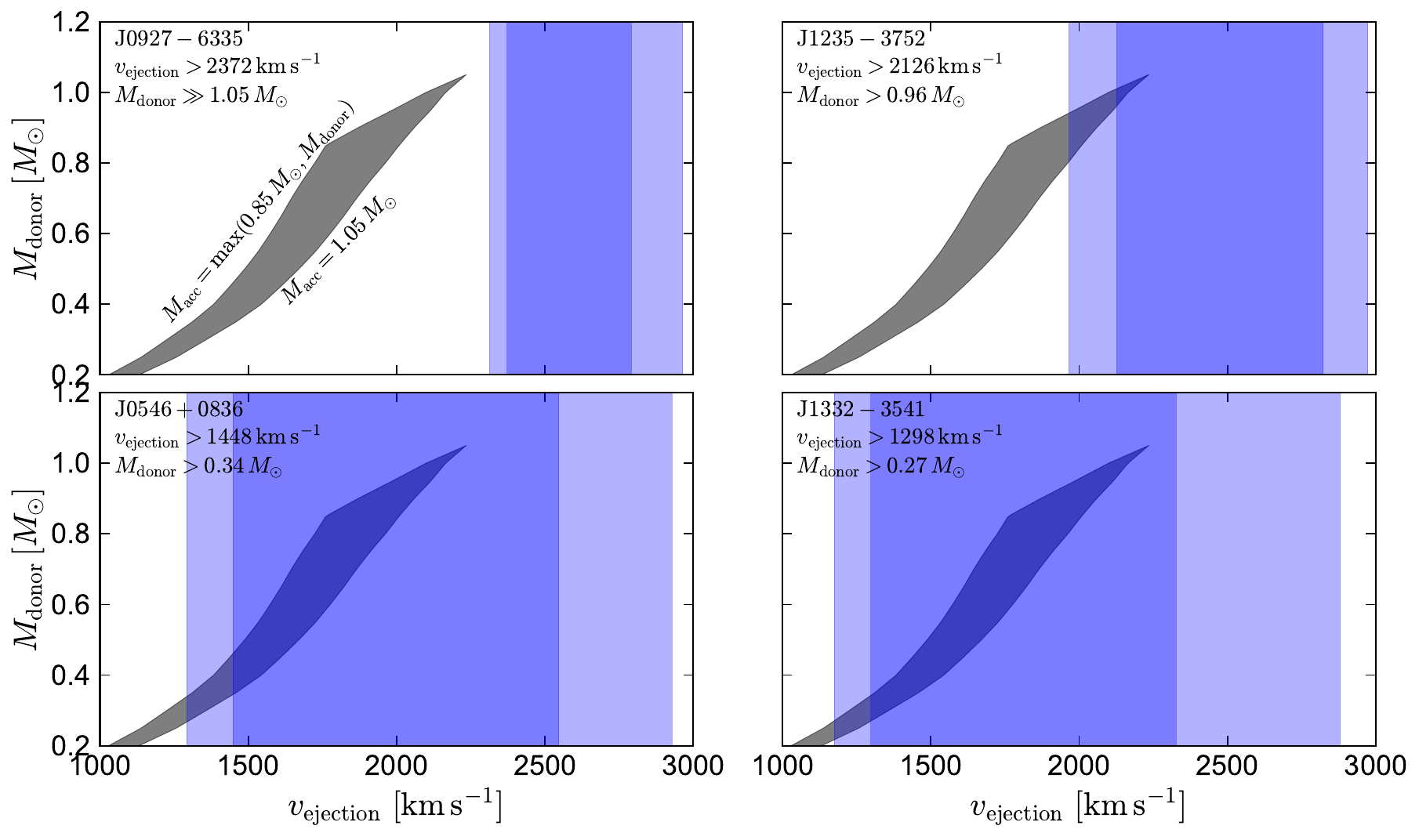}
    \caption{Implied pre-explosion masses of the runaway WDs in the four new $\rm D^6$ star candidates, as inferred from their velocities. Grey shaded region shows the predicted velocity at birth (abscissa) for a range of donor masses (ordinate). At fixed $M_{\rm donor}$, the grey shaded region encloses the most plausible range of accretor masses in the $\rm D^6$ scenario, ranging from $0.85\,M_{\odot}$ (left boundary; subject to the constraint that the accretor is at least as massive as the donor) to $1.05\,M_{\odot}$ (right boundary; the maximum expected mass for a CO~WD). Blue shaded regions show the 1$\sigma$ and 2$\sigma$ constraints on $v_{\rm ejection}$ for each system. The inferred minimum ejection velocities of J0927-6335 and J1235-3752 are uncomfortably high for the $\rm D^6$ scenario: J0927-6335 is too fast to be explained for any donor mass with $M_{\rm acc}\leq 1.05\,M_{\odot}$ (i.e., it likely requires at least one WD more massive than $1.05\,M_{\odot}$); J1235-3752 can only be explained if both WDs had masses above $1.0\,M_{\odot}$. The other two objects are compatible with a wide range of donor masses.}
    \label{fig:vkick_m2}
\end{figure*}

Figure~\ref{fig:vkick_m2} compares our constraints on the measured ejection velocities of the four new $\rm D^6$ candidates to theoretical predictions. The post-SN velocity of the runaway donor is expected to be close to its orbital velocity at the time of the explosion, 
\begin{equation}
    \label{eq:vorb}
    v_{{\rm orb}}=\sqrt{\frac{GM_{{\rm acc}}}{\left(1+q\right)a}}\, ,
\end{equation}
where $M_{\rm acc}$ is the mass of the more massive WD, $q=M_{\rm donor}/M_{\rm acc}$ is the mass ratio, and $a$ is the orbital separation. The orbital separation is set by the fact that the donor fills its Roche lobe,
\begin{equation}
     \label{eq:Roche}
     a=\frac{R_{{\rm donor}}}{0.49}\left[0.6+q^{-2/3}\ln\left(1+q^{1/3}\right)\right]\, ,
\end{equation}
where we use the Roche-lobe radius approximation from \citet{Eggleton1983}. Finally, the radius of the donor is set by a mass-radius relation, which we take from the models of \citet{Bedard2020}. 

In the $\rm D^6$ scenario, the accreting WD is expected to have a CO core. For WDs formed by single-star evolution, this would imply $M_{\rm acc} \lesssim 1.05\, M_{\odot}$ \citep[e.g.,][]{Siess2007}, as higher-mass WDs are expected to have ONe cores that are difficult to detonate owing to the long oxygen-burning lengthscales. We adopt a minimum accretor mass of $0.85\,M_{\odot}$ because double detonation is less likely to occur in lower-mass WDs, whose cores have lower densities \citep[e.g.,][]{Shen2014}. 


Figure~\ref{fig:vkick_m2} shows the range of donor and accretor masses that could produce a given ejection velocity $v_{\rm ejection}$ with grey shading. These calculations are similar to those presented by \citet{Bauer2021}.\footnote{We adopt a maximum accretor mass of $1.05\,M_{\odot}$ here, while they adopted a maximum mass of $1.15\,M_{\odot}$. We consider higher-mass accretors below.} The grey shading is the same in each panel. In blue, we show the constraints on $v_{\rm ejection}$ for the four individual objects, with a different object shown in each panel, and dark and light shading showing $1\sigma$ and $2\sigma$ constraints.

For J0927-6335, the inferred ejection velocity is too high to plausibly be explained in the $\rm D^6$ scenario, even with two $1.05\,M_{\odot}$ WDs. If we adopt the 2$\sigma$ lower limit on $v_{\rm ejection}$ of $\rm 2315\,\rm km\,s^{-1}$, this is still comfortably above the maximum velocity that can be achieved with two $1.05\,M_{\odot}$ WDs, which in our calculations is $\sim 2235\,\rm km\,s^{-1}$. For J1235-3752, the observationally inferred ejection velocity can plausibly be explained, but only if both WDs had masses near $1.05\,M_{\odot}$, meaning that the total mass of the binary was $\gtrsim 2.0\,M_{\odot}$.

\begin{figure}
    \centering
    \includegraphics{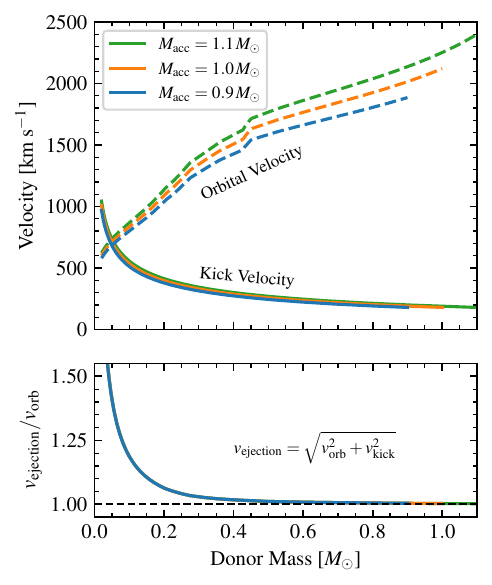}
    \caption{Orbital and kick velocities for WD donors in D$^6$ systems. The dashed lines in the upper panel show orbital velocities for Roche-lobe-filling donors assuming the mass-radius relation for $T_{\rm eff} = 10,000\, \rm K$ DA~WDs following \citet{Bauer2021}, while the solid lines show kick velocities calculated according to Equation~\eqref{eq:vkick}. The lower panel shows that the total ejection velocity is nearly identical to the final orbital velocity for higher donor masses corresponding to higher velocities.}
    \label{fig:vkick_vs_vorb}
\end{figure}

It is in principle possible that the binary progenitors of these systems had a higher velocity before the explosion than assumed in our modeling, in which case our constraints on $v_{\rm ejection}$ could be overestimated. However, this seems unlikely, since our modeling already assumes a boost of $\sim 200\,\rm km\,s^{-1}$ due to Galactic rotation. 
While the donor WD can receive an additional kick from SN ejecta interaction to boost its total velocity \citep{Bauer2019}, the overall magnitude of this kick is likely negligible compared to the orbital velocity, particularly for the highest-mass systems needed to produce velocities greater than $2000\, \rm km\,s^{-1}$. The kick velocity can be estimated as the momentum imparted by SN ejecta according to
\begin{equation}
\label{eq:vkick}
v_{\rm kick} = \eta \frac{\pi R_{\rm don}^2}{4\pi a^2} \frac{p_{\rm ej}}{M_{\rm don}}\, ,
\end{equation}
where $p_{\rm ej}$ represents the total momentum carried by SN ejecta, $\pi R_{\rm don}^2/4\pi a^2$ is the fraction of the solid angle intersecting the donor for ejecta interaction, and $\eta \approx 1/3$--1/2 is a momentum transfer efficiency factor \citep{Hirai2018}. While simulations suggest that $\eta \approx 1/3$ may be a typical value for nondegenerate donor stars \citep{Hirai2018,Bauer2019}, the relatively higher internal pressure of a WD donor makes it likely that it will be somewhat more efficient in capturing momentum from the ejecta, which would lead to an efficiency closer to $\eta \approx 1/2$. Figure~\ref{fig:vkick_vs_vorb} shows the resulting kick velocities assuming $\eta = 1/2$ and $p_{\rm ej} = M_{\rm acc} \times 10{,}000\,\rm km\, s^{-1}$, with the WD mass-radius relation taken from \citet{Bauer2021} and the orbital velocities obeying Equations~\eqref{eq:vorb} and \eqref{eq:Roche}. The kick velocity should be nearly perpendicular to the orbital velocity of the donor, so the final ejection velocity $v_{\rm ejection} \approx \sqrt{v_{\rm orb}^2 + v_{\rm kick}^2}$ is very similar to $v_{\rm orb}$ for all but the lowest-mass (and slowest) donor WDs. The velocity boost due to SN kicks is negligible in the regime of high velocities ($v \gtrsim 1500\,\rm km\, s^{-1}$).
The high observed velocities of J1235-3752 and particularly J0927-6335 are thus difficult to explain with canonical-mass CO~WD accretors with $M<1.05\,M_{\odot}$, even with possible boosts from supernova ejecta taken into account.

As an alternate explanation, we consider the possibility that the accretors had masses above $1.05\,M_{\odot}$. This would not be expected for CO~WDs formed in isolation, but it can occur as a result of stable mass transfer from a He-star companion, which can lead to steady He burning on the surface of the CO~WD \citep[e.g.,][]{Yoon2003, Piersanti2014, Brooks2016}.  Some population-synthesis models predict that this scenario is fairly common, occurring in $\sim 40$\% of double-degenerate SN~Ia progenitors \citep{Ruiter2013}, though these predictions depend sensitively on several uncertain processes in binary evolution.

\begin{figure*}
    \centering
    \includegraphics[width=\textwidth]{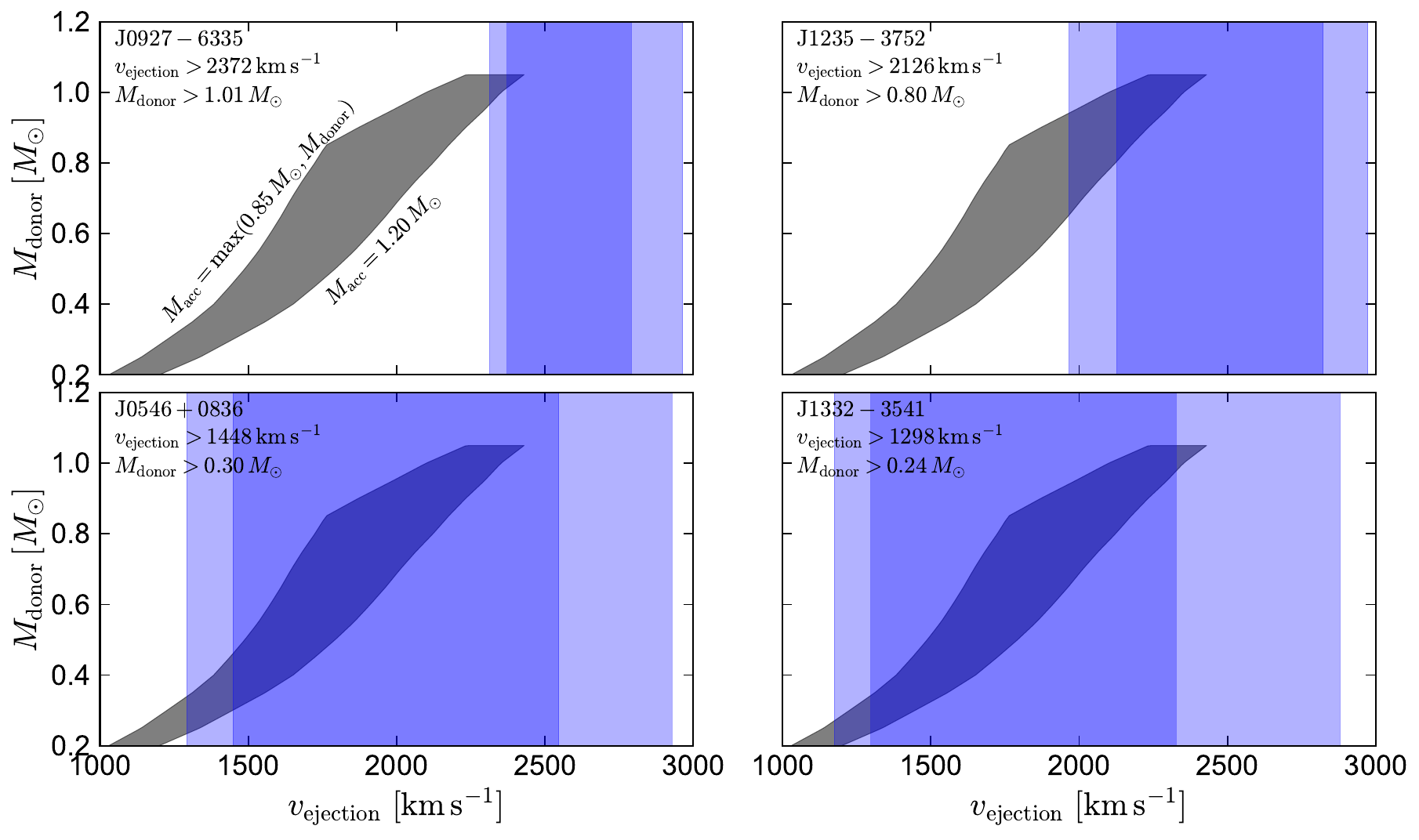}
    \caption{Same as Figure~\ref{fig:vkick_m2}, but now assuming a maximum accretor mass of $1.20\,M_{\odot}$. This represents a scenario where a CO~WD accretor grew via stable mass transfer prior to the formation of the second WD. We still assume $M_{\rm donor} < 1.05\,M_{\odot}$. In this case the high velocities of J0927-6335 and J1235-3752 can be explained, but they still require the donor to have been another high-mass WD, with $M_{\rm donor}>1.01\,M_{\odot}$ for J0927-6335 and $M_{\rm donor}>0.8\,M_{\odot}$ and J1235-3752. }
    \label{fig:vkick_m2_1.2}
\end{figure*}

Figure~\ref{fig:vkick_m2_1.2} compares our constraints on the observed ejection velocities to predictions in the $\rm D^6$ scenario for a maximum accretor mass of $1.20\,M_{\odot}$. A higher-mass accretor leads to moderately higher predicted $v_{\rm ejection}$ at fixed $M_{\rm donor}$, such that the observed velocities of the two fastest stars can plausibly be reproduced. This still requires the donor to have been quite massive, however, with $M_{\rm donor} \gtrsim 1\,M_{\odot}$ for the two fastest stars.

\section{Discussion}
\label{sec:disc}

\subsection{ Where are the lower-mass runaway donors?} Our kinematic modeling implies ejection velocities for two of the four new $\rm D^6$ stars that can only be explained by two high-mass WDs, with total binary masses of $\gtrsim 2.0\,M_{\odot}$. Similarly, \citet{Bauer2021} found that two of the three original $\rm D^6$ stars from \citetalias{Shen2018} must have had total binary masses of $\gtrsim 1.6\,M_{\odot}$. There is one $\rm D^6$ star whose velocity conclusively requires a low-mass donor: D6-2, whose inferred $v_{\rm ejection}\sim 1040\,{\rm km\,s^{-1}}$ implies a donor mass of 0.15--0.5\,$M_{\odot}$, depending on its temperature at the time of merger \citep{Bauer2021}. We suspect that J1332-3541 is likely also low-mass, since it is the only object with a helium-dominated atmosphere, but its velocity is consistent with a wide range of donor masses.

Binaries containing two massive WDs are probably intrinsically rare compared to those containing a lower-mass WD, both because high-mass WDs are disfavored by the initial-mass function (IMF) and because mass loss due to a companion tends to reduce the mass of the WD produced by a star of a given initial mass. Observationally, the intrinsic mass distribution of binary WDs is quite uncertain. Among $\sim 200$ known double-degenerate close binaries in the Milky Way \citep[e.g.,][]{Brown2020, Burdge2020, El-Badry2021b, Kosakowski2023}, none are known to have a total mass exceeding the Chandrasekhar limit. However, observational biases strongly favor low-mass WDs, and only small numbers of massive WDs have been subject to high-resolution RV monitoring \citep[e.g.,][]{Napiwotzki2001, Napiwotzki2020}. Attempts to infer the WD+WD binary merger rate from RV surveys \citep[e.g.,][]{Badenes2012, Maoz2017, Maoz2018} have  concluded that the merger rate of binaries with total mass $\gtrsim 1.4\,M_{\odot}$ is about an order of magnitude lower than the merger rate of all WD+WD binaries.

Population-synthesis simulations also generally predict that the merger rate of WD+WD binaries containing at least one low-mass WD exceeds the merger rate of massive WD+WD binaries \citep[e.g.,][]{Yungelson2017}. Here we are interested in particular in the relative merger rates of (a) binaries containing a massive CO~WD with mass $\gtrsim 0.85\,M_{\odot}$ and a low-mass WD, and (b) binaries containing two massive WDs. This is not a quantity many population-synthesis simulations have specifically predicted, but \citet[][their Figure 9]{Toonen2012} show that even among double CO~WD binaries with $M_1> 0.85\,M_{\odot}$, the predicted merger rate of systems with $M_2 < 0.6\,M_{\odot}$ exceeds that of systems where $M_2 > 0.8\,M_{\odot}$. The quantitative predictions depend significantly on the adopted common envelope model, which is uncertain \citep[e.g.,][]{Nelemans2001, vanderSluys2006, Toonen2012}. 

If the coalescence of binaries containing a lower-mass WD also led to a double detonation and the formation of a hypervelocity WD, we would expect to detect $\rm D^6$ stars with ejection velocities of $\rm 1000-1500\,\rm km\,s^{-1}$ in larger numbers than those with $v_{\rm ejection} > 2000\,\rm km\,s^{-1}$. Our search included  all sources with tangential velocities above $600\,\rm km\,s^{-1}$, and thus should have been similarly sensitive to fast and ``slow'' $\rm D^6$ stars, if their luminosities and lifetimes are similar. 

These considerations suggest that either (a) WD+WD mergers with low-mass donors do not produce surviving hypervelocity WDs (either there is no double detonation, or the donor is destroyed), or (b) the hypervelocity WDs produced by lower-mass donors are fainter and/or shorter-lived than those produced by more-massive donors. There is some support for option (b) in the population of $\rm D^6$ stars discovered thus far: the two nearest $\rm D^6$ stars, D6-2 and J1332-3541, are also the slowest (Figure~\ref{fig:cmd}). D6-2 is likely the youngest among the known $\rm D^6$ stars: it is associated with a supernova remnant, to which it has a flight time of $\sim 10^5$\,yr, while most of the other $\rm D^6$ stars have kinematic ages of order $\rm 10^6$\,yr. D6-2 and J1332-3541 are less luminous than any of the other $\rm D^6$ stars. These considerations suggest that there are likely more objects like D6-2 and J1332-3541 with larger distances and/or older ages, that have not been detected yet because they are faint. 
Quantitative models for the thermal evolution of $\rm D^6$ stars are required to test whether a scenario where the low-mass stars are fainter and have shorter lifetimes is tenable.



\subsection{How does the birth rate of D$^6$ stars compare to the SN~Ia rate?}
\label{sec:rates}

\begin{figure*}
    \centering
    \includegraphics[width=\textwidth]{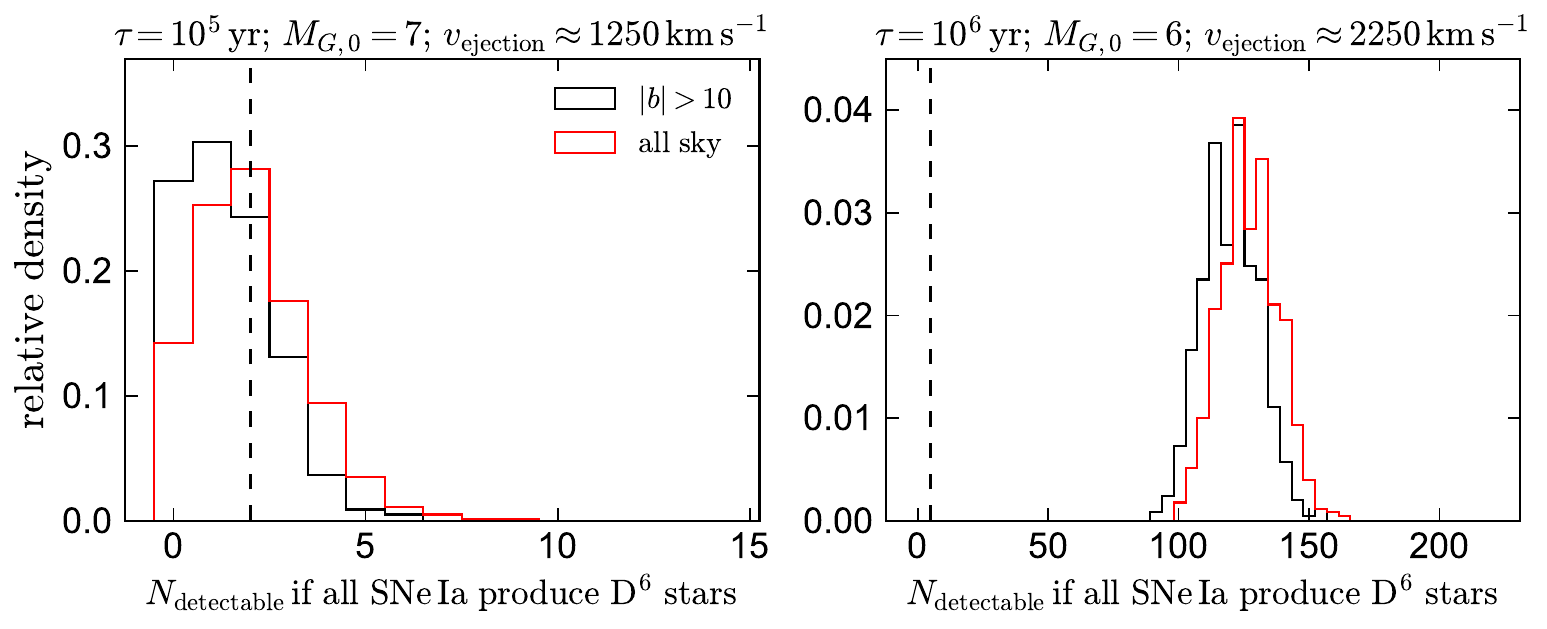}
    \caption{Expected number of $\rm D^6$ stars detectable by our search if all Galactic SNe~Ia produce $\rm D^6$ stars. We consider two possible models for $\rm D^6$ stars. The left panel shows objects with $M_{G,0} = 7$\,mag, a luminous lifetime of $10^5$\,yr, and a birth velocity of $1250\,\rm km\,s^{-1}$, perhaps similar to the lower-mass D$^6$ stars D6-2 and J1332-3541. Here, our detection of two objects (dashed line) is consistent with a scenario where all SNe~Ia produce a D$^6$ star. The right panel shows objects with  $M_{G,0} = 6$\,mag, a luminous lifetime of $10^6$\,yr, and a birth velocity of $2250\,\rm km\,s^{-1}$, perhaps representative of the other D$^6$ stars. Detection of five such objects is consistent with only 3--5\% of SNe~Ia producing such a D$^6$ star. }
    \label{fig:rate_estimate}
\end{figure*}

To test whether the observed population of $\rm D^6$ stars is consistent with a scenario where most SNe~Ia produce one, we performed simple Monte Carlo simulations and applied an approximate selection function for our search.
 We approximate the stellar content of the Milky Way as a disk with exponential radial scale length of 2600\,pc and exponential scale height of 300\,pc \citep{Bland2016}. SNe~Ia are assumed to be distributed evenly throughout the disk in a mass-weighted sense, with one $\rm D^6$ star launched every 300\,yr on average, traveling in a straight line with no deceleration.

We considered two populations of D$^6$ stars. For the first, representing low-mass ($M\lesssim 0.4\,M_{\odot}$) stars, we assume a uniform ejection velocity distribution between 1000 and 1500\,$\rm km\,s^{-1}$ and an absolute magnitude of $M_{G,0}=7$. We ``turn off'' these stars after $10^5$\,yr, assuming they will become too faint to detect. This lifetime is motivated by the $10^5$\,yr flight time of D6-2 to its associated remnant (\citetalias{Shen2018}), but the estimate is admittedly uncertain. For the second, we assume a uniform ejection velocity distribution between 2000 and 2500\,$\rm km\,s^{-1}$ and absolute magnitude of $M_{G,0}=6$, with a lifetime of $10^6$\,yr. These properties are motivated by the velocities, absolute magnitudes, and flight times to the disk midplane of the faster $\rm D^6$ stars. 

For stars with Galactic latitudes $|b| < 10^\circ$, we assume extinction $A_V = 1\,\rm mag\,kpc^{-1}$, with $A_G = 0.85\,A_V$. No extinction is assumed for stars with $|b| > 10^\circ$. We consider a simulated $\rm D^6$ star ``detected'' if it satisfies $G < 20$\,mag, $\mu > 50\,\rm mas\,yr^{-1}$, and $v_\perp > 600\,\rm km\,s^{-1}$. The results of this experiment are shown in Figure~\ref{fig:rate_estimate}. We find that the discovery of two stars like D6-2 and J1332-3541 is consistent with a production rate that is $\sim 100\%$ of the Milky Way's SN~Ia rate, subject to small-number statistics.  The population representing the five other D$^6$ stars is consistent with a production rate of $\sim 3-5\%$ of the SN~Ia rate.  These are very rough numbers owing to the many uncertainties both in the detection efficiency and in the approximations used in the theoretical calculation, but they are at least consistent with the possibility that a large fraction, and perhaps all, SNe~Ia are produced via the D$^6$ scenario, with most occurring in binaries with low-mass donors and a small fraction with high-mass donors.

\citet{Igoshev2023} recently carried out a search for hypervelocity WDs among {\it Gaia} sources with $\varpi/\sigma_{\varpi} > 4$. Finding no new sources with peculiar spectra and high velocities, they concluded that the birth rate of $\rm D^6$ stars is ``at least two orders of magnitudes less than the inferred SN~Ia rate,'' in conflict with our estimates above. However, their work did not include simulations to estimate which $\rm D^6$ stars could actually be detected. The detection efficiency and birth rate implied by the observed population depends critically on the luminosity and lifetime of all  $\rm D^6$ stars (not just those that are detected), which is still very poorly constrained. Robust evolutionary models are needed to overcome these uncertainties.

\subsection{What is the evolutionary link between hot and cool $\rm D^6$ stars?} The three $\rm D^6$ stars disovered by \citetalias{Shen2018} fall in a tight clump in the CMD, suggesting temperatures of 6000--7000\,K and radii of 0.2--0.4\,$R_{\odot}$. Our newly discovered objects fall in a bluer clump in the {\it Gaia} CMD, with temperatures ranging from $\sim 20,000$ to $>100,000$\,K. It is natural to identify the two populations of $\rm D^6$ stars with different stages on an evolutionary track. 

In their evolutionary models for SN~Iax postgenitors, \citet{Zhang2019} predict that after energy is injected into the envelope of a WD, it will first puff up and cool before contracting and heating up, eventually moving back down a cooling track. In this scenario, the new hot $\rm D^6$ stars could simply be older than the cooler and puffier objects. These models can reproduce some aspects of both the hot and cool D$^6$ stars. Their models with $M=0.15\,M_{\odot}$ remain cool and puffy for a few Myr before contracting and have temperatures and luminosities comparable to the cool D$^6$ stars discovered by \citetalias{Shen2018}. Their models with $M=0.3\,M_{\odot}$ have temperatures, radii, and evolutionary lifetimes more similar to the hotter observed objects, reaching a maximum temperature of $T_{\rm eff}\approx 10^5\,\rm K$ at an age of a few $\times 10^5$\,yr before evolving down the cooling track. 

The main challenge to applying these models to the observed $\rm D^6$ stars lies in the high observed velocities, which imply masses of order $1\,M_{\odot}$ for several objects. The high-mass models constructed by \citet{Zhang2019} evolve much more quickly than the low-mass models; those with $M=1.0\,M_{\odot}$ heat up within at most a few thousand years and reach the WD cooling track within $<10^5$\,yr, spending negligible time in the cool and puffy state. This behavior is in contrast with the trend suggested by the observed population, within which the lowest-mass object is most likely the youngest. In addition, the fact that two of the cool $\rm D^6$ stars (D6-1 and D6-3) have high inferred masses \citep{Bauer2021} and are likely closer than most of the hotter stars suggests that $\rm D^6$ stars spend more time in the cool and puffy phase than predicted by the high-mass models. 


The flight times we infer for the new $\rm D^6$ stars are of order 1\,Myr, suggesting that the typical lifetime of the hot phase is at least this long. In particular, J1235-3752 has an inferred flight time of $1.9_{-0.3}^{+0.9}$\,Myr to the disk and is almost certainly more than 1\,kpc above the midplane.\footnote{Of course, we do not really know the scale height of the progenitors before explosion, but unless the $\rm D^6$ scenario is exclusively a low-metallicity phenomenon, there is little reason to expect these stars to be born more than a few 100\,pc from the midplane (see Section~\ref{sec:thickdisk}).} J1235-3752 is the coolest and puffiest of the new $\rm D^6$ stars. This suggests that the star is less evolved than the other hot $\rm D^6$ stars (i.e., it is likely still contracting and heating up) and thus that evolutionary timescales vary within the hot population. Our kinematic modeling yields comparable flight times to the disk midplane for hot and cool $\rm D^6$ stars; that is, there is no clear trend between temperature and inferred flight time.

It is of course also possible that the hot and cool $\rm D^6$ stars represent different evolutionary outcomes, with some runaway WDs remaining cool and puffy for long periods, and others quickly becoming hot and compact. However, both the known hot and cool $\rm D^6$ stars include objects traveling fast enough that they must have formed as massive runaway donors from double-degenerate binaries, so it is not clear what the additional variables could be.

\subsection{Are $\rm D^6$ stars preferentially born from the thick disk?}
\label{sec:thickdisk}
Of the seven $\rm D^6$ stars listed in Table~\ref{tab:all_systems}, two have negative $t_{\rm disk}$, meaning that they are traveling toward the disk. One of these is D6-2, which is associated with a supernova remnant about 300\,pc from the midplane. The other is J0927-6335, with a current elevation of $z=-0.70^{+0.27}_{-0.35}\,\rm kpc$ and vertical velocity $v_z = 519^{+69}_{-56}\,\rm km\,s^{-1}$. If this object was born 1\,Myr ago (a typical kinematic age for the other $\rm D^6$ stars), it would have been launched from $z\approx -1.25\,\rm kpc$. Even if it was born very recently, the 1$\sigma$ lower limit on its birth height is $z=-0.43\,\rm kpc$. This is somewhat farther from the midplane than expected for typical double-degenerate mergers, 50\% (75\%) of which are expected to occur within 1\,Gyr (3\,Gyr) of an episode of star formation \citep[e.g.,][]{Ruiter2009}. In the solar neighborhood, the scale heights of stars younger than 1\,Gyr and younger than 3\,Gyr are respectively $h_z = 100$ and $h_z \approx 220$\,pc \citep{Sollima2019}. We conclude that it is somewhat unexpected for one of the two $\rm D^6$ stars for which a birth height can be inferred to have been born so far from the disk. This could potentially hint at a progenitor channel that favors the thick disk (e.g., low metallicity), but the number of $\rm D^6$ stars known so far is still too small to rule out a statistical fluke.

\section{Summary and conclusions}
\label{sec:conclusions}

We have carried out a search for hypervelocity WDs that are runaways from thermonuclear explosions. Our search extends previous efforts (e.g., \citetalias{Shen2018}; \citealt{Scholz2018}; \citealt{Bromley2018}; \citealt{Bromley2018}; \citealt{Ruffini2019}; \citealt{Raddi2019}; \citealt{Igoshev2023}; \citealt{Heber2023}) by searching to larger distances, where precise parallaxes are not available. We distinguish genuine thermonuclear runaways from false positives via their unusual spectra and high RVs, ultimately identifying six high-confidence candidates with hydrogen-free spectra (Figures~\ref{fig:spec6156}-\ref{fig:speclp40}). The success of our search demonstrates that runaway WDs can still be efficiently identified even at large distances by selecting sources with high proper motions and blue colors. 

In contrast to most of the previously identified hypervelocity WDs, several of our new discoveries are hot and compact, including three objects with effective temperatures above 60,000\,K and one with effective temperature above 100,000\,K. Four objects have RVs faster than $1000\,\rm km\,s^{-1}$. The properties of these objects (i.e., temperatures, luminosities, and surface abundances) are different from those of any previously discovered hypervelocity WDs, but bear some resemblance to the predictions of previously calculated evolutionary models \citep{Zhang2019, Bauer2019}. On the basis of their high velocities, we classify them as $\rm D^6$ stars: runaway donors from double-WD binaries whose companions exploded  \citep{Shen2018b}. Although the new objects are hotter and smaller than the previously identified candidates (\citetalias{Shen2018}), we know of no other viable model to explain the WDs' high velocities and hydrogen-free spectra. This suggests that the hot and compact hypervelocity stars most likely represent a later evolutionary state of the cool and puffy $\rm D^6$ stars. An alternative possibility is that the hot objects are not produced via the D$^6$ scenario at all, but rather through some other kind of thermonuclear event, such as a Type Iax SN or an electron-capture SN. The predicted velocities for this scenario are more uncertain, but are generally lower than in the $\rm D^6$ scenario \citep{Jordan2012,  Kromer2013, Jones2019}.

Assuming D$^6$ stars are launched from the Galactic disk, we constrained their past trajectories, birth velocities, and past and future trajectories (Figures \ref{fig:trajectory6156}-\ref{fig:trajectory6164}), finding typical kinematic ages of $\sim 1$\,Myr. The past trajectories are at present too uncertain to trace these objects to an SN remnant, and any remnants would most likely have dissipated by now anyway. In the future, all of the D$^6$ stars will leave the Milky Way, streaming outward isotropically and traveling $\sim 20$\,Mpc in a Hubble time. If a significant fraction of SNe~Ia produce a D$^6$ star, the Galaxy has likely launched more than $10^7$ of them into intergalactic space. An interesting corollary is that there should be large numbers of faint, nearby D$^6$ stars launched from galaxies all throughout the local volume passing through the Solar neighborhood.

Two of the objects we have discovered -- J1235-3752 and J0927-6335 -- currently have faster RVs than any known stars.\footnote{There are several possible contenders for the title ``fastest stellar RV''. The classical hypervelocity stars thought to be ejected from the Galactic Center \citep{Brown2014, Koposov2020} all have $\rm |RV| < 1050\,\rm km\,s^{-1}$. The star S0-2 in the Galactic Center has been observed with a total velocity of +4000\,$\rm km\,s^{-1}$ near periastron \citep{GRAVITYCollaboration2018, Do2019}; faster than any of the objects studied in this work. However, the star's $\rm |RV|$ only exceeds that of J0927-6335 for $\sim 1$ year every 16 years, and it will next do so in 2034. Its most negative velocity, $-1800\,\rm km\,s^{-1}$, is slower than J0927-6335. The RVs of other S-stars that have been measured so far are all $\rm |RV| < 1400\,\rm km\,s^{-1}$ \citep{Chu2023}. The hypervelocity globular cluster HVGC-1 has an RV of $-1026\,\rm km\,s^{-1}$, and an inferred RV of $\sim 2100-2300\,\rm km\,s^{-1}$ with respect to M87, from which it may have been launched \citep{Caldwell2014}. However, this object is a globular cluster, not a star, and its provenance in M87 is not unambiguous. There are several detected resolved star candidates with presumably large redshifts in lensed galaxies \citep[e.g.][]{Welch2022}, but these do not represent physical velocities.} Although their distances are not well-constrained -- making it difficult to infer tangential velocities -- their RVs alone place stringent lower limits on the velocities. These in turn yield lower limits on the masses of their progenitor WDs (Figures~\ref{fig:vkick_m2} and~\ref{fig:vkick_m2_1.2}). Surprisingly, these limits are quite high, requiring that both WDs in the progenitor binary had masses $\gtrsim 1\,M_{\odot}$. These results suggest that either most surviving SN runaways come from (highly) super-Chandrasekhar-mass binaries or runaways from lower-mass binaries are fainter and/or shorter-lived. Intriguingly, the two nearest D$^6$ stars are also the faintest and likely have the lowest mass, supporting the possibility that there may be significant numbers of fainter, relatively nearby ($d\lesssim 2\,\rm kpc$) $\rm D^6$ stars from lower-mass donors yet to be discovered. Such objects may be efficiently discovered by the Rubin Observatory \citep{LSSTScienceCollaboration2009, Ivezic2019}.

There is now a sizable population of hypervelocity stars associated with thermonuclear SNe. Table~\ref{tab:all_systems} lists the properties of the 16 such objects characterized to date, including 7 that are likely $\rm D^6$ stars and 9 in related classes. Modeling this population will ultimately make it possible to infer the formation rate of thermonuclear runaways and ultimately, the fraction of SNe~Ia formed through the double-degenerate channel. Our estimate of the birth rate of $\rm D^6$ stars is consistent with a scenario in which most SNe~Ia produce a hypervelocity runaway WD but the observed population is dominated by the most massive and brightest runaways (Figure~\ref{fig:rate_estimate}). Models for the thermal evolution of $\rm D^6$ stars are needed for more robust estimates of their birth rate.

\section*{acknowledgments}
We thank Charlie Conroy, JJ Hermes, Ruediger Pakmor, Stephan Geier, Ulrich Heber, and Roberto Raddi for useful discussions, and the anonymous referee for a constructive report. We are grateful to Robert Kurucz for making his programs and databases for spectral synthesis publicly available. The Kavli Institute for Theoretical Physics (KITP) hosted the program, ``White Dwarfs as Probes of the Evolution of Planets, Stars, the Milky Way, and the Expanding Universe,'' during which this project was initiated.

This research was supported in part by the U.S. National Science Foundation (NSF) under grants PHY-1748958 and AST-2107070.
Financial support for K.J.S. was in part provided by NASA/ESA {\it Hubble Space Telescope} programs \#15871 and \#15918. 
J.S. acknowledges support from the Packard Foundation. 
R.P.N. acknowledges support for this work provided by NASA through Hubble Fellowship grant HST-HF2-51515.001-A awarded by the Space Telescope Science Institute, which is operated by the Association of Universities for Research in Astronomy, Inc., under NASA contract NAS5-26555.
A.V.F.'s group at U.C. Berkeley has been supported by the Christopher R. Redlich Fund, Alan Eustace (W.Z. is a Eustace Specialist in Astronomy), Frank and Kathleen Wood (T.G.B. is a Wood Specialist in Astronomy), and numerous other donors. This project has received funding from the European Research Council (ERC) under the European Union’s Horizon 2020 research and innovation program (grant agreement \#101020057).

We thank the staffs of the various observatories at which data were obtained.
Partially based on observations obtained at the Southern Astrophysical Research (SOAR) telescope, which is a joint project of the Minist\'{e}rio da Ci\^{e}ncia, Tecnologia e Inova\c{c}\~{o}es (MCTI/LNA) do Brasil, the US NSF's NOIRLab, the University of North Carolina at Chapel Hill (UNC), and Michigan State University (MSU). 
Some of the data presented herein were obtained at the W. M. Keck Observatory, which is operated as a scientific partnership among the California Institute of Technology, the University of California, and NASA; the observatory was made possible by the generous financial support of the W. M. Keck Foundation.
A major upgrade of the Kast spectrograph on the Shane 3\,m telescope at Lick Observatory, led by Brad Holden, was made possible through gifts from the Heising-Simons Foundation, William and Marina Kast, and the University of California Observatories. Research at Lick Observatory is partially supported by a generous gift from Google.

\newpage

\newpage

\bibliographystyle{mnras}

\appendix

\section{Follow-up Spectroscopy}
\label{sec:appendix}

\subsection{Additional candidates observed}

In addition to the objects in Table~\ref{tab:all_systems}, we obtained follow-up spectra of 21 candidates listed in Table~\ref{tab:query2}. Compared to the candidates in Table~\ref{tab:query}, these objects have either lower proper motions ($\mu = 30-50\,\rm mas\,yr^{-1}$ instead of $\mu > 50\,\rm mas^{-1}$) or $\varpi/\sigma_{\varpi}> 5$. None of these objects turned out to be good candidates: they all have $\rm |RV| < 500\,\rm km\,s^{-1}$ and almost all show hydrogen lines in their spectra.

\begin{table*}
\begin{tabular}{llllllllll}
{\it Gaia} DR3 Source ID & $G$ & $G_{\rm BP}-G_{\rm RP}$ & $\varpi$ & $\mu$ & $4.74\mu/\varpi$ & $\frac{4.74\mu}{\varpi+\sigma_{\varpi}}$ & RV & verdict & instrument \\
 & [mag] & [mag] & [mas] & [$\rm mas\,yr^{-1}$] & [$\rm km\,s^{-1}$] & [$\rm km\,s^{-1}$] & [$\rm km\,s^{-1}$]  &  &  \\

\hline
6753874560165966848 & 16.15 & -0.32 & $0.24 \pm 0.06$ & 31.6 & 633 & 509  & $70\pm 30$ & sdO/B & SOAR \\
6762185287532572928 & 16.35 & -0.09 & $0.28 \pm 0.07$ & 36.0 & 612 & 492  & $-8\pm 10$ & sdO/B & MagE \\
4187869370201203840 & 16.56 & -0.17 & $0.16 \pm 0.07$ & 36.1 & 1047 & 745 & $-246\pm 10$ & sdO/B & MagE \\
1293316877344107776 & 16.61 & -0.54 & $0.24 \pm 0.06$ & 31.3 & 624 & 501  & $-174 \pm 20$ & sdO/B & DBSP \\
6368583523760274176 & 16.87 & -0.37 & $0.31 \pm 0.07$ & 36.6 & 564 & 465  & $-80\pm 30$ & sdO/B & SOAR \\
6110450457358521856 & 17.15 & -0.41 & $0.36 \pm 0.09$ & 33.2 & 442 & 356  & $220\pm 30$ & sdO/B & SOAR \\
6080702345638173056 & 17.19 & -0.23 & $0.29 \pm 0.08$ & 36.5 & 589 & 462  & $34\pm 10$ & sdO/B & MagE \\
4185588433336966272 & 17.43 & -0.07 & $0.34 \pm 0.11$ & 33.1 & 464 & 348  & $-140\pm 30$ & sdO/B & SOAR \\
3537042874067950336 & 17.49 & -0.38 & $1.33 \pm 0.10$ & 123.8 & 439 & 408 & $360\pm 30$ & DA WD & SOAR \\
6247253167253957632 & 17.57 & -0.15 & $0.35 \pm 0.11$ & 41.3 & 558 & 429  & $88\pm 10$ & sdO/B & MagE \\
6866912945436640384 & 17.67 & -0.22 & $0.29 \pm 0.10$ & 34.2 & 552 & 415  & $10\pm 30$ & sdO/B & SOAR \\
6640949596389193856 & 17.88 & 0.29 & $1.74 \pm 0.16$ & 193.2 & 525 & 482  & $270\pm 50$ & DA WD & SOAR \\
5921767076544699648 & 18.44 & 0.22 & $0.53 \pm 0.18$ & 52.8 & 468 & 349   & $446\pm 20$ & DA WD & MagE \\
269073928658921984 & 18.46 & -0.12 & $0.67 \pm 0.17$ & 83.7 & 593 & 472   & $-190\pm 50$ & DA WD & Kast \\
4304107918421908352 & 18.60 & -0.28 & $0.36 \pm 0.18$ & 45.9 & 611 & 408  & $-170\pm 30$ & sdO/B & LRIS \\
4180185566327070208 & 19.20 & 0.74 & $-0.10 \pm 0.46$ & 37.7 & -1880 &488& $-20\pm 30$ &  MS star & SOAR \\
2432538119174778368 & 19.35 & 0.02 & $0.13\pm 0.32$ & 32.1 & 1162 & 334  & $-100\pm 50$ & DA WD & SOAR \\
6314180989790944128 & 19.42 & -0.23 & $0.45 \pm 0.35$ & 60.8 & 637 & 360 & $-120\pm 50$ & DA WD & SOAR \\
3751658090584489344 & 19.44 & 0.90 & $-0.01 \pm 0.53$ & 33.1 & -18776 & 301 & $330\pm 30$ & MS star & SOAR \\
643759866174825088 & 19.48 & 0.86 & $1.64 \pm 0.42$ & 141.5 & 409 & 325 & $-36\pm 50$ & DA WD & Kast \\
4043027714240237440 & 19.49 & 0.64 & $0.48 \pm 0.39$ & 55.9 & 554 & 305 & $200\pm 30$ & MS star & SOAR \\

\end{tabular}
\caption{Additional candidates we observed. Compared to those in Table~\ref{tab:query}, these have lower proper motions and/or higher-significance parallaxes. None turned out to be strong thermonuclear runaway candidates. }
\label{tab:query2}
\end{table*}

\subsection{Summary of observations and data reduction}

\subsubsection{MagE}
\label{sec:MagE}
We observed 9 candidates using the Magellan Echellette spectrograph \citep[MagE;][]{Marshall2008}  on the 6.5\,m Magellan Baade telescope at Las Campanas Observatory. All observations were carried out with the $0.7''$-wide slit, yielding a spectral resolution $R \approx 5500$ and wavelength coverage of 3500--11,000\,\AA. 

We reduced the spectra using \texttt{PypeIt} \citep[][]{Prochaska_2020}, which performs bias and flat-field correction, cosmic-ray removal, wavelength calibration, sky subtraction, extraction of 1D spectra, merging of spectral orders, and heliocentric RV corrections. We verified the stability of the wavelength solution at the $\sim 5\,\rm km\,s^{-1}$ level using observations of RV standards and telluric absorption lines.

\subsubsection{LRIS}
We observed 9 candidates using the Low Resolution Imaging Spectrometer \citep[LRIS;][]{Oke1995} on the 10\,m Keck-I telescope on Maunakea. Some of the observations used the 600/7500 grating on the red side and the 600/4000 grism on the blue side, resulting in a full width at half-maximum intensity (FWHM) of 4.7\,\AA\ on the red side and 4.0\,\AA\ on the blue side, or a typical resolution $R\approx 1500$. Other observations used the 400/8500 grating on the red side and the 400/3400 grism on the blue side, resulting in FWHM $\approx 6.9$\,\AA\ on the red side and 6.5\,\AA\ on the blue side, or a typical resolution  $R\approx 1000$. LRIS is equipped with an atmospheric dispersion corrector.

We reduced the data using the LRIS automated reduction pipeline \citep[LPIPE;][]{Perley2019}, which performs bias and flat-field corrections, cosmic-ray removal, wavelength calibration and flexure corrections using night-sky lines, extraction of 1D spectra, telluric corrections, and flux calibration using a standard star. We adopted a minimum RV uncertainty of $\rm 10\,km\,s^{-1}$. 

We observed one source, J0546+0836, for a total of 6000\,s because it is faint and has very weak absorption lines. We obtained 10 exposures, each with an exposure time of 600\,s, and coadded them to improve the SNR. 

\subsubsection{GMOS}
\label{sec:GMOS}
We observed three candidates using the Gemini Multi-Object Spectrograph \citep[GMOS;][]{Hook2004} on the 8.1\,m Gemini-South telescope at Cerro Pachón (program GS-2023A-FT-107). We used the B1200\_G5321 grating with a $0.75''$-wide slit and central wavelength of 4500\,\AA, leading to wavelength coverage from 3700 to 5300\,\AA\ and resolution $R\approx 3000$. We obtained a spectrum of a CuAr comparison lamp for wavelength calibration on-sky immediately after each 900\,s science exposure. The data were reduced using \texttt{PypeIt}. 

\begin{table*}
\centering
\begin{tabular}{lllll}
Target & Instrument & Resolution & MJD & Exposure time (s)  \\
\hline
J1235-3752 & MagE & 5500 & 59983.26 & 1800 \\
J1332-3541 & MagE & 5500 & 59983.28 & 1800 \\
J1332-3541 & LRIS & 1000 & 60054.52 & 900 \\
J0927-6335 & MagE & 5500 & 59933.34 & 900 \\
J0927-6335 & MagE & 5500 & 59974.21 & 1800 \\
J0546+0836 & LRIS & 1500 & 59959.35 & $10\times 600$ \\ 
J0546+0836 & ESI & 8000 & 59990.36 & $3\times 1200$ \\ 
J1311-1846 & LRIS & 1000 & 60028.51 & 600 \\ 
J1109+0001 & LRIS & 1500 & 59906.59 & 1200 \\ 

\end{tabular}
\caption{Observing log for the six targets confirmed to be hydrogen-free.}
\label{tab:obslog}
\end{table*}

\subsubsection{ESI}
\label{sec:ESI}
We observed J0546+0836 with the Echellette Spectrograph and Imager \citep[ESI;][]{Sheinis2002} on the 10\,m Keck-II telescope on Maunakea. We used the $0.5''$-wide slit, yielding a resolution $R\approx 8000$, with a useful wavelength coverage of 3900--10,000\,\AA. We reduced the data using the MAuna Kea Echelle Extraction (MAKEE) pipeline, which performs bias subtraction, flat fielding, wavelength calibration, and sky subtraction. The pipeline also carries out a linear shift to the wavelength solution using night-sky emission lines, which we have verified to yield RVs stable to $\sim 3\,\rm km\,s^{-1}$. 

\subsubsection{SOAR}
We observed 19 candidates with the Goodman Spectrograph \citep{Clemens2004} on the SOAR telescope.
All SOAR observations used a 1.2\arcsec\ slit and a 400\,line\,mm$^{-1}$ grating, giving an approximate wavelength coverage of 4000--7850\,\AA\ with a FWHM resolution of $\sim 6.6$\,\AA. For each candidate we obtained a single exposure of length 1200--1800\,s, using longer exposures for fainter targets. All spectra were reduced and optimally extracted in the standard manner. The wavelength calibration was performed with comparison-lamp spectra obtained immediately after each science exposure, and a first-order flux calibration was applied to each spectrum.

\subsubsection{DBSP}
We observed four candidates with the Double Spectrograph \citep[DBSP;][]{Oke1982} on the 5\,m Hale telescope at Palomar Observatory. The 600/4000 grating was used on the blue side and the 316/7500 grating on the red side, together with the D55 dichroic. We employed the $1.0''$ or $1.5''$ slit, depending on the seeing, yielding an average resolution $R\approx 1500$ and wavelength coverage of 3000--10,700\,\AA. 

We reduced the spectra using \texttt{PypeIt} \citep[][]{Prochaska_2020} with the DBSP\_DRP wrapper \citep{Mandigo-Stoba2022}. This performs bias and flat-field corrections, cosmic-ray removal, wavelength calibration, extraction of 1D spectra, telluric corrections, and flux calibration using a standard star. We corrected for flexure using telluric absorption lines, as described by \citet{Nagarajan2023}.

\subsubsection{Kast}
We observed two candidates with the Kast double spectrograph \citep{Miller1994} mounted on the Shane 3\,m telescope at Lick Observatory. These observations used the $2''$-wide slit, 600/4310 grism, and 300/7500 grating. This instrument configuration has a combined wavelength range of 3500–-10,500\,\AA, and a spectral resolving power of $R\approx 800$.  To minimize slit losses caused by atmospheric dispersion \citep{Filippenko1982}, the slit was oriented at or near the parallactic angle. The data were reduced following standard techniques for CCD processing and spectrum extraction \citep{Silverman2012} utilizing IRAF \citep{Tody1986} routines and custom Python and IDL codes.\footnote{https://github.com/ishivvers/TheKastShiv}  Low-order polynomial fits to comparison-lamp spectra were used to calibrate the wavelength scale, and small adjustments derived from night-sky lines in the target frames were applied. The spectra were flux calibrated and telluric corrected using spectra of appropriate spectrophotometric standard stars observed on the same night, at similar airmasses, and with an identical instrument configuration.

\section{Constraints with different priors}
\label{sec:appendix_priors}

We list results of kinematic modeling with alternative distance priors in Tables~\ref{tab:all_systems_EDSD} and~\ref{tab:all_systems_flat_distanc}.

\begin{table*}
\begin{tabular}{llllllllll}
Name & {\it Gaia} DR3 Source ID &  $G$  & $G_{\rm BP}-G_{\rm RP}$ & $\rm RV$  & 
$d$ & $v_{\rm tot}$ & $v_{\rm ejection}$ & $z$ & $t_{\rm disk} = z/v_z$ \\
  &   &  [mag] &  [mag] & $\rm [km\,s^{-1}]$  & 
$[\rm kpc]$ & $[\rm\,km\,s^{-1}]$ & $[\rm\,km\,s^{-1}]$ & $[\rm kpc]$ & $[\rm Myr]$ \\
\hline
\hline 
\multicolumn{10}{l}{\underline{Suspected $\rm D^6$ stars}}   \\ 
D6-1 & 5805243926609660032 & 17.4 & 0.48 & $1200 \pm 40$ & $1.97^{+0.30}_{-0.24}$ & $2094^{+266}_{-205}$ & $2303^{+263}_{-202}$ & $-0.59^{+0.07}_{-0.09}$ & $0.64^{+0.04}_{-0.04}$ \\ 
D6-2 & 1798008584396457088 & 17.0 & 0.41 & $80 \pm 10$ & $0.85^{+0.05}_{-0.04}$ & $1156^{+59}_{-52}$ & $1057^{+62}_{-55}$ & $-0.27^{+0.02}_{-0.02}$ & $-0.72^{+0.01}_{-0.01}$ \\ 
D6-3 & 2156908318076164224 & 18.2 & 0.43 & $-20 \pm 80$ & $2.34^{+0.31}_{-0.35}$ & $2326^{+309}_{-349}$ & $2480^{+342}_{-387}$ & $0.96^{+0.12}_{-0.14}$ & $2.24^{+0.19}_{-0.16}$ \\ 
{\bf J1235} & 6156470924553703552 & 19.0 & -0.28 & $-1694 \pm 10$ & $4.14^{+0.94}_{-1.09}$ & $2693^{+316}_{-326}$ & $2495^{+327}_{-326}$ & $1.76^{+0.39}_{-0.46}$ & $1.85^{+0.68}_{-0.23}$ \\ 
{\bf J0927} & 5250394728194220800 & 19.4 & -0.32 & $-2285 \pm 20$ & $4.35^{+1.89}_{-1.41}$ & $2736^{+225}_{-121}$ & $2501^{+225}_{-120}$ & $-0.67^{+0.22}_{-0.30}$ & $-1.29^{+0.35}_{-0.38}$ \\ 
{\bf J0546} & 3335306915849417984 & 19.1 & -0.25 & $1200 \pm 20$ & $4.17^{+1.89}_{-1.47}$ & $1740^{+563}_{-362}$ & $1906^{+571}_{-382}$ & $-0.71^{+0.26}_{-0.33}$ & $0.61^{+0.04}_{-0.06}$ \\ 
{\bf J1332} & 6164642052589392512 & 19.4 & -0.55 & $1090 \pm 50$ & $2.40^{+0.94}_{-0.93}$ & $1932^{+624}_{-553}$ & $2065^{+597}_{-529}$ & $1.09^{+0.42}_{-0.41}$ & $0.74^{+0.07}_{-0.12}$ \\ 
\multicolumn{10}{l}{\underline{Suspected LP 40-365 stars}}   \\ 
LP 40-365 & 1711956376295435520 & 15.6 & 0.23 & $498 \pm 5$ & $0.61^{+0.01}_{-0.01}$ & $837^{+5}_{-5}$ & $607^{+5}_{-5}$ & $0.43^{+0.01}_{-0.01}$ & $4.88^{+0.38}_{-0.33}$ \\ 
J1603 & 5822236741381879040 & 17.8 & 0.16 & $-480 \pm 5$ & $2.17^{+0.53}_{-0.35}$ & $846^{+70}_{-42}$ & $621^{+82}_{-48}$ & $-0.37^{+0.06}_{-0.10}$ & $1.55^{+0.11}_{-0.10}$ \\ 
J0905 & 688380457508503040 & 19.6 & 0.24 & $300 \pm 50$ & $3.76^{+2.49}_{-1.61}$ & $424^{+343}_{-162}$ & $641^{+347}_{-201}$ & $2.43^{+1.59}_{-1.03}$ & $-18.30^{+54.31}_{-33.81}$ \\ 
J1825 & 6727110900983876096 & 13.3 & -0.02 & $-47 \pm 5$ & $0.95^{+0.03}_{-0.02}$ & $429^{+18}_{-17}$ & $662^{+18}_{-17}$ & $-0.17^{+0.00}_{-0.01}$ & $1.60^{+0.02}_{-0.02}$ \\ 
{\bf J1311} & 3507697866498687232 & 18.3 & 0.35 & $55 \pm 10$ & $2.30^{+1.34}_{-0.71}$ & $1120^{+527}_{-278}$ & $1007^{+562}_{-300}$ & $1.61^{+0.93}_{-0.49}$ & $3.97^{+0.17}_{-0.19}$ \\ 
{\bf J1109} & 3804182280735442560 & 19.1 & -0.09 & $100 \pm 10$ & $3.34^{+1.83}_{-1.27}$ & $1561^{+760}_{-525}$ & $1359^{+790}_{-534}$ & $2.70^{+1.47}_{-1.02}$ & $3.99^{+0.19}_{-0.29}$ \\ 
\multicolumn{10}{l}{\underline{Suspected runaway helium star donors}}   \\ 
US 708 & 815106177700219392 & 18.9 & -0.44 & $917 \pm 7$ & $8.54^{+0.99}_{-0.99}$ & $996^{+10}_{-10}$ & $830^{+17}_{-15}$ & $6.29^{+0.73}_{-0.73}$ & $11.67^{+1.82}_{-1.69}$ \\ 
\multicolumn{10}{l}{\underline{Other/unknown}}   \\ 
J1240 & 1682129610835350400 & 18.4 & -0.29 & $-177 \pm 10$ & $0.43^{+0.02}_{-0.02}$ & $241^{+21}_{-19}$ & $449^{+22}_{-19}$ & $0.35^{+0.02}_{-0.02}$ & $-$ \\ 
J1637 & 1327920737357113088 & 20.3 & -0.13 & $300 \pm 50$ & $3.65^{+2.29}_{-1.54}$ & $1314^{+692}_{-433}$ & $1134^{+921}_{-420}$ & $2.45^{+1.52}_{-1.03}$ & $4.43^{+0.77}_{-0.97}$ \\ 

\end{tabular}
\caption{Same as Table~\ref{tab:all_systems}, but inferred with an exponentially decreasing space density distance prior with $L=1.35$\,kpc. }
\label{tab:all_systems_EDSD}
\end{table*}

\begin{table*}
\begin{tabular}{llllllllll}
Name & {\it Gaia} DR3 Source ID &  $G$  & $G_{\rm BP}-G_{\rm RP}$ & $\rm RV$  & 
$d$ & $v_{\rm tot}$ & $v_{\rm ejection}$ & $z$ & $t_{\rm disk} = z/v_z$ \\
  &   &  [mag] &  [mag] & $\rm [km\,s^{-1}]$  & 
$[\rm kpc]$ & $[\rm\,km\,s^{-1}]$ & $[\rm\,km\,s^{-1}]$ & $[\rm kpc]$ & $[\rm Myr]$ \\
\hline
\hline 
\multicolumn{10}{l}{\underline{Suspected $\rm D^6$ stars}}   \\ 
D6-1 & 5805243926609660032 & 17.4 & 0.48 & $1200 \pm 40$ & $1.94^{+0.30}_{-0.23}$ & $2064^{+267}_{-199}$ & $2273^{+263}_{-196}$ & $-0.58^{+0.07}_{-0.09}$ & $0.64^{+0.04}_{-0.04}$ \\ 
D6-2 & 1798008584396457088 & 17.0 & 0.41 & $80 \pm 10$ & $0.84^{+0.05}_{-0.04}$ & $1151^{+59}_{-52}$ & $1052^{+62}_{-55}$ & $-0.27^{+0.02}_{-0.02}$ & $-0.72^{+0.01}_{-0.01}$ \\ 
D6-3 & 2156908318076164224 & 18.2 & 0.43 & $-20 \pm 80$ & $2.31^{+0.32}_{-0.36}$ & $2297^{+324}_{-359}$ & $2448^{+357}_{-396}$ & $0.95^{+0.13}_{-0.14}$ & $2.24^{+0.19}_{-0.16}$ \\ 
{\bf J1235} & 6156470924553703552 & 19.0 & -0.28 & $-1694 \pm 10$ & $4.42^{+0.81}_{-1.17}$ & $2781^{+278}_{-360}$ & $2586^{+288}_{-366}$ & $1.87^{+0.34}_{-0.49}$ & $1.77^{+0.57}_{-0.17}$ \\ 
{\bf J0927} & 5250394728194220800 & 19.4 & -0.32 & $-2285 \pm 20$ & $5.40^{+1.82}_{-1.99}$ & $2851^{+251}_{-201}$ & $2617^{+250}_{-200}$ & $-0.84^{+0.32}_{-0.29}$ & $-1.50^{+0.44}_{-0.33}$ \\ 
{\bf J0546} & 3335306915849417984 & 19.1 & -0.25 & $1200 \pm 20$ & $5.11^{+1.76}_{-2.02}$ & $2012^{+551}_{-546}$ & $2184^{+556}_{-565}$ & $-0.88^{+0.36}_{-0.31}$ & $0.64^{+0.03}_{-0.07}$ \\ 
{\bf J1332} & 6164642052589392512 & 19.4 & -0.55 & $1090 \pm 50$ & $2.21^{+1.08}_{-0.97}$ & $1808^{+709}_{-545}$ & $1947^{+678}_{-521}$ & $1.00^{+0.48}_{-0.43}$ & $0.72^{+0.09}_{-0.15}$ \\ 
\multicolumn{10}{l}{\underline{Suspected LP 40-365 stars}}   \\ 
LP 40-365 & 1711956376295435520 & 15.6 & 0.23 & $498 \pm 5$ & $0.61^{+0.01}_{-0.01}$ & $837^{+5}_{-5}$ & $607^{+5}_{-5}$ & $0.43^{+0.01}_{-0.01}$ & $4.87^{+0.39}_{-0.33}$ \\ 
J1603 & 5822236741381879040 & 17.8 & 0.16 & $-480 \pm 5$ & $2.17^{+0.56}_{-0.36}$ & $845^{+73}_{-43}$ & $620^{+86}_{-49}$ & $-0.37^{+0.06}_{-0.10}$ & $1.55^{+0.12}_{-0.10}$ \\ 
J0905 & 688380457508503040 & 19.6 & 0.24 & $300 \pm 50$ & $8.03^{+4.71}_{-4.64}$ & $1031^{+712}_{-652}$ & $1252^{+709}_{-658}$ & $5.17^{+3.02}_{-2.97}$ & $-15.45^{+3.54}_{-10.85}$ \\ 
J1825 & 6727110900983876096 & 13.3 & -0.02 & $-47 \pm 5$ & $0.95^{+0.03}_{-0.03}$ & $429^{+18}_{-17}$ & $662^{+18}_{-17}$ & $-0.17^{+0.01}_{-0.01}$ & $1.60^{+0.02}_{-0.02}$ \\ 
{\bf J1311} & 3507697866498687232 & 18.3 & 0.35 & $55 \pm 10$ & $2.24^{+1.75}_{-0.73}$ & $1095^{+688}_{-285}$ & $981^{+732}_{-307}$ & $1.57^{+1.21}_{-0.50}$ & $3.96^{+0.19}_{-0.21}$ \\ 
{\bf J1109} & 3804182280735442560 & 19.1 & -0.09 & $100 \pm 10$ & $3.96^{+2.08}_{-1.80}$ & $1820^{+863}_{-745}$ & $1627^{+897}_{-763}$ & $3.20^{+1.67}_{-1.45}$ & $4.07^{+0.16}_{-0.33}$ \\ 
\multicolumn{10}{l}{\underline{Suspected runaway helium star donors}}   \\ 
US 708 & 815106177700219392 & 18.9 & -0.44 & $917 \pm 7$ & $8.54^{+0.98}_{-1.00}$ & $996^{+10}_{-10}$ & $831^{+17}_{-15}$ & $6.29^{+0.72}_{-0.73}$ & $11.68^{+1.80}_{-1.71}$ \\ 
\multicolumn{10}{l}{\underline{Other/unknown}}   \\ 
J1240 & 1682129610835350400 & 18.4 & -0.29 & $-177 \pm 10$ & $0.43^{+0.02}_{-0.02}$ & $241^{+21}_{-19}$ & $448^{+22}_{-19}$ & $0.35^{+0.02}_{-0.02}$ & $-$ \\ 
J1637 & 1327920737357113088 & 20.3 & -0.13 & $300 \pm 50$ & $4.93^{+2.70}_{-2.57}$ & $1695^{+837}_{-748}$ & $1567^{+1055}_{-785}$ & $3.30^{+1.80}_{-1.71}$ & $4.90^{+0.61}_{-1.24}$ \\

\end{tabular}
\caption{Same as Table~\ref{tab:all_systems}, but inferred with a flat distance prior. }
\label{tab:all_systems_flat_distanc}
\end{table*}

\end{document}